\def\AA            {{\!A}}
\def\Aeps          {{\rm A}_\varepsilon}
\def\alg           {algebra}
\newcommand\apppicture[2] {\put(#2,0) {\begin{picture}(0,0)(0,0)
                   \scalebox{.29}{\includegraphics{app#1.eps}} \end{picture}}}

\def\bc            {boundary condition}

\def\BE            {\begin{equation}}
\def\be            {\begin{equation}}
\def\bea           {\begin{equation}\begin{array}l}
\def\bearl         {\begin{array}{l}}
\def\bearll        {\begin{array}{ll}}

\def\C             {\ensuremath{\mathcal C}}
\def\cat           {category}
\def\cats          {categories}
\def\cblk          {\Gamma^{\rm bulk}}
\def\cbnd          {\Gamma^{\rm bnd}}
\def\cbndd         {\Upsilon^{\rm bnd}}

\def\Cbnd          {{\rm c}^{\rm bnd}}

\def\Cbulk         {{\rm c}^{\rm bulk}}
\def\cft           {conformal field theory}

\def\cfts          {conformal field theories}

\def\cir           {\,{\circ}\,}

\def\complex       {\ensuremath{\mathbb C}}

\def\Con           {Conformal }
\def\Corfu         {Correlation function}
\def\corfu         {correlation function}

\def\dim           {{\rm dim}}
\def\dsty          {\displaystyle }
\def\EE            {\end{equation}}

\def\ee            {\end{equation}}
\def\eear          {\end{array}}
\newcommand\eev[1] {{{}^{\vee\!}}{#1}}
\def\End           {{\rm End}}
\newcommand\epicture[2] {\end{picture}\\{}\\[#1.#2em]\end{array}}
\def\eps           {\varepsilon}

\def\eq            {\,{=}\,}
\newcommand\erf[1] {(\ref{#1})}
\newcommand\erp[2] {(\ref{V#1#2})}

\newcommand\Frac[2]{\mbox{\large$\frac{#1}{#2}$}}

\def\Hom           {{\rm Hom}}
\def\HomA          {{\rm Hom}_{\!A}}
\def\HomAA         {{\rm Hom}_{\!A|A}}

\newcommand\hsp[1] {\mbox{\hspace{#1 em}}}

\def\I             {{\mathcal I}}
\def\ib            {{\bar\imath}}
\def\id            {\mbox{\sl id}}

\def\iN            {\,{\in}\,}

\newcommand\includeourbeautifulpicture[2] {{\begin{picture}(0,0)(0,0)
                   \scalebox{.38}{\includegraphics{V-#1#2.eps}} \end{picture}}}
\newcommand\Includeourbeautifulpicture[3] {{\begin{picture}(0,0)(0,0)
                   \scalebox{.38}{\includegraphics{V-#1#2#3.eps}} \end{picture}}}
\newcommand\INcludeourbeautifulpicture[4] {{\begin{picture}(0,0)(0,0)
                   \scalebox{.38}{\includegraphics{V-#1#2#3#4.eps}} \end{picture}}}

\newcommand\includeournicelargepicture[2] {{\begin{picture}(0,0)(0,0)
            \scalebox{.28}{\includegraphics{V-#1#2.eps}} \end{picture}}}
\newcommand\Includeournicelargepicture[3] {{\begin{picture}(0,0)(0,0)
            \scalebox{.28}{\includegraphics{V-#1#2#3.eps}} \end{picture}}}
\newcommand\includeournicemediumpicture[2] {{\begin{picture}(0,0)(0,0)
            \scalebox{.32}{\includegraphics{V-#1#2.eps}} \end{picture}}}
\newcommand\Includeournicemediumpicture[3] {{\begin{picture}(0,0)(0,0)
            \scalebox{.32}{\includegraphics{V-#1#2#3.eps}} \end{picture}}}
\newcommand\includeournicesmallpicture[2] {{\begin{picture}(0,0)(0,0)
            \scalebox{.5}{\includegraphics{V-#1#2.eps}} \end{picture}}}

\def\Itemize       {\begin{itemize}}
\def\iX            {\ensuremath{\imath(\X)}}
\def\J             {\mbox{$\JJ$}}
\def\jb            {{\bar\jmath}}

\newcommand\labl[1]{\label{#1}\ee }
\newcommand\labp[2]{\label{V#1#2}\ee  }

\def\lhs           {left hand side}

\def\Llb           {\mbox{\Large(}}

\def\Lrb           {\mbox{\Large)}}

\def\M             {{\dot M}}
\newcommand\mcll   {\multicolumn2{|l|}}
\newcommand\mclll  {\multicolumn3{|l|}}
\newcommand\mclo   {\multicolumn1{l|}}

\def\Meps          {{\rm M\ddot o}_\varepsilon}
\def\Mf            {\ensuremath{{\rm M}_{f}}}

\def\MX            {\ensuremath{{\rm M}_{\rm X}}}

\def\NX            {\ensuremath{{\rm N}_{\rm X}}}
\def\nxt           {\raisebox{.08em}{\rule{.44em}{.44em}}\hsp{.4}}
\def\Nxt           {\raisebox{.08em}{\rule{.44em}{.44em}}}

\def\obj           {{\mathcal O}bj}
\def\objc          {{\mathcal O}bj(\C)}
\def\one           {{\bf1}}

\def\oro           {\ensuremath{{\rm or}_1}}
\def\ort           {\ensuremath{{\rm or}_2}}

\def\otA           {\,{\otimes}_{\!A}^{}\,}

\def\OtA           {{\otimes}_{\!A}^{}}
\newcommand\ot[1]  {\,{\otimes^{#1}}\,}
\newcommand\oT[1]  {\,{\otimes^{#1}}}
\def\oti           {\,{\otimes}\,}
\def\Oti           {{\otimes}}
\def\otic          {\,{\otimes_{\mathbb C}}\,}
\def\parfu         {partition function}
\def\q             {quantum }
\def\Q             {Quantum }
\def\r             {{\rho}}

\def\reals         {\ensuremath{\mathbb R}}

\def\Reps          {{\rm R}_\varepsilon}

\def\rhs           {right hand side}
\def\rmM           {{\rm M}}

\newcommand\sect[1]{\section{#1}\setcounter{equation}0%
                    \setcounter{theorem}0\setcounter{rmtheorem}0}

\def\sse           {\scriptsize}
\def\sss           {\scriptscriptstyle}
\def\StIbox        {{\begin{picture}(0,0)(0,0)
                   \scalebox{.38}{\includegraphics{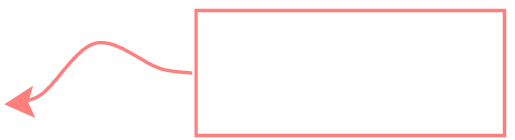}}
                   \put(-31.3,3.3){\small$S^2{\times}I$} \end{picture}}}
\def\Stn           {S^3{}_{\!\!\!\! {\rm n}.}}
\def\Stnbox        {{\begin{picture}(0,0)(0,0)
                   \scalebox{.38}{\includegraphics{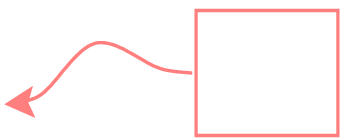}}
                   \put(-15.1,3.3){\small$\Stn$} \end{picture}}}
\def\StSebox       {{\begin{picture}(0,0)(0,0)
                   \scalebox{.38}{\includegraphics{box_for_S2S1.eps}}
                   \put(-33.3,3.3){\small$S^2{\times}S^1$} \end{picture}}}

\def\TAA           {{\rm T}_{\!A,A}}

\def\tft           {topological field theory}

\def\Times         {\,{\times}\,}
\def\To            {\,{\to}\,}

\def\Tr            {{\rm tr}}

\def\twodim        {two-di\-men\-si\-o\-nal}

\def\Vect          {{\mathcal V}\mbox{\sl ect}}
\def\Vee           {^{\vee}}

\newcommand\void[1]{}
\def\vPhi          {\varPhi}
\def\vPsi          {\varPsi}

\def\X             {{\ensuremath{\rm X}}}
\def\xcut          {{\cdot{\sss|}\cdot}}
\def\Xi            {\ensuremath{{\rm X}^\xcut_{i,\jb}}}

\def\Xh            {\ensuremath{\widehat{\X}}}

\def\Yh            {\ensuremath{\widehat{\rm Y}}}
\def\zet           {\mbox{$\mathbb Z$}}

\newcommand\dtRemark          {\begin{remark}}
\newcommand\dtlCorollary[1]   {\begin{corollary}\label{#1}}
\newcommand\dtlDefinition[1]  {\begin{definition}\label{#1}}
\newcommand\dtlLemma[1]       {\begin{lemma}\label{#1}}
\newcommand\dtlProposition[1] {\begin{proposition}\label{#1}}
\newcommand\dtlRemark[1]      {\begin{remark}\label{#1}}
\newcommand\dtlTheorem[1]     {\begin{theorem}\label{#1}}
\newcommand\ENDC    {\end{corollary}}
\newcommand\ENDD    {\end{definition}}
\newcommand\ENDL    {\end{lemma}}
\newcommand\ENDP    {\end{proposition}}
\newcommand\ENDR    {\end{remark}}
\newcommand\ENDT    {\end{theorem}}
\newcommand\Proof   {\proof}
\newcommand\qed     {\endproof}

\documentclass{tac}

\usepackage{latexsym,amsfonts,amsmath,amssymb,epsf}
\usepackage[mathscr]{eucal}
\usepackage[dvips]{graphics}
\usepackage{rotating}

\newcounter{theorem}
\newcounter{rmtheorem}

\begin{document}

\author{Jens Fjelstad, J\"urgen Fuchs, Ingo Runkel,\\ and Christoph Schweigert}
\address{Fachbereich Mathematik, Universit\"at Hamburg
         Schwerpunkt Algebra und Zahlentheorie,
     \\  and Zentrum f\"ur mathematische Physik,
         Bundesstra\ss e 55, D\,--\,20\,146\, Hamburg
\\[5pt]  Institutionen f\"or fysik, Karlstads Universitet,
     \\  Universitetsgatan 5, S\,--\,651\,88 Karlstad
\\[5pt]  Max-Planck-Institut f\"ur Gravitationsphysik,
         Albert-Einstein-Institut,
     \\  Am M\"uhlenberg 1, D\,--\,14\,476 Golm
\\[5pt]  Fachbereich Mathematik, Universit\"at Hamburg
         Schwerpunkt Algebra und Zahlentheorie,
     \\  and Zentrum f\"ur mathematische Physik,
         Bundesstra\ss e 55, D\,--\,20\,146 Hamburg}

\title{TFT CONSTRUCTION OF RCFT CORRELATORS
  \\   V: PROOF OF MODULAR INVARIANCE AND FACTORISATION}

\copyrightyear{2005}

\amsclass{81T40,18D10,18D35,81T45}

\eaddress{jens.fjelstad@kau.se\\jfuchs@fuchs.tekn.kau.se\\
          ingo.runkel@kcl.ac.uk\\schweigert@math.uni-hamburg.de}

\maketitle

\begin{abstract}
The correlators of two-dimensional rational conformal field theories that are 
obtained in the TFT construction of \cite{fuRs4,fuRs8,fuRs10} are shown to be 
invariant under the action of the relative modular group and to obey bulk and 
boundary factorisation constraints. We present results both for conformal 
field theories defined on oriented surfaces and for theories defined on 
unoriented surfaces. In the latter case, in particular the so-called cross 
cap constraint is included.
\end{abstract} 

\tableofcontents 

\sect{Introduction}

Quantum field theories have, deservedly, the reputation of being
both mathematically rich and difficult to address at a conceptual
level. This is, in fact, a major reason for the importance of
``exactly solvable'' models in quantum field theory. However, the notion 
of exact solvability needs some reflection: at a naive level, one might 
aim at computing ``all'' correlation functions -- i.e.\ for all fields in 
the theory and on all spaces on which the theory makes sense -- explicitly. 
It is not hard to see that this hope is unrealistic, even for a model
apparently as simple as the \twodim\ massless free boson on an arbitrary
conformal surface. (In this model one would need, for example, a rather 
detailed explicit knowledge of higher genus theta functions, and of the 
dependence of determinants on the moduli of the surface.)

A more realistic aim is to achieve the following: 
\\[.3em]
\nxt To establish the {\em existence\/} of a consistent collection of
     correlation functions, for all fields in the theory and all spaces on
     which the theory is supposed to make sense. 
\\[.3em]
\nxt To identify those features of correlation functions which at the same
     time are interesting and can be controlled.
\\
Specifically, one should have at one's disposal sufficiently powerful 
mathematical tools to establish general properties of these characteristics, 
and one should be able to set up efficient algorithms for computing them 
explicitly in specific models.

\medskip

These are the two goals for which the TFT approach to the correlators of 
(two-dimensional, euclidean) \cft\ -- the topic of the present series of 
papers --  has been designed. 
The TFT approach makes it possible to formulate and reach these goals in a 
purely categorical setting. In particular, all relevant information on the
problem is encoded in appropriate category theoretic structures, and thus
algebro-geometric and functional-analytic issues that must be dealt with
in other approaches to \cft\ are taken into account by assumptions on 
these category theoretic structures. 
Among the interesting and controllable features 
that can be computed in this approach are in particular the coefficients of the 
correlation functions in an expansion in a chosen basis of conformal blocks.
   
\medskip

Our approach to CFT correlators applies to all chiral conformal field theories 
for which the representation category of the underlying chiral algebra is a 
modular tensor category. The structure of a modular tensor category
on the representation category arises from the properties of 
(half-)monodromies of conformal blocks on surfaces of genus zero and one.
A modular tensor category, in turn, allows one to construct a three-dimensional 
topological field theory (TFT), which furnishes a tensor functor from a suitable 
cobordism category to a category of vector spaces 
(for some perinent details about TFT, see appendix \ref{app:3dTFT-MTC}).
In particular, it provides 
(projective) representations of the mapping class groups for surfaces of higher
genus. A central assumption in the TFT approach, which is known to be met in a 
large class of models, is that these representations are indeed the ones arising 
from the monodromies of conformal blocks on these surfaces.

Another essential ingredient in our construction is the principle of holomorphic
factorisation: the correlator $C(\X)$ on a two-dimensional world sheet \X\
is an element of the vector space $\mathcal H(\Xh)$ that the TFT
assigns to the double \Xh\ of \X: 
  $$ C(\X) \in \mathcal H(\Xh) \, . $$
(Here it should be kept in mind that the vector space $\mathcal H(\Xh)$ is
actually a space of multivalued functions of the marked points and of the complex 
structure of \Xh.) To this end, the surface \Xh\ must come with enough structure
so that it indeed constitutes an object in the cobordism category, i.e.\ an 
extended surface in the terminology of \cite{TUra}. In the TFT approach, the 
structure of \X\ that makes it into a valid world sheet encodes precisely the 
right amount of information such that its double \Xh\ has a canonical 
structure of an extended surface.

Two different types of local two-dimensional conformal field theories are 
covered by the TFT approach:
\\[.3em]
\nxt theories that are defined on oriented world sheets only (the surfaces
     are allowed to have non-empty boundary); and
\\[.3em]
\nxt theories that are defined on world sheets without orientation (the 
     surfaces are allowed to have non-empty boundary, and are not required to
     be orientable).
\\[.3em]
Apart from a modular tensor category \C, further input is required, and this
additional input is different for the two types of theories. In the oriented 
case, a {\em symmetric special Frobenius algebra\/} $A$ in \C\ 
contains sufficient information to construct all correlators \cite{fuRs,fuRs4}.
Morita equivalent algebras yield the same correlators.
For unoriented world sheets, as shown in \cite{fuRs8}, in order to have enough 
input for constructing all correlators one must require that the algebra is even 
a {\em Jandl algebra\/} $\tilde A$. (The structure of a Jandl algebra generalises
the notion of a complex algebra with involution;
relevant information about algebras in monoidal categories is collected in
appendix \ref{Astuff}.)

\medskip

With regard to the two aims for understanding a quantum field theory formulated
above, the objectives of the papers in this series are the following: 
\\[.2em]
\nxt In the papers \cite{fuRs4}, \cite{fuRs8} and \cite{fuRs10}, crucial 
     quantities of conformal quantum field theories -- partition functions
     \cite{fuRs4,fuRs8} and operator products \cite{fuRs10} -- were extracted
     and algorithms for computing them were presented.
\\[.2em]
\nxt The purpose of the paper \cite{fuRs9} was to establish a systematic
     procedure, based on abelian group cohomology, to obtain symmetric special
     Frobenius  algebras, and to specialise results from \cite{fuRs4,fuRs10}
     to the \cfts\ that are obtained by this procedure.
\\[.2em]
\nxt The aim of the present paper is to prove that the correlators obtained by
     our prescription satisfy the relevant consistency conditions, and thereby
     to establish the existence of a consistent set of correlators.

\medskip

Let us describe the consistency conditions for correlators in some detail. 
First of all, correlators must obey the so-called chiral Ward identities. 
In the present 
approach, these are taken into account by the postulate that the correlation 
function on \X\ is an element of the space $\mathcal H(\Xh)$ of conformal 
blocks on the double. (For the construction of such spaces of multivalued 
functions from a conformal vertex algebra, see e.g.\ \cite{FRbe}.)

There are, however, also two other requirements that must be imposed:
\\[.3em] \centerline{
{\em modular invariance\/} \hsp{1.4} and \hsp{1.4} {\em factorisation\/}.\hsp2}

\bigskip

By the axioms of TFT, the space $\mathcal H(\Xh)$ of conformal blocks
carries a projective representation of the mapping class group ${\rm Map}(\Xh)$
of \Xh\ \cite[chap.\,IV:5]{TUra}.
The double \Xh\ comes with an orientation reversing involution $\tau$
such that \X\ is the quotient of \Xh\ by the action of the group 
$\langle \tau\rangle$ generated by $\tau$. The mapping class group 
${\rm Map}(\X)$ of \X\ can be identified with the subgroup of ${\rm Map}(\Xh)$ 
that commutes with $\tau$. The latter group -- termed {\em relative modular
group\/} in \cite{bisa2} -- acts genuinely on the vector space $\mathcal H(\Xh)$.
The requirement of modular invariance is the postulate that $C(\X)$ should be 
invariant under the action of this group. (For the precise statements see 
section \ref{sec:inv-map}. Actually, modular invariance can be generalised
to a covariance property of correlation functions, see the theorems
\ref{thm:iso-or} and \ref{thm:iso-unor}.)

The modular invariance property not only implies that the torus partition 
function is modular invariant in the ordinary sense. It also implies that the
correlator $C(\X)$ (a vector in the space $\mathcal H(\Xh)$ of multivalued 
functions) is in fact a single-valued function,
both of the insertion points and (up to the Weyl anomaly) of the conformal 
structure of \X. This is indeed the motivation to impose this constraint.

As emphasised above, we distinguish between two different types of conformal 
field theories, depending on whether oriented or unoriented surfaces are 
considered. This difference, too, is relevant for the issue of modular 
invariance: in the oriented case, only orientation preserving maps from \X\ to 
\X\ are admitted, while for the unoriented theory this requirement is dropped.
As a consequence, modular invariance for the oriented torus without 
field insertions is invariance of the torus partition function under the action
of ${\rm SL}(2,\zet)$, whereas in the unoriented case, in addition the symmetry
of the partition function under exchange of left- and right-movers is implied.

\medskip

Up to now we have only discussed constraints that involve a single world sheet
\X. Factorisation constraints relate world sheets of different topology, and 
are thus more subtle and more powerful. There are two types of factorisation 
constraints, boundary factorisation (present when $\partial\X\,{\ne}\,\emptyset$)
and bulk factorisation.  We formulate boundary factorisation as follows:
Consider an interval in \X\ with end points on the boundary $\partial\X$.  
Cut \X\ along this interval, and glue to the two newly created boundary 
segments the chords of two semi-disks with one boundary field insertion each.
This yields a new world sheet with two additional boundary insertions. Boundary
factorisation relates the correlator for this new world sheet and (the inverse 
of) the two-point function on the disk to the correlator for the original 
world sheet. (For the precise statements, see section \ref{sec:facbdry}, 
in particular theorems \ref{thm:bnd-or} and \ref{thm:bnd-unor}.)

Bulk factorisation works in a similar manner. Consider a
circle $S^1$ embedded in \X\ admitting an oriented neighbourhood.
Cut \X\ along this circle, and glue to the two newly created boundary
circles the equators of two hemispheres with one bulk field insertion each.
This yields a new world sheet with two additional bulk insertions. Bulk 
factorisation relates the correlator for this new world sheet and (the inverse 
of) the two-point function on the sphere to the correlator of the original world 
sheet. (For the precise statements, see section \ref{sec:facbulk}, in particular 
theorems \ref{thm:bulk-or} and \ref{thm:bulk-unor}.)
 
For unoriented world sheets \X, it can happen that an embedded circle 
$S^1\,{\subset}\,\X$ does not admit an oriented neighbourhood.
Cutting along such a circle and gluing in a hemisphere produces a world 
sheet with just a single additional bulk field insertion. We deduce from the 
previous results that there is an identity relating the 
correlator for this new world sheet, (the inverse of) the two-point
function on the sphere and the one-point function on the cross cap
$\mathbb{RP}^2$ to the correlator of the original world sheet. As a particular 
case, this implies the cross cap constraint studied in \cite{fips}.

\medskip

As an important consequence of the factorisation constraints, one can
cut every world sheet into a few simple building blocks and reconstruct
the correlators from a few fundamental correlators on these particular surfaces. 
This explains on the one hand the importance of the correlators considered 
in \cite{fuRs10}, and on the other hand it proves that they indeed solve the 
sewing constraints that were formulated in \cite{sono2,sono3,cale,lewe3}.

Factorisation constraints can be motivated by a Hamiltonian formulation of the
theory: locally around the circle or interval along which \X\ is cut, the 
world sheet looks like $S^1\,{\times}\, (-\epsilon,\epsilon)$ and as $[-1,1]\,
{\times}\,(-\epsilon,\epsilon)$, respectively. In a Hamiltonian description, 
one would like to associate dual topological vector spaces 
$\mathcal H_{{\rm bulk}}^\pm$ to circles $S^1$ with opposite orientations, and 
$\mathcal H_{{\rm bnd}}^\pm$ to intervals $[-1,1]$ 
with opposite orientations. Very much like when defining 
a path integral for a particle using a Wiener measure, one should then
be allowed to insert a complete system of states at a fixed ``time''
$t\iN(-\epsilon,\epsilon)$ and sum over it, without changing the result.

Summing over intermediate states is closely related to the concept of an 
operator product expansion: Suppose that two insertions, say in the bulk of \X, 
are close. Then they can be encircled by an $S^1$ along which we apply bulk 
factorisation. This way, the situation is related to a world sheet with one 
insertion less and to the three-point correlator on the sphere, which determines 
the OPE coefficients. This statement can be regarded as the best possible 
counterpart in the TFT approach of 
the fact that the OPE controls the short distance behaviour in all situations.

\medskip

This paper is organised as follows. Section \ref{labelsec2} contains precise 
statements of our main results: invariance under the mapping class group, and 
bulk and boundary factorisation. In sections 
\ref{sec:modinv}\,--\,\ref{sec:bulkfac} detailed proofs of these three types of
consistency conditions are presented, both for oriented and for unoriented 
world sheets. 
In appendix \ref{variousoldstuff} some basic information about modular
tensor categories, about algebras in monoidal categories, and about
topological field theory is recalled.
Appendix \ref{The ass} is devoted to the definition of
the vector $C(\X)\iN\mathcal H(\Xh)$ assigned to the correlator for a world 
sheet \X, using suitable cobordisms from $\emptyset$ to \Xh. Non-degeneracy 
of the two-point correlators on a disk and on a sphere enters crucially in the 
formulation of boundary and bulk factorisation, respectively; it is established 
in appendix \ref{labelappB}. Finally, appendix \ref{sec:morphprop} provides a 
few specific properties of certain morphisms which are employed in the proofs.

When referring to \cite{fuRs4,fuRs8,fuRs9,fuRs10}, we use the numbering of
sections, theorems, equations etc.\ as in those papers, preceded by a prefix
``I:" for \cite{fuRs4}, `II:" for \cite{fuRs8}, etc.; thus e.g.\ ``lemma IV:2.2"
refers to lemma 2.2 in \cite{fuRs10}, while (II:2.26) stands for equation (2.26)
in \cite{fuRs8}.

\medskip

\noindent
{\bf Acknowledgements.}
We are indebted to N.\ Potylitsina-Kube for her skillful help with the 
illustrations. J.Fu.\ is supported by VR under project no.\ 621--2003--2385,
and C.S.\ is supported by the DFG project SCHW 1162/1-1.


\sect{Consistency conditions for CFT correlators} \label{labelsec2}

Let us select once and for all a modular tensor category \C\ as
well as the following data:
\\[2pt]
\nxt a symmetric special Frobenius algebra $A$ in \C;
\\[2pt]
\nxt a symmetric special Frobenius algebra with reversion (a Jandl algebra)
$\tilde A$ in \C.
\\[2pt]
An explanation of these terms can e.g.\ be found in definitions I:3.3, I:3.4 
and II:2.1; 
for convenience, we repeat these definitions in appendix \ref{variousoldstuff}.
Also note that, by omitting the reversion, every Jandl algebra
provides us with a symmetric special Frobenius algebra, but not
every symmetric special Frobenius algebra admits a reversion.

The structure of a symmetric special Frobenius algebra is needed to study 
correlators on oriented world sheets, while the structure of a Jandl algebra is 
needed to investigate correlators on unoriented world sheets. More specifically,
the TFT construction of correlation functions that was proposed in \cite{fuRs} 
and developed in \cite{fuRs4,fuRs8,fuRs10} provides us with assignments
  \be
  C_\AA : \quad \X \mapsto C_\AA(\X) \qquad {\rm and} \qquad
  C_{\!\tilde A} : \quad  \tilde{\rm X} \mapsto C_{\!\tilde A}(\tilde{\rm X})
  \labl C
which take an oriented world sheet $\X$ to a vector $C_\AA(\X)\iN\mathcal H
(\Xh)$ in the space of blocks on the double \Xh\ of \X, and an unoriented world 
sheet $\tilde{\rm X}$ to a vector $C_{\!\tilde A}(\tilde{\rm X})\iN\mathcal{H}
(\widehat{\tilde{\rm X}})$. The precise definition of oriented and unoriented 
world sheets that is to be used here, as well as the notion of the double of 
a world sheet, is recalled in appendix \ref{app:worldsheet}. The assignments 
\erf{C} are summarised in appendix \ref{A2}. When we want to describe `one and 
the same \cft' both on oriented and on unoriented world sheets, then the 
relevant symmetric special Frobenius algebra $A$ is the one obtained from the 
Jandl algebra $\tilde A$ by forgetting the reversion, so that if an unoriented 
world sheet $\tilde\X$ is obtained from an oriented world sheet \X\ by 
forgetting the orientation, then $C_{\!\tilde A}(\tilde\X) \eq C_\AA(\X)$.
       
In the present section we formulate the assertions that the systems of
correlators $C_\AA$ and $C_{\!\tilde A}$ are invariant under the
action of the mapping class groups and that they are consistent with
factorisation. The proofs of these claims are presented in sections
\ref{sec:modinv}\,--\,\ref{sec:bulkfac}.

To be precise, we consider here correlators with any number of bulk and 
boundary field insertions, and with arbitrary boundary conditions preserving 
the underlying chiral symmetry.
On the other hand, the formalism of \cite{fuRs4,fuRs8,fuRs9,fuRs10} allows in
addition for the treatment of conformal defects and defect fields. While
the proofs given here are for correlators without such defects, our
methods allow one to establish the consistency of correlators in the presence of
defects as well, albeit various aspects of the consistency conditions 
and their proofs become a lot more involved. In particular, one must introduce 
defect fields (in the bulk and on the boundary) from which an arbitrary 
number of defect lines is emanating -- see the discussion at the 
beginning of section IV:3.4, and in particular figure (IV:3.31) -- and 
introduce bases for these fields, as well as compute their two-point functions.

The tensor unit $\one$ in any modular tensor category \C\ is a Jandl algebra
(and thus in particular a symmetric special Frobenius algebra). The CFT obtained
from $A\,{=}\,\one$, or from $\tilde A\,{=}\,\one$, has as its torus 
partition function the charge conjugation modular invariant. This CFT is often 
referred to as the `Cardy case'. In the Cardy case, the TFT formulation of 
correlators was accomplished in \cite{fffs2,fffs3}, where also consistency with 
factorisation and the action of the mapping class group was established. The 
corresponding statements and proofs for the general case given in the present 
paper simplify considerably when setting $A\,{=}\,\one$ (respectively, $\tilde 
A\,{=}\,\one$), and then indeed reduce to those given in \cite{fffs2,fffs3}.

\subsection{Invariance under the mapping class group}\label{sec:inv-map}

$~$

\noindent
We will now formulate the statements concerning covariance properties of 
correlators, and in particular invariance under the action of the mapping 
class group of the world sheet.

Let $E$ and $E'$ be extended surfaces 
(for the notion of an extended surface, see appendix \ref{app:worldsheet}).
To an isomorphism $f{:}~E\To E'$ of extended surfaces,\,\footnote{~%
An isomorphism $f{:}~E\To E'$ of extended surfaces is an orientation
preserving homeomorphism $f$ such that for the respective Lagrangian subspaces
we have $f_*(\lambda_E)\eq\lambda_{E'}$, and such that each marked point
$(p,[\gamma],U,\eps)$ of $E$ gets mapped to a marked point
$(f(p),[f \cir \gamma],U,\eps)$ of $E'$.}
we associate a cobordism $\Mf$ from $E$ to $E'$ as
  \be
  \Mf := \Llb E \Times [-1,0] \sqcup E' \Times [0,1] \Lrb/{\sim} \,,
  \ee
where `$\sim$' is an equivalence relation that relates boundary points in the two
components $E_<\,{:=}\, E\Times[-1,0]$ and $E_>\,{:=}\,E'\Times[0,1]$ of the
disjoint union $E_<\,{\sqcup}\,E_>$. Denoting points in $E_<$ and $E_>$ by 
$(p,t)_{<}$ and $(p,t)_{>}$, respectively, the equivalence relation is given by
  \be
  (p,0)_{<} \sim (f(p),0)_{>} \,.
  \ee
We abbreviate the homomorphism that the TFT assigns to the cobordism by $f_\sharp$,
  \be
  f_\sharp := Z(\Mf,E,E'): \quad \mathcal{H}(E)\To\mathcal{H}(E') \,.
  \ee
For $E=E'$, the assignment
  \be
  [f] \,\mapsto f_\sharp
  \labl{eq:rep-Map}
furnishes a projective action of the mapping class group
${\rm Map}_{\rm or}(E)$ on $\mathcal{H}(E)$ (see e.g.\ chapter IV.5 of \cite{TUra}).

\medskip

Let now \X\ and ${\rm Y}$ be unoriented world sheets and
$[f] \iN {\rm Map}(\X,{\rm Y})$, i.e.\ $[f]$ is a homotopy class of
homeomorphisms $f{:}\ \X\To{\rm Y}$ that map each marked arc
of \X\ to an arc of ${\rm Y}$ with the same marking, and map each boundary
segment of $\X$ to a boundary segment of ${\rm Y}$ with the same \bc.
Denote by $\pi_\X$ the canonical projection from \Xh\ to \X, and by
$\pi_{\rm Y}$ the one from \Yh\ to ${\rm Y}$. The homeomorphism $f$
has a unique lift $\hat f{:}\ \Xh\To\Yh$ such that $\hat f$ is an orientation 
preserving homeomorphism and $\pi_{\rm Y} \cir \hat f \eq f \cir \pi_\X$.

\medskip

For oriented world sheets, invariance of the correlators under the action of 
the mapping class group is then formulated as follows.

\dtlTheorem{thm:iso-or}
{\em Covariance of oriented correlators\/}:
For any two oriented world sheets \X\ and ${\rm Y}$, and any orientation 
preserving $[f]\iN{\rm Map}(\X,{\rm Y})$ we have
  \be
  C_\AA({\rm Y}) = \hat f_\sharp(C_\AA(\X)) \,.
  \labl{eq:map}
\ENDT

\dtlCorollary{cor:map-or}
{\em Modular invariance of oriented correlators\/}:
For any oriented world sheet \X\ and any $[f] \iN {\rm Map}_{\rm or}
(\X)$ we have
  \be
  C_\AA(\X) = \hat f_\sharp(C_\AA(\X)) \,.
  \labl{eq:cor:map}
\ENDC

For unoriented world sheets, the corresponding statements are:

\dtlTheorem{thm:iso-unor}
{\em Covariance of unoriented correlators\/}: For any two unoriented
world sheets \X\ and ${\rm Y}$, and any $[f]\iN{\rm Map}(\X,{\rm Y})$, we have
  \be
  C_{\!\tilde A}({\rm Y}) = \hat f_\sharp(C_{\!\tilde A}(\X)) \,.
  \labl{eq:map-un}
\ENDT

\dtlCorollary{cor:map-unor}
{\em Modular invariance of unoriented correlators\/}:
For any unoriented world sheet \X\ and any $[f] \iN {\rm Map}(\X)$ we have
  \be
  C_{\!\tilde A}(\X) = \hat f_\sharp(C_{\!\tilde A}(\X)) \,.
  \labl{eq:cor:map-un}
\ENDC

For the partition function (i.e., zero-point correlator) $C_\AA({\rm T};
\emptyset)$ on an oriented torus, the assertion of corollary \ref{cor:map-or}
is just what is usually referred to as modular invariance of the
torus \parfu. When the torus is unoriented, then this is replaced by
corollary \ref{cor:map-unor} for $C_{\!\tilde A}({\rm T};\emptyset)$;
the fact that $\tilde A$ is a Jandl algebra implies 
that the torus partition function is symmetric, compare remark II:2.5(i).


\subsection{Cutting and gluing of extended surfaces}\label{sec:ex-cut-glue}

$~$

\noindent
To discuss the factorisation properties of correlators, we must first analyse
the corresponding cutting and gluing procedures at the chiral level, i.e.\
for extended surfaces.

Let $S^2$ be the unit sphere embedded in $\reals^3$ with Cartesian coordinates
$(x,y,z)$, oriented by the inward pointing normal. Denote by
$S^2_{(2)}$ the unit sphere $S^2$ with two embedded arc germs $[\gamma^\pm]$
that are centered at the points $p^+ \eq (0,0,1)$ and $p^- \eq (0,0,-1)$, with
arcs given by $\gamma^+(t) \eq ( -\sin t, 0 , \cos t )$ and
$\gamma^-(t) \eq {-} \gamma^+(t)$. Pictorially,
  \bea \begin{picture}(200,74)(0,40)
  \put(60,0)         {\includeourbeautifulpicture 91
    \setlength{\unitlength}{.38pt}
    \put(98,207)        {\tiny $1$}
    \put(55,197)        {\tiny $2$}
    \put(153,229)       {\scriptsize $p^+$}
    \put(153,56)        {\scriptsize $p^-$}
    \setlength{\unitlength}{1pt}
    }
  \put(0,52)         {$ S^2_{(2)} \;= $}
  \put(174,54)	   {\scriptsize $x$}
  \put(136,77)	   {\scriptsize $y$}
  \put(114,114)	   {\scriptsize $z$}
  \epicture15 \labp 91
Further, let 
  \BE
  \Aeps := \{(x,y,z) \iN S^2 \,|\, |z|\,{<}\,\eps\}
  \EE
be an
open annulus around the equator, and understand by $S^1 \,{\subset}\, \Aeps$
the unit circle in the $x$-$y$-plane. Given an extended surface $E$, a 
continuous orientation preserving embedding $f{:}\ \Aeps \To E$
and an object $U\iN\objc$, we will define another extended surface, to be
denoted by $\Gamma_{\!f,U}(E)$. In brief, $\Gamma_{\!f,U}(E)$ is obtained by
cutting $E$ along $f(S^1)$ and then closing the resulting holes by gluing
two half-spheres with marked points $(U,+)$ and $(U,-)$, respectively.
In the case of unoriented world sheets, an additional datum is necessary, namely
a prescription to which part of the cut surface the upper and lower hemispheres 
are to be glued. The procedure of first embedding the `collar' $\Aeps$ and then 
cutting along the circle $f(S^1)\,{\subset}\, f(\Aeps)$ provides a natural 
choice of this additional datum.

\medskip

In more detail, $\Gamma_{\!f,U}(E)$ is constructed as follows. For $U\iN\objc$,
let $P_U$ be the extended surface given by $S^2_{(2)}$ with
the arc germ $[\gamma^+]$ marked by $(U,+)$ and $[\gamma^-]$ marked by
$(U,-)$. We denote by $P^+_{U,\eps}$ and $P^-_{U,\eps}$
the two hemispheres of $P_U$ consisting of those points of $P_U$ for which
$z\,{\ge}\, {-}\eps$ and $z \,{\le}\, \eps$, respectively, so that
$\Aeps\eq P^+_{U,\eps} \,{\cap}\, P^-_{U,\eps}$.
Also, let $\tilde E \,{:=}\, E\,{\setminus}\,f(S^1)$; then we set
  \be
  \Gamma_{\!f,U}(E) := \big(\, \tilde E
  \sqcup P^+_{U,\eps} \sqcup P^-_{U,\eps} ~ \big) \,/{\sim } \,,
  \ee
where the lower hemisphere $P^-_{U,\eps}$ is joined to $\tilde E$ via
$f(q) \,{\sim}\, q$ for $q \iN \{ (x,y,z) \iN P^-_{U,\eps} \,|\, z\,{>}\,0 \}$,
while the upper hemisphere $P^+_{U,\eps}$ is joined via
$f(q) \,{\sim}\, q$ for $q \iN \{ (x,y,z) \iN P^+_{U,\eps} \,|\, z\,{<}\,0 \}$.
This procedure is illustrated in the following figure:
  \begin{eqnarray} \begin{picture}(420,215)(10,0)
  \put(0,19)         {\Includeourbeautifulpicture 92a 
    \setlength{\unitlength}{.38pt} 
    \put(72,27)       {\tiny $1$}
    \put(38,0)        {\tiny $2$}
    \put(110,305)     {\scriptsize $f(\Aeps)$}
    \put(120,278)     {\scriptsize $\downarrow$}
    \setlength{\unitlength}{1pt}
    }
  \put(165,4)        {\Includeourbeautifulpicture 92b 
    \setlength{\unitlength}{.38pt}
    \put(82,367)        {\tiny $1$}
    \put(61,338)        {\tiny $2$}
    \put(87,177)        {\tiny $1$}
    \put(53,151)        {\tiny $2$}
    \setlength{\unitlength}{1pt}
    }
  \put(330,0)        {\Includeourbeautifulpicture 92c 
    \setlength{\unitlength}{.38pt}
    \put(73,443)        {\tiny $1$}
    \put(48,407)        {\tiny $2$}
    \put(77,131)        {\tiny $1$}
    \put(45,105)        {\tiny $2$}
    \put(102,343)       {\scriptsize $(U,-)$}
    \put(97,195)        {\scriptsize $(U,+)$}
    \setlength{\unitlength}{1pt}
    }
  \put(125,103)      {$ \longmapsto $}
  \put(290,103)      {$ \longmapsto $}
  \put(88,190)       {$ E $}
  \put(248,199)      {$ \tilde E $}
  \put(375,205)      {$ \Gamma_{\!f,U}(E) $}
  \end{picture} \nonumber\\~ \label{eq:pic-i2}
  \end{eqnarray}

To turn the two-manifold $\Gamma_{\!f,U}(E)$ into an extended surface we must also
specify a La\-gran\-gi\-an subspace $\lambda\,{\subset}\,H_1(\Gamma_{\!f,U}(E),
\reals)$. To this end we employ the cobordism
${\rm M}_{f,U}(E){:} 
    $\linebreak[0]$
\Gamma_{\!f,U}(E) \To E$ that is obtained by taking the
cylinder over $\Gamma_{\!f,U}(E)$ and identifying points on its boundary:
  \be
  {\rm M}_{f,U}(E) = \Llb \Gamma_{\!f,U}(E) \Times [0,1] \Lrb / {\sim} \,.
  \labl{MfUE}
The equivalence relation is given by
  \be
  (\iota^+(p)\,,1) \,{\sim}\, (\iota^-(\tau(p))\,,1)
  \quad{\rm for}~~ p \iN \{ (x,y,z) \iN S^2 \,|\, z\,{\ge}\,0 \} \,,
  \ee
where $\tau{:}\ S^2 \To S^2$ is the involutive map $\tau(x,y,z)\,{:=}\,
(x,y,-z)$, while $\iota^\pm{:}\ P^\pm_{U,\eps} \To \Gamma_{\!f,U}(E)$ 
are the natural embeddings implied by the construction of $\Gamma_{\!f,U}(E)$.
Close to the images of $P^-_{U,\eps}$ and $P^+_{U,\eps}$, the manifold 
${\rm M}_{f,U}(E)$ looks as follows\,\footnote{~%
   That $(U,\pm)$ occur at opposite places in \erf{eq:pic-i2} and
   \erf{eq:pic-V49} is due to the fact that it is $-\Gamma_{\!f,U}(E)$
   rather than $\Gamma_{\!f,U}(E)$ that appears in
   $\partial{\rm M}_{f,U}(E) \eq E\,{\sqcup}\,\big({-}\Gamma_{\!f,U}(E)\big)$.}
  \bea \begin{picture}(233,116)(0,48)
  \put(80,0)        {\includeourbeautifulpicture 49 }
  \put(128,78)      {\scriptsize $U$}
  \put(127.8,121)   {\scriptsize $(U,+)$}
  \put(127.8,40)    {\scriptsize $(U,-)$}
  \epicture20 \labl{eq:pic-V49}

\medskip\noindent
Here the inner boundary components (which are connected by the $U$-ribbon) are
part of $\Gamma_{\!f,U}(E) \Times \{0\}$, while the outer boundary is part of
$\{ [p,1] \iN \partial {\rm M}_{f,U}(E) \} \,{\cong}\, E$.
Given the cobordism ${\rm M}_{f,U}(E){:}\ \Gamma_{\!f,U}(E) \To E$
we obtain the Lagrangian subspace $\lambda$ of the extended surface
$\Gamma_{\!f,U}(E)$ from the Lagrangian subspace $\lambda_E$
of $E$ as $\lambda \,{:=}\,{\rm M}_{f,U}(E)^* \lambda_E$, i.e.\ as the
set of those cycles $x$ in $\Gamma_{\!f,U}(E)$ for which there exists an
$x' \iN \lambda(E)$ such that $x{-}x'$ is homologous to zero in the
three-manifold ${\rm M}_{f,U}(E)$; as shown in section IV.4.2 of \cite{TUra},
this yields indeed a Lagrangian subspace of $H_1( \Gamma_{\!f,U}(E) , \reals)$.

The TFT assigns to ${\rm M}_{f,U}(E)$ a linear map
  \be
  g_{f,U}(E) := Z({\rm M}_{f,U}(E) , \Gamma_{\!f,U}(E) , E)
  :\quad \mathcal{H}(\Gamma_{\!f,U}(E)) \to \mathcal{H}(E) \,.
  \ee
Morphisms obtained from gluing constructions similar to the above
will be referred to as {\em gluing homomorphisms\/}.
	
\medskip

For the description of cutting and gluing world sheets further on,
it is helpful to iterate the above procedure, thus giving rise to an operation
of {\em cutting twice\/}, and to restrict one's attention to simple objects $U$.
This is done as follows. Given an extended
surface $E$, a map $f{:}\ \Aeps \To E$, and a label $k\iN\I$ of a simple
object of \C, let $\tilde f{:}\ \Aeps \To \Gamma_{\!f,U_k}(E)$ be the map
$\tilde f(x,y,z) \,{:=}\, \iota^-(-x,y,-z)$.

Then we define a new extended surface $\hat{\Gamma}_{\!f,k}(E)$ as
  \be
  \hat{\Gamma}_{\!f,k}(E)
  := \Gamma_{\!\tilde f,U_{\bar k}}(\Gamma_{\!f,U_k}^{}(E)) \,.
  \ee
In pictures we have
  \begin{eqnarray} \begin{picture}(420,257)(10,0)
  \put(0,18)         {\Includeourbeautifulpicture 92a 
    \setlength{\unitlength}{.38pt} 
    \put(73,27)         {\tiny $1$}
    \put(38,0)          {\tiny $2$}
    \put(110,305)       {\scriptsize $f(\Aeps)$}
    \put(120,278)       {\scriptsize $\downarrow$}
    \put(201,436)       {$ E $}
    \setlength{\unitlength}{1pt}
    }
  \put(165,16)       {\Includeourbeautifulpicture 93b 
    \setlength{\unitlength}{.38pt}
    \put(67,511)        {\tiny $1$}
    \put(41,477)        {\tiny $2$}
    \put(77,131)        {\tiny $1$}
    \put(44,108)        {\tiny $2$}
    \put(110,504)       {\scriptsize $\tilde f(\Aeps)$}
    \put(120,477)       {\scriptsize $\downarrow$}
    \put(102,360)       {\scriptsize $(U_k,-)$}
    \put(97,195)        {\scriptsize $(U_k,+)$}
    \put(113,572)       {$ \Gamma_{\!f,U_k}^{}(E) $}
    \setlength{\unitlength}{1pt}
    }
  \put(330,0)        {\Includeourbeautifulpicture 93c 
    \setlength{\unitlength}{.38pt}
    \put(67,594)        {\tiny $1$}
    \put(41,561)        {\tiny $2$}
    \put(72,96)         {\tiny $1$}
    \put(38,71)         {\tiny $2$}
    \put(92,495)        {\scriptsize $(U_{\bar k},+)$}
    \put(88,374)        {\scriptsize $(U_{\bar k},-)$}
    \put(94,281)        {\scriptsize $(U_k,-)$}
    \put(92,161)        {\scriptsize $(U_k,+)$}
    \put(25,678)        {$\Gamma_{\!\tilde f,U_{\bar k}}(\Gamma_{\!f,U_k}^{}(E))$}
    \setlength{\unitlength}{1pt}
    }
  \put(125,121)      {$ \longmapsto $}
  \put(290,121)      {$ \longmapsto $}
  \end{picture} \nonumber\\~ \label{eq:pic-i3}
  \end{eqnarray}
Analogously as for $\Gamma_{\!f,U}(E)$ we define a cobordism
$\hat {\rm M}_{f,k}{:}\ \hat{\Gamma}_{\!f,k}(E) \To E$ by composing the
cobordism ${\rm M}_{\tilde f,U_{\bar k}}{:}\ \hat{\Gamma}_{\!f,k}(E) \To \Gamma
_{\!f,U_k}(E)$ with ${\rm M}_{f,U_k}{:}\ \Gamma_{\!f,U_k}(E)\To E$. The part of 
the cobordism $\hat{\rm M}_{f,k}$ that is analogous to \erf{eq:pic-V49} looks as
  \bea \begin{picture}(210,106)(0,59)
  \put(60,0)        {\includeourbeautifulpicture 41 }
  \put(110,113)     {\scriptsize $\bar{k}$}
  \put(110,47)      {\scriptsize ${k}$}
  \put(74,122)      {\scriptsize $(U_{\bar{k}},-)$}
  \put(117,104)     {\scriptsize $(U_{\bar{k}},+)$}
  \put(74,53)       {\scriptsize $(U_{k},+)$}
  \put(117,36)      {\scriptsize $(U_{k},-)$}
  \epicture37 \labl{eq:picV-41}
We also set
  \be
  \hat g_{f,k}(E) := Z(\hat {\rm M}_{f,k}(E) , \hat{\Gamma}_{\!f,k}(E) , E)
  : \quad \mathcal{H}(\hat{\Gamma}_{\!f,k}(E)) \to \mathcal{H}(E) \,.
  \labl{eq:glue2}
This linear map is another example of a gluing homomorphism.

This finishes our discussion of cutting and gluing operations
at the chiral level.

\subsection{Boundary factorisation} \label{sec:facbdry}

\subsubsection*{Two-point function on the disk}

Denote by $D \,{\equiv}\, D(M_l,M_r,k,\psi^+,\psi^-)$ the world sheet for the
standard two-point function on the unit disk, i.e.\ (see definition \ref{def:ws}
for the conventions regarding the world sheet and boundary labels)
  \bea \begin{picture}(310,85)(0,27)
  \put(145,0)        {\includeourbeautifulpicture 95 
    \setlength{\unitlength}{.38pt}  
    \put(195,159)       {\tiny $1$}
    \put(164,192)       {\tiny $2$}
    \put( 38,165)       {\scriptsize $M_l$}
    \put(236,165)       {\scriptsize $M_r$}
    \put(158,237)       {\scriptsize $\vPsi^+$}
    \put(158, 36)       {\scriptsize $\vPsi^-$}
    \setlength{\unitlength}{1pt}
    }
  \put(0,53)         {$ D(M_l,M_r,k,\psi^+,\psi^-) \;= $}
  \epicture11 \labp 95
Here the boundary and bulk are oriented as indicated, the boundary condition
is $M_l$ and $M_r$ for $x\,{<}\,0$ and $x\,{>}\,0$, respectively, and
$k \iN \I$ is the label of a simple object. The two field insertions are
  \bea
  \vPsi^+ = (M_l,M_r,U_k,\psi^+, p^+, [\gamma^+])
  \qquad {\rm and} \\{}\\[-.6em]
  \vPsi^- = (M_r,M_l,U_{\bar k},\psi^-, p^-, [\gamma^-]) \,, \eear
  \labl{217}
respectively, where $p^+\eq (0,1)$ and the arc germ $[\gamma^+]$ is given by
$\gamma^+(t) \eq (-\sin t, \cos t)$, while $p^- \eq {-}p^+$ and
$\gamma^-(t) \eq {-}\gamma^+(t)$.
Accordingly, the morphisms $\psi^\pm$ are elements of the spaces
$\Hom_A(M_l\oti U_k, M_r)$ and $\Hom_A(M_r\oti U_{\bar k}, M_l)$, respectively.

\medskip

The double $\hat D$ of $D$ can be identified with $S^2_{(2)}$, where
$[\gamma^+]$ is now marked by $(U_k,+)$ and $[\gamma^-]$ by
$(U_{\bar k},+)$. The corresponding space $\mathcal{H}(\hat D)$ of conformal
blocks is one-dimensional; a basis is given by applying $Z$ to the cobordism
$B^+_{k\bar k}{:}\ \emptyset \To \hat D$ that is given by
  \bea \begin{picture}(220,110)(0,46)
  \put(56,0)         {\includeourbeautifulpicture 97 }
  \put(0,76)         {$ B^+_{k\bar k} \; := $}
  \put(88,80)        {\scriptsize \begin{turn}{90}$\lambda_{k\bar{k}}$\end{turn}}
  \put(102,151)      {\scriptsize $(U_k,+)$}
  \put(102,3)        {\scriptsize $(U_{\bar k},+)$}
  \put(98,116)       {\tiny $1$}
  \put(81,112)       {\tiny $2$}
  \epicture26 \labl{i7}
(with the black side of the ribbons facing the reader).
The morphism $\lambda_{k\bar k} \iN \Hom(U_k\oti U_{\bar k},\one)$ is the
non-zero morphism introduced in (I:2.29), where it was denoted by
$\lambda_{(k,\bar k),0}$ instead. For the morphism dual to $\lambda_{i\ib}$,
which was denoted by $\Upsilon^{(i,\ib),0}$ in \cite{fuRs4}, we use here an
abbreviated notation, too, namely $\bar{\lambda}^{i\ib}$. Also, in the sequel,
when drawing pictures in blackboard framing (using lines in place of ribbons),
these morphisms will be represented graphically as
  \bea \begin{picture}(120,123)(0,13)
 \put(0,80){
  \put(50,0)         {\Includeourbeautifulpicture 06a }
  \put(0,24)         {$ \lambda_{i\ib} \;= $}
  \put(46,-8)        {\scriptsize $U_i$}
  \put(63,-8)        {\scriptsize $U_{\ib }$}
 }
 \put(-130,0){
  \put(180,0)        {\Includeourbeautifulpicture 06b }
  \put(130,24)       {$ \bar\lambda^{i\ib}_{} \;= $}
  \put(177,54)       {\scriptsize $U_i$}
  \put(193,54)       {\scriptsize $U_{\ib }$}
 }
  \epicture-1 \labl{fig:llbar}
The normalisation of $\lambda_{k\bar k}$, and hence the choice of basis 
\erf{i7}, is arbitrary, but will be kept fixed throughout the paper.

In the chosen basis the structure constant $c^{\rm bnd} \iN \complex$
of the two-point function on the disk is defined by
  \be
  C(  D(M_l,M_r,k,\psi^+,\psi^-) )
  =: c^{\rm bnd}_{M_l,M_r,k ; \psi^+,\psi^-}\, B^+_{k \bar k} \,.
  \labl{Dscdef}
Here $C$ stands for either $C_\AA$ or $C_{\tilde A}$. In the first case, $D$
is understood as an oriented world sheet, in the latter case as an unoriented
world sheet, i.e.\ the 2-orientation of the disk is omitted. Given bases 
$\{ \psi^+_\alpha \}$ of $\Hom_A(M_l \oti U_k, M_r)$ and $\{ \psi^-_\beta \}$ 
of $\Hom_A(M_r \oti U_{\bar k}, M_l)$, we define a complex-valued matrix 
$c^{\rm bnd}_{M_l,M_r,k}$ by
  \be
  {(c^{\rm bnd}_{M_l,M_r,k})}^{}_{\alpha\beta}
  := c^{\rm bnd}_{M_l,M_r,k ; \psi^+_\alpha,\psi^-_\beta} \,.
  \ee
It is shown in appendix~\ref{sec:2ptD} that
the boundary two-point function is non-degenerate, in the sense that the matrix 
$c^{\rm bnd}_{M_l,M_r,k}$ is invertible for fixed $M_l$, $M_r$ and $k$.

\subsubsection*{Cutting world sheets along an interval}
         
In the sequel we use several subsets of $D$, in particular the strip
  \BE
  \Reps := \{(x,y)\iN D \,|\, |y|\,{<}\,\eps\} \,, 
  \EE
as well as
  \BE
  \bearl
  D^+_\eps := \{(x,y)\iN D \,|\, y\,{>}\,{-}\eps\} \qquad{\rm and}
  \\{}\\[-.4em]
  D^-_\eps := \{(x,y)\iN D \,|\, y\,{<}\,\eps\} \,, \eear
  \EE
such that
$\Reps \eq D^+_\eps \,{\cap}\, D^-_\eps$. The boundary of $\Reps$
satisfies $\partial\Reps\,{\cap}\,\Reps \eq \Reps\,{\cap}\,\partial D$.
Let \X\ be a world sheet (oriented or unoriented) with non-empty boundary.
Suppose we are given a continuous injection $f{:}\ \Reps \To \X$ such that
$f(\partial \Reps{\cap}\Reps) \,{\subset} 
     $\linebreak[0]$
\partial \X$, and such that its 
restriction $f{:}\ \partial\Reps{\cap}\Reps \To \partial\X$ is orientation 
preserving, and  $f(1,0)$ lies on a boundary segment of \X\ labelled by $M_r$, 
while $f(-1,0)$ lies on a (not necessarily different) boundary segment
labelled by $M_l$. We define a new world sheet $\cbnd_{f,k,\psi^+,\psi^-}(\X)$,
which is obtained from \X\ by cutting along the image of $[-1,1]\Times\{0\}$
under $f$ and gluing half disks with boundary insertions $\psi^+$, $\psi^-$ to
the cuts. This procedure is used to formulate the boundary factorisation of
correlators in theorems \ref{thm:bnd-or} and \ref{thm:bnd-unor}.

In more detail, to obtain $\cbnd_{f,k,\psi^+,\psi^-}(\X)$ we start from
$\tilde{\X} \eq \X\,{\setminus}\,f([-1,1]\Times\{0\})$ and define
  \be
  \cbnd_{f,k,\psi^+,\psi^-}(\X) := \Llb \, \tilde{\X}
  \,\sqcup\, D^+_{\eps} \,\sqcup\, D^-_{\eps} ~ \Lrb \,/\,{\sim } \,,
  \labl{bndglue}
where the lower half disk gets joined to $\tilde{\X}$ via
$f(p) \sim p$ for $p \iN \{ (x,y) \iN D^-_{\eps} \,|\, y\,{>}\,0 \}$,
while the upper half disk gets joined via
$f(p) \sim p$ for $p \iN \{ (x,y) \iN D^+_{\eps} \,|\, y\,{<}\,0 \}$.
This procedure is illustrated in the following figure:
  \begin{eqnarray} \begin{picture}(420,215)(21,0)
  \put(0,22)         {\Includeourbeautifulpicture 96a
    \setlength{\unitlength}{.38pt}
    \put(152,280)       {\scriptsize $f({\rm R}_{\eps})$}
    \put(166,255)       {\scriptsize $\downarrow$}
    \put(79,113)        {\scriptsize $M_l$}
    \put(249,113)       {\scriptsize $M_r$}
    \setlength{\unitlength}{1pt}
    }
  \put(172,6)        {\Includeourbeautifulpicture 96b 
    \setlength{\unitlength}{.38pt}
    \put( 76,434)       {\scriptsize $M_l$}
    \put(250,434)       {\scriptsize $M_r$}
    \put( 79,120)       {\scriptsize $M_l$}
    \put(244,120)       {\scriptsize $M_r$}
    \setlength{\unitlength}{1pt}
    }
  \put(346,0)        {\Includeourbeautifulpicture 96c 
    \setlength{\unitlength}{.38pt}
    \put( 84,434)       {\scriptsize $M_l$}
    \put(230,434)       {\scriptsize $M_r$}
    \put( 84,120)       {\scriptsize $M_l$}
    \put(230,120)       {\scriptsize $M_r$}
    \put(160,234)       {\scriptsize $\psi^+$}
    \put(161,324)       {\scriptsize $\psi^-$}
    \setlength{\unitlength}{1pt}
    }
  \put(148,104)      {$ \longmapsto $}
  \put(314,104)      {$ \longmapsto $}
  \end{picture} \nonumber\\
  \label{eq:pic-i6} {} \end{eqnarray}
If \X\ is an oriented world sheet then so is $\cbnd_{f,k,\psi^+,\psi^-}(\X)$.
If \X\ is an unoriented world sheet, then so is $\cbnd_{f,k,\psi^+,\psi^-}(\X)$.

\dtlRemark{rem:bnd-missing}
Note that  $\Reps$ has two boundary components, both of which are oriented by 
the inward pointing normal, i.e.\ the orientation of $\partial\Reps\,{\cap}\,
\Reps \eq \Reps\,{\cap}\,\partial D$ is inherited from $\partial D$.
\\
To show that a system of correlators is consistent with boundary factorisation
we have to prove a factorisation rule for every way to cut a world sheet $\X$
along a non-selfintersecting curve $\gamma$ connecting two points on the
boundary $\partial\X$.
For oriented world sheets, this cutting procedure is directly implemented by
the construction of $\cbnd_{f,k,\psi^+,\psi^-}(\X)$ just described, because for
any $\gamma$ we can find an orientation preserving embedding $f$ of $\Reps$
into $\X$ such that the image of $[-1,1]\Times \{0\}$ is $\gamma$. Since
$\partial\X$ and $\partial \Reps$ are oriented by the inward pointing normals,
$f$ is automatically consistent with the boundary orientation.
\\
In the unoriented case, on the other hand, one can still find an embedding $f$
of $\Reps$ into $\X$ such that $f([-1,1]\Times \{0\})$ is equal to $\gamma$,
but now $f$ need not preserve the boundary orientation. However, we are free to
choose an equivalent labelling of boundary conditions with possibly different
orientation of $\partial X$ (see appendix \ref{app:equiv-lab}), and we can use 
this freedom to ensure that $f$ is orientation preserving on the left
boundary component (the one with negative $x$-values) of $\Reps$.
Afterwards, instead of $\Reps$ we consider $\Reps'$, defined
to be equal to $\Reps$ except that the orientation of the
right boundary component of $\Reps'$ is reversed as compared to $\Reps$. Given
an embedding $f{:}\ \Reps'\To\X$ preserving the boundary orientation, we can
again construct a new world sheet $\cbnd_{f,k,\psi^+,\psi^-}(\X)$ obtained by
cutting $\X$ along $f([-1,1]\Times \{0\})$ and gluing half disks. However, owing
to our labelling conventions for boundary conditions, the details of
the construction of $\cbnd_{f,k,\psi^+,\psi^-}(\X)$ are slightly more involved
than in the situation described
above; the details are given in section \ref{sec:bnd-proof-unor} below.
\ENDR

\subsubsection*{Gluing homomorphism between spaces of blocks}

We abbreviate
  \be
  \X'_k := \cbnd_{f,k,\psi^+,\psi^-}(\X) \,.
  \ee
To compare the correlators for the world sheets $\X'_k$ and $\X$, we need to
construct a map between the spaces of blocks on their doubles.
This is again provided by an appropriate gluing homomorphism
  \be
  G^{\rm bnd}_{f,k} :\quad \mathcal{H}(\Xh'_k) \to \mathcal{H}(\Xh) \,,
  \labl{eq:Gbnd-spaces}
which we proceed to define. Let the extended surface
$H^-_{\bar k k}$ be given by $S^2_{(2)}$ with $[\gamma^+]$ marked by
$(U_{\bar k},-)$ and $[\gamma^-]$ marked by $(U_k,-)$. Consider
the cobordism $B^-_{\bar k k}{:}\ \emptyset \To H^-_{\bar k k}$ given by
  \bea \begin{picture}(220,107)(0,51)
  \put(56,0)         {\includeourbeautifulpicture 94 }
  \put(0,76)         {$ B^-_{\bar k k} \; := $}
  \put(88,93)        {\scriptsize \begin{turn}{270}$\bar\lambda^{k\bar k}$\end{turn}}
  \put(102,151)      {\scriptsize $(U_{\bar k},-)$}
  \put(102,3)        {\scriptsize $(U_k,-)$}
  \put(98,116.3)     {\tiny $1$}
  \put(82,112)       {\tiny $2$}
  \epicture29 \labp 94
where $\bar\lambda^{k\bar k}\iN\Hom(\one,U_k\oti U_{\bar k})$ is the morphism 
introduced in figure \erf{fig:llbar}.

Given $f{:}\ \Reps \To \X$, we can construct an embedding
$\hat f{:}\ \Aeps\To\Xh$. On a point $(x,y,z) \iN \Aeps$ it is defined as
  \be
  \hat f(x,y,z) := \left\{ \bearll
  [f(x,z) , -{\rm or}_2(\X)] ~ & {\rm for}~ y\,{\le}\,0 \,, \\{}\\[-.4em]
  [f(x,z) , {\rm or}_2(\X)]    & {\rm for}~ y\,{\ge}\,0 \,.  \eear \right.
  \ee
(Note that for points with $y{=}0$ we have $(x,z)\iN\partial D$, so that by
construction of the double \Xh\ we have the equality
$[f(x,z) , -{\rm or}_2(\X)] \eq [f(x,z) , {\rm or}_2(\X)]$.)
Comparing \erf{eq:pic-i3} and \erf{eq:pic-i6}, it is not
difficult to see that there is a canonical isomorphism between the extended
surfaces $\Xh'_k \,{\sqcup}\, H^-_{\bar k k}$ and 
$\hat{\Gamma}_{\!\hat f,k}(\Xh)$. Thus the gluing homomorphism
\erf{eq:glue2} provides us with a map
  \be
  \hat g_{\hat f,k}(\Xh) :\quad \mathcal{H}(\Xh') \otic
  \mathcal{H}(H^-_{\bar k k}) \to \mathcal{H}(\Xh) \,.
  \ee
Using the basis element
$B^-_{\bar k k} \iN \mathcal{H}(H^-_{\bar k k})$ we finally define
the gluing morphism \erf{eq:Gbnd-spaces} by
  \be
  G^{\rm bnd}_{f,k}(v) :=
  \hat g_{\hat f,k}(\Xh) (v \Oti B^-_{\bar k k}) \,.
  \labl{eq:Gbnd-def}
for $v \iN \mathcal{H}(\Xh'_k)$.
Note that, strictly speaking, in \erf{eq:Gbnd-def} we should write
$Z(B^-_{\bar k k})$ instead of $B^-_{\bar k k}$. However, to simplify
notation we will use the same symbol for a cobordism and its invariant
when the meaning is clear from the context.

\subsubsection*{The boundary factorisation}

We are now ready to formulate the statements about boundary factorisation of
the correlators.

\dtlTheorem{thm:bnd-or}
{\em Boundary factorisation for oriented world sheets\/}:
Let \X\ be an oriented world sheet with non-empty boundary, and let
$f{:}\ \Reps \To \X$ be injective, continuous and 2-orientation preserving
and satisfy $f(\partial\Reps{\cap}\Reps) \,{\subset}\, \partial \X$.
Let the boundary components containing $f(1,0)$ and $f(-1,0)$ be labelled by
$M_r$ and $M_l$, respectively. Then
  \be
  C_\AA(\X) = \sum_{k\in\I} \sum_{\alpha,\beta} \dim(U_k) \,
  {(c^{\rm bnd~~~~-1}_{M_l,M_r,k})}^{}_{\beta\alpha}\,
  G^{\rm bnd}_{f,k}\big( C_\AA( \cbnd_{f,k,\alpha,\beta}(\X) ) \big) \,,
  \labl{eq:bnd-or}
where $\alpha$ and $\beta$ run over bases of $\,\Hom_A(M_l \oti U_k, M_r)$
and of $\,\Hom_A(M_r \oti U_{\bar k}, M_l)$, respectively.
\ENDT

\dtlTheorem{thm:bnd-unor}
{\em Boundary factorisation for unoriented world sheets\/}:
Let \X\ be an unoriented world sheet with non-empty boundary.
Suppose we are given either of the following data:
\\[2pt]
{\rm(i)}~\,A map $f{:}\ \Reps \To \X$ that is injective and continuous, such that
$f(\partial\Reps{\cap}\Reps) \,{\subset}\, \partial \X$, $f$ preserves the 
boundary orientation, and the boundary components containing $f(1,0)$ and 
$f(-1,0)$ are labelled by $M_r$ and $M_l$, respectively.
\\[2pt]
{\rm(ii)}~A map $f{:}\ \Reps' \To \X$ that is injective and continuous, such that
$f(\partial\Reps'{\cap}\Reps') \,{\subset}\, \partial \X$, $f$ preserves the 
boundary orientation, and the boundary components containing 
$f(1,0)$ and $f(-1,0)$ are labelled by $M_r^\sigma$ and $M_l$, respectively.
\\
Then we have
  \be
  C_{\!\tilde A}(\X) = \sum_{k\in\I} \sum_{\alpha,\beta} \dim(U_k) \,
  {(c^{\rm bnd~~~~-1}_{M_l,M_r,k})}^{}_{\beta\alpha}\,
  G^{\rm bnd}_{f,k}\big( C_{\!\tilde A}( \cbnd_{f,k,\alpha,\beta}(\X) ) \big) \,,
  \labl{eq:bnd-unor}
where $\alpha$ and $\beta$ run over bases of $\,\Hom_{\tilde A}(M_l\oti U_k,M_r)$
and of $\,\Hom_{\tilde A}(M_r \oti U_{\bar k}, M_l)$, respectively.
\ENDT

\noindent 
Here $\Reps'$ is as introduced in remark \ref{rem:bnd-missing} above, and
$M_r^\sigma$ denotes the $\tilde A$-module conjugate to $M_r$, as described
in definition II:2.6.


\subsection{Bulk factorisation} \label{sec:facbulk}

\subsubsection*{Two-point function on the sphere}

Denote by $S(i,j,\phi^+,\phi^-)$ the world sheet for the standard two-point
correlator on the unit sphere, given by $S^2_{(2)}$ with field insertions
  \bea
  \vPhi^+ = (i,j,\phi^+, p^+, [\gamma^+], {\rm or}_2)
  \qquad {\rm and} \\{}\\[-.4em]
  \vPhi^- = (\ib, \jb, \phi^-, p^-, [\gamma^-], {\rm or}_2) \,.
  \eear\labl{230}
Here ${\rm or}_2$ denotes the orientation of $S^2_{(2)}$, and $\phi^\pm$ are
bimodule morphisms $\phi^+ \iN \HomAA(U_i 
      $\linebreak[0]$
\oT+ A \ot- U_j, A)$ and
$\phi^- \iN \HomAA(U_{\bar\imath} \oT+ A \ot- U_{\bar\jmath}, A)$ (see formula
(IV:2.17) for the definition of $\ot{\pm}$).

The double $\hat S$ of $S$ can be identified with
$S^2_{(2)} \,{\sqcup}\, S^2_{(2)}$ by taking $p\,{\mapsto}\, (p,{\rm or}_2)$
for the first copy and $p \,{\mapsto}\, (\tau'(p), -{\rm or}_2)$ for
the second copy, where $\tau'(x,y,z) \,{:=}\, (x,-y,z)$. Note that this
identification respects the orientation and is
compatible with the embedded arcs. On the first copy, to be denoted by $S_i$,
we mark $[\gamma^+]$ and $[\gamma^-]$ by $(U_i,+)$ and $(U_{\bar\imath},+)$,
while on the second copy, to be denoted by $S_j$, we mark the two arcs by
$(U_j,+)$ and $(U_{\bar\jmath},+)$, respectively. The corresponding space
$\mathcal{H}(\hat S) \,{\cong}\, \mathcal{H}(S_i) \otic \mathcal{H}(S_j)$
of conformal blocks is one-dimensional; a basis is given by
$B^+_{i\bar\imath} \oti B^+_{j\bar\jmath}$, with $B^+_{k\bar k}$ as defined
in \erf{i7}. Accordingly, the structure constant $\Cbulk \iN \complex$
of the two-point correlator on the sphere is defined by
  \be
  C(  S(i,j,\phi^+,\phi^-) )
  =: \Cbulk_{i,j,\phi^+,\phi^-} \, B^+_{i\bar\imath} \oti B^+_{j\bar\jmath} \,.
  \labl{eq:S2sc}
Given bases $\{ \phi^+_\alpha \}$ of $\HomAA(U_i \oT+ A \ot- U_j, A)$ and
$\{\phi^-_\beta\}$ of $\HomAA(U_{\bar\imath} \oT+ A \ot- U_{\bar\jmath},A)$,
we use the two-point function on the sphere to define a matrix $\Cbulk_{i,j}$ by
  \be
  {(\Cbulk_{i,j})}^{}_{\alpha\beta}
  := \Cbulk_{i,j, \phi^+_\alpha,\phi^-_\beta} \,.
  \ee
As shown in appendix~\ref{sec:2ptS2}, the bulk two-point function is 
non-degenerate, in the sense that for fixed $i$, $j$
the matrix $\Cbulk_{i,j}$ is invertible.

\subsubsection*{Cutting world sheets along a circle}
         
Let $\Aeps$ be the equatorial annulus on the sphere that we introduced in 
section \ref{sec:ex-cut-glue}, and denote by $S^\pm_\eps$ the subsets of 
$S \,{\equiv}\, S(i,j,\phi^+,\phi^-)$ given by 
  \BE
  \bearl
  S^+_\eps := \{(x,y,z)\iN S \,|\, z\,{>}\,{-}\eps\} \qquad{\rm and}
  \\{}\\[-.4em]
  S^-_\eps := \{(x,y,z) \iN S \,|\, z\,{<}\,\eps\} \,, \eear
  \EE 
such that $\Aeps \eq S^+_\eps \, {\cap}\, S^-_\eps$.

Let \X\ be an (oriented or unoriented) world sheet and $f{:}~\Aeps \To \X$ \
an injective, continuous map. Further, let $i,j \iN \I$ be labels of simple
objects, and let $\phi^\pm$ be bimodule morphisms
$\phi^+ \iN \HomAA(U_i \oT+ A \ot- U_j, A)$,
$\phi^- \iN \HomAA(U_{\bar\imath} \oT+ A \ot- U_{\bar\jmath}, A)$.
Similar to the procedure described in section \ref{sec:facbdry},
we define a new world sheet $\cblk_{f,i,j,\phi^+,\phi^-}(\X)$,
which is obtained from \X\ by cutting along $f(S^1)$ and
gluing hemispheres with bulk insertions $\phi^+$, $\phi^-$ to the cuts.
We set $\tilde {\X}\,{:=}\,\X \,{\setminus}\, f(S^1)$ and define
  \be
  \cblk_{f,i,j,\phi^+,\phi^-}(\X)
  := \Llb \tilde {\X} \sqcup S^+_\eps \sqcup S^-_\eps \Lrb / {\sim} \,.
  \labl{233}
Here the lower hemisphere gets joined to $\tilde\X$ via
$f(p) \,{\sim}\, p$ for $p \iN \{ (x,y,z) \iN S^-_{\eps} \,|\, z\,{>}\,0 \}$,
while the upper hemisphere gets joined via $f(p) \,{\sim}\, p$ for $p \iN 
\{ (x,y,z) \iN S^+_{\eps} \,|\, z\,{<}\,0 \}$. Again, if $\X$ is an oriented 
world sheet, then so is $\cblk_{f,i,j,\phi^+,\phi^-}(\X)$. If $\X$ is an 
unoriented world sheet, then so is $\cblk_{f,i,j,\phi^+,\phi^-}(\X)$.

\dtlRemark{rem:bulk-missing}
In theorems \ref{thm:bulk-or} and \ref{thm:bulk-unor} below we formulate
factorisation identities which are obtained when cutting a world sheet $\X$
along a closed curve $\gamma$ that is obtained as the image $f(S^1)$ under an
embedding $f{:}\ \Aeps \To \X$.
\\[.2em]
On the other hand, given a closed non-self\-intersecting curve $\gamma$
in \X, in the spirit of remark \ref{rem:bnd-missing} we can wonder 
whether there is a factorisation identity associated to cutting $\X$ along
$\gamma$. For oriented world sheets we can find an embedding $f{:}\ \Aeps\To\X$
such that $f(S^1)$ is equal to $\gamma$, since any small neighbourhood of 
$\gamma$ is oriented and hence has the topology of an annulus.
In this case we find the identity described in theorem \ref{thm:bulk-or}.
For unoriented world sheets, on the other hand, a small neighbourhood of 
$\gamma$ can have either the topology of an annulus or of a M\"obius strip.
In the former case, there is again
an embedding $f{:}\ \Aeps \To \X$ such that $f(S^1)$ is equal to 
$\gamma$, and one obtains the identity given in theorem \ref{thm:bulk-unor}.
\\[.2em]
Regarding the second case, let $h{:}\ \Meps \To \X$ be an embedding
of a M\"obius strip $\Meps$ of width $\eps$ into \X, such that
$h(S^1) \eq \gamma$ (where $S^1$ is the circle forming the `middle' of the 
M\"obius strip). Note that cutting $\X$ along $\gamma$ introduces only a 
single additional boundary component. Accordingly, cutting along $\gamma$ 
does not locally decompose the world sheet into two disconnected parts, 
as is the case for any embedding $\Aeps \To \X$. 
\\
Nonetheless it turns out that an embedding $h{:}\ \Meps \To \X$ leads to an 
identity for correlators, too, even though it is not a factorisation identity 
because, as just described, the world sheet is not cut into two locally 
disconnected parts. In words, one can recover the correlator $C(X)$ by gluing
a hemisphere with one bulk insertion to the cut boundary of
$\X \,{\setminus}\, \gamma$, and summing over a basis of bulk fields,
weighting each term with the one-point function on the cross cap $\mathbb{RP}^2$
(multiplied by the inverse of the two-point function on the sphere).
In this way one obtains in particular the {\em cross cap constraint\/} that
has been studied in \cite{fips}.
\\
In slightly more detail, given an embedding $h{:}\ \Meps \To \X$, consider 
the curve $\gamma'$ that is given by the boundary $\partial h(\Meps)$.
For small $\eps$, the curve $\gamma'$ is running close to $\gamma$ and
has a small neighbourhood with the topology of an annulus. Thus for $\gamma'$
we can find an embedding $f{:}\ \Aeps \To \X$ such that $f(S^1)\eq\gamma'$.
The world sheet cut along $\gamma'$ and with the half-spheres $S^\pm_\eps$
glued to the holes is then given by $\cblk_{f,i,j,\phi^+,\phi^-}(\X)$. Note that
gluing a half-sphere to the component of $\X\,{\setminus}\,f(S^1)$ that contains 
$\gamma$ produces a cross cap. The above mentioned identity for correlators 
is now obtained by applying theorem \ref{thm:bulk-unor} to the embedding $f$.
\ENDR

\subsubsection*{Gluing homomorphism between spaces of blocks}

We abbreviate
  \be
  \X'_{ij} := \cblk_{f,i,j,\phi^+,\phi^-}(\X) \,.
  \ee
To formulate the bulk factorisation constraint, we need a map
$G^{\rm bulk}_{f,ij}$ between the spaces of conformal blocks on the
doubles of $\X'_{ij}$ and $\X$. The embedding $f{:}\ \Aeps \To \X$ gives rise
to two embeddings $f_i$ and $f_j$ of $\Aeps$ into the double $\Xh$, given by
  \bea
  f_i(x,y,z) = [f(x,y,z),{\rm or}_2]  \qquad {\rm and} \\{}\\[-.4em]
  f_j(x,y,z) = [f(x,-y,z),-{\rm or}_2] \eear
  \labl{eq:fifj-def}
for $(x,y,z)\iN \Aeps$. 
We can now apply the `cutting twice' procedure \erf{eq:pic-i3} to both 
embeddings $f_i$ and $f_j$. This results in an extended surface
  \be
  {\rm Y}_{ij} := \hat{\Gamma}_{\!f_i,i}( \hat{\Gamma}_{\!f_j,j}( \Xh )) \,.
  \labl{Yij}
This extended surface comes together with an isomorphism ${\rm Y}_{ij}\,{\cong}
\, \Xh'_{ij}\,{\sqcup}\,H^-_{\bar\imath i}\,{\sqcup}\,H^-_{\bar\jmath j}$,
with $H^-_{\bar k k}$ as defined after formula \erf{eq:Gbnd-spaces},
so that the gluing homomorphism $\hat g$ provides a map
  \be
  \hat g_{f_i,i} \circ \hat g_{f_j,j} :\quad
  \mathcal{H}(\Xh'_{ij}) \otic \mathcal{H}(H^-_{\bar\imath i})
  \otic \mathcal{H}(H^-_{\bar\jmath j}) \to \mathcal{H}(\Xh) \,.
  \ee
We use this map to obtain the desired map
$G^{\rm bulk}_{f,ij}{:}~\mathcal{H}(\Xh'_{ij})\To\mathcal{H}(\Xh)$ as
  \be
  G^{\rm bulk}_{f,ij}(v) := \hat g_{f_i,i} \cir \hat g_{f_j,j}(
  v \oti B^-_{\bar\imath i} \oti B^-_{\bar\jmath j}) \,.
  \ee

\subsubsection*{The bulk factorisation}

We have now gathered all the ingredients to formulate the consistency of the
assignment of correlators $\X \,{\mapsto}\, C(\X)$ with bulk factorisation.

\dtlTheorem{thm:bulk-or}
{\em Bulk factorisation for oriented world sheets\/}:
Let \X\ be an oriented world sheet and let $f{:}\ \Aeps \To \X$
be injective, continuous and orientation preserving. Then
  \be
  C_\AA(\X) = \sum_{i,j \in\I} \sum_{\alpha,\beta} \dim(U_i) \dim(U_j)
  ({\Cbulk_{i,j}}^{\,\scriptstyle -1})^{}_{\beta\alpha}\,
  G^{\rm bulk}_{f,ij}\big( C_\AA( \cblk_{f,i,j,\alpha,\beta}(\X) ) \big) \,,
  \labl{eq:bulk-or}
where $\alpha$ and $\beta$ run over bases of $\,\HomAA(U_i \oT+ A \ot- U_j, A)$ 
and $\,\HomAA(U_{\bar\imath} \oT+ A \ot- U_{\bar\jmath}, A)$, respectively.
\ENDT

\dtlTheorem{thm:bulk-unor}
{\em Bulk factorisation for unoriented world sheets\/}:
Let \X\ be an unoriented world sheet and let $f{:}\ \Aeps \To \X$
be injective and continuous. Then
  \be
  C_{\!\tilde A}(\X) = \sum_{i,j \in\I} \sum_{\alpha,\beta} \dim(U_i) \dim(U_j)
  ({\Cbulk_{i,j}}^{\,\scriptstyle -1})^{}_{\beta\alpha}\, G^{\rm bulk}
  _{f,ij}\big( C_{\!\tilde A}( \cblk_{f,i,j,\alpha,\beta}(\X) ) \big)\,, 
  \labl{eq:bulk-unor}
where $\alpha$ and $\beta$ run over bases of
$\,\Hom_{\!\tilde A |\tilde A}(U_i \oT+ \tilde A \ot- U_j, \tilde A)$ and
$\,\Hom_{\!\tilde A |\tilde A}(U_{\bar\imath} \oT+ \tilde A \ot- U_{\bar\jmath},
\tilde A)$, respectively.
\ENDT


\sect{Proof of modular invariance} \label{sec:modinv}

This section presents the proof that the oriented correlators $C_\AA(\X)$ are 
invariant under the action of the mapping class group ${\rm Map}_{\rm or}(\X)$, 
and that the unoriented correlators $C_{\!\tilde A}(\X)$ are invariant under 
the action of ${\rm Map}(\X)$. To this end, we will prove the more general 
covariance results in theorems \ref{thm:iso-or} and \ref{thm:iso-unor} which, 
when specialised to $\X\eq{\rm Y}$, imply invariance under the action of the 
mapping class group via the corollaries \ref{cor:map-or} and \ref{cor:map-unor}.

Our strategy is to reduce the covariance property to the statement that the 
construction of the correlators is independent of the choice of 
triangulation.\,\footnote{~%
  As in appendix \ref{A2}, by a triangulation we shall mean a cell decomposition
  in which each vertex is three-valent and a face can have arbitrarily many
  edges.}
We will treat the case of oriented world sheets in detail 
and then point out the differences in the unoriented case.

\subsection{Oriented world sheets}

$~$

\noindent
Let us first establish the independence of our prescription of the choice of
triangulation:

\dtlProposition{prop:triangulation}
Let $\X$ be an oriented world sheet.
The correlator $C_\AA(\X)$ given by the prescription in appendix \ref{The ass}
does not depend on the particular choice of triangulation of the embedded
world sheet $\imath_{\X}(\X)\,{\subset}\,\MX$.
\ENDP

The proof of this statement is contained in the lemmas
\ref{lem:fusion}\,--\,\ref{lem:bdy-bubble} that will be established below.
To prepare the stage for these lemmas, we first recall that
any two triangulations of a surface can be connected by a series of only two
types of moves, the {\em bubble\/} move and the ({\sl2},{\sl2}) or {\em fusion\/}
move (also known as the \twodim\ Matveev moves), see e.g.\ \cite{tuvi,fuhk,kamo}. 
These look as follows:
  \bea \begin{picture}(368,65)(0,30)
  \put(0,0)       {\includeourbeautifulpicture 31 }
  \put(66.3,50.6) {\small fusion}
  \put(69.6,39.1) {$ \longleftrightarrow $}
  \put(214,39.1)  {and}
  \put(303,50.6)  {\small bubble}
  \put(308,39.1)  {$ \longleftrightarrow $}
  \epicture17 \labl{dual-moves}

Now as described in the appendix, the construction of the ribbon graph for
the correlator $C_\AA(\X)$ requires to place $A$-ribbons along the interior edges
of a chosen triangulation of the world sheet \X\ and to connect them suitably
at the vertices of the triangulation. This procedure involves several arbitrary
choices.  The correlator, which is the invariant of the ribbon graph, does, 
however, not depend on these choices (see again the appendix), so that it is 
sufficient to
consider just one specific possibility for each choice involved. Concretely,
at each vertex of the triangulation that lies in the interior of $\X$, there are
three different possibilities of joining the three outgoing $A$-ribbons (either
using the ribbon graph fragment displayed in figure \erf{fig:vertex-graph}, or
using the graphs obtained from that figure by a $2\pi/3$- or $4\pi/3$-rotation
of the part contained in the dashed circle), and we choose any one of them.
Similarly, for joining two incoming $A$-ribbons along an edge that connects two
interior vertices we select one of the two possibilities to insert the left 
ribbon graph fragment in figure \erf{edge-graph}.
Finally, any boundary segment is covered by a ribbon labelled by an $A$-module
$M$; at a vertex of the triangulation lying on that segment of $\partial\X$, we
place the graph \erf{boundary-graph}.

When we want to make the ribbon graph $R$ embedded in a cobordism $\rmM$ 
explicit, we will use the notation $\rmM[R]$. For a given triangulation $T$ of 
\X, let $R_T$ be the ribbon graph in $\MX$ constructed as above. To show that 
$Z(\MX[R_T]) \eq Z(\MX[R_{T'}])$ for any two triangulations $T$ and $T'$ of \X, 
it suffices to consider triangulations $T$ and $T'$ that differ only by a 
single move, either a fusion or a bubble move. Let us start with the fusion 
move. We cut out a three-ball in the three-manifold $\MX$ that contains the 
part in which $T$ and $T'$ differ. We can project the ribbon graph fragment
contained in that three-ball in a non-singular
manner to a plane (for instance, to a plane patch of $\imath(\X)$).
The resulting graph can then be interpreted as a morphism in the category \C.
Let us for the moment restrict our attention to the case that
only interior vertices of the triangulation are involved. Then from $\MX[R_T]$ 
and $\MX[R_{T'}]$ we arrive at the two sides of the picture
  \bea \begin{picture}(380,153)(0,37)
  \put(3,0)        {\Includeourbeautifulpicture 35a }
  \put(201,0)      {\Includeourbeautifulpicture 35b }
  \put(0,-9)       {\scriptsize $ A^\vee $}
  \put(28,-9)      {\scriptsize $ A^\vee $}
  \put(112,187)    {\scriptsize $ A $}
  \put(85,187)     {\scriptsize $ A $}
  \put(195,-9)     {\scriptsize $ A^\vee $}
  \put(289,-9)     {\scriptsize $ A^\vee $}
  \put(240,187)    {\scriptsize $ A $}
  \put(360,187)    {\scriptsize $ A $}
  \put(143.3,96.6) {\small fusion}
  \put(146.6,85.1) {$\longleftrightarrow$}
  \epicture23 \labl{A-fusion-move}
Similarly, in the case of the bubble move we obtain
  \bea \begin{picture}(310,319)(0,56)
  \put(0,0)        {\Includeourbeautifulpicture 36a }
  \put(240,0)      {\Includeourbeautifulpicture 36b }
  \put(14,-9)      {\scriptsize $ A $}
  \put(250,-9)     {\scriptsize $ A $}
  \put(250,371)    {\scriptsize $ A^\vee $}
  \put(113,371)    {\scriptsize $ A^\vee $}
  \put(172,190.6)  {\small bubble}
  \put(177,179.1)  {$\longleftrightarrow$}
  \epicture39 \labl{A-bubble-move}
We now proceed to prove that the two sides of \erf{A-fusion-move}, as well
as those of \erf{A-bubble-move}, are actually equal.

\dtlLemma{lem:fusion}
The two sides of the fusion move \erf{A-fusion-move} coincide.
\ENDL

\Proof
We start by noticing that
  \begin{eqnarray} \begin{picture}(420,168)(0,0)
  \put(3,0)        {\Includeourbeautifulpicture 37a }
  \put(159,0)      {\Includeourbeautifulpicture 37b }
  \put(323,0)      {\Includeourbeautifulpicture 37c }
  \put(0,-9)       {\scriptsize $ A^\vee $}
  \put(28,-9)      {\scriptsize $ A^\vee $}
  \put(114,167)    {\scriptsize $ A $}
  \put(85,167)     {\scriptsize $ A $}
  \put(153,-9)     {\scriptsize $ A^\vee $}
  \put(181,-9)     {\scriptsize $ A^\vee $}
  \put(237,167)    {\scriptsize $ A $}
  \put(266.3,167)  {\scriptsize $ A $}
  \put(318,-9)     {\scriptsize $ A^\vee $}
  \put(345,-9)     {\scriptsize $ A^\vee $}
  \put(375,167)    {\scriptsize $ A $}
  \put(403,167)    {\scriptsize $ A $}
  \put(138,79)     {$ = $}
  \put(292,79)     {$ = $}
  \end{picture} \nonumber\\~ \label{V37}
  \end{eqnarray} 
Here the first equality is a simple deformation of the graph, while in the 
second equality 
the Frobenius property, together with the unit and counit properties, is used.
\\
Next we perform again a slight deformation so as to arrive at the first graph in
the following sequence of equalities:
  \begin{eqnarray} \begin{picture}(420,168)(5,0)
  \put(3,0)        {\Includeourbeautifulpicture 39a }
  \put(180,0)      {\Includeourbeautifulpicture 39b }
  \put(322,0)      {\Includeourbeautifulpicture 39c }
  \put(-2,-9)      {\scriptsize $ A^\vee $}
  \put(25,-9)      {\scriptsize $ A^\vee $}
  \put(100,167)    {\scriptsize $ A $}
  \put(128,167)    {\scriptsize $ A $}
  \put(207,-9)     {\scriptsize $ A^\vee $}
  \put(174,-9)     {\scriptsize $ A^\vee $}
  \put(247,167)    {\scriptsize $ A $}
  \put(276,167)    {\scriptsize $ A $}
  \put(317,-9)     {\scriptsize $ A^\vee $}
  \put(345,-9)     {\scriptsize $ A^\vee $}
  \put(387,167)    {\scriptsize $ A $}
  \put(420,167)    {\scriptsize $ A $}
  \put(153,79)     {$ = $}
  \put(297,79)     {$ = $}
  \end{picture} \nonumber\\~ \label{V39}
  \end{eqnarray}
Here the first equality follows by using coassociativity twice, while the second
is again a direct consequence of the Frobenius, unit and counit properties.
The last graph of the previous picture can be deformed to equal
the first graph in
  \begin{eqnarray} \begin{picture}(420,168)(20,0)
  \put(3,0)        {\Includeourbeautifulpicture 40a }
  \put(199,0)      {\Includeourbeautifulpicture 40b }
  \put(352,0)      {\Includeourbeautifulpicture 40c }
  \put(-2,-9)      {\scriptsize $ A^\vee $}
  \put(30,-9)      {\scriptsize $ A^\vee $}
  \put(68,167)     {\scriptsize $ A $}
  \put(100,167)    {\scriptsize $ A $}
  \put(193,-9)     {\scriptsize $ A^\vee $}
  \put(225,-9)     {\scriptsize $ A^\vee $}
  \put(297,167)    {\scriptsize $ A $}
  \put(264,167)    {\scriptsize $ A $}
  \put(391,-9)     {\scriptsize $ A^\vee $}
  \put(432,-9)     {\scriptsize $ A^\vee $}
  \put(393,167)    {\scriptsize $ A $}
  \put(434,167)    {\scriptsize $ A $}
  \put(175,79)     {$ = $}
  \put(329,79)     {$ = $}
  \end{picture} \nonumber\\[.3em]
  {} \label{fig:V-40} \end{eqnarray}
The first of these equalities follows by coassociativity and the symmetry 
property (compare (I:5.9)), and the second by coassociativity together 
with another deformation of the graph.
Finally, from the last picture in \erf{fig:V-40}, the right hand side of
the fusion move \erf{A-fusion-move} is obtained by use of the symmetry
property of $A$.
\qed

\dtlLemma{lem:bubble}
The two sides of the bubble move \erf{A-bubble-move} coincide.
\ENDL

\Proof
Consider the following moves.
  \begin{eqnarray}
  \begin{picture}(420,273)(13,0)
  \put(0,0)        {\Includeourbeautifulpicture 38a }
  \put(162,0)      {\Includeourbeautifulpicture 38b }
  \put(340,0)      {\Includeourbeautifulpicture 38c }
  \put(0,-9)       {\scriptsize $ A^\vee $}
  \put(189,-9)     {\scriptsize $ A^\vee $}
  \put(468,272)    {\scriptsize $ A $}
  \put(297,272)    {\scriptsize $ A $}
  \put(115,272)    {\scriptsize $ A $}
  \put(367,-9)     {\scriptsize $ A^\vee $}
  \put(140,130)    {$ = $}
  \put(315,130)    {$ = $}
  \end{picture}
  \nonumber
     \end{eqnarray}\begin{eqnarray}
  \begin{picture}(420,300)(13,0)
  \put(135,0)      {\Includeourbeautifulpicture 38d }
  \put(287,0)      {\Includeourbeautifulpicture 38e }
  \put(385,0)      {\Includeourbeautifulpicture 38f }
  \put(163,-9)     {\scriptsize $ A^\vee $}
  \put(315,-9)     {\scriptsize $ A^\vee $}
  \put(316,272)    {\scriptsize $ A $}
  \put(234.5,272)  {\scriptsize $ A $}
  \put(396.4,272)  {\scriptsize $ A $}
  \put(394,-9)     {\scriptsize $ A^\vee $}
  \put(106,130)    {$ = $}
  \put(260,130)    {$ = $}
  \put(355,130)    {$ = $}
  \end{picture}
  \nonumber \\[.3em] \label{bubble-proof}
  \end{eqnarray}
The first equality is a simple deformation of the graph; the second equality
follows by the Frobenius property, while the third requires Frobenius, symmetry
and the counit property; the next-to-last step follows by specialness and the
Frobenius property, and the last by the symmetry property.
\\
Finally, by composing the first and last graph in this sequence both from the
bottom and from the top with the \rhs\ of \erf{A-bubble-move}\,%
 \footnote{~When starting directly with the
 \lhs\ of \erf{A-bubble-move}, rather than with the \lhs\ of \erf{bubble-proof},
 it is actually convenient to perform the various manipulations in a slightly
 different order from what we have done here.}
after a few simple manipulations we arrive at the graphs on the
two sides of the bubble move \erf{A-bubble-move}, respectively.
\qed

When also vertices of the triangulation that lie on $\partial\X$ are involved,
then instead of \erf{A-fusion-move} and \erf{A-bubble-move} the fusion and bubble
move amount to
  \bea \begin{picture}(360,140)(0,43)
  \put(3,0)        {\Includeourbeautifulpicture 44a }
  \put(198,0)      {\Includeourbeautifulpicture 44b }
  \put(-1,182)     {\scriptsize $ A $}
  \put(27,182)     {\scriptsize $ A $}
  \put(97,182)     {\scriptsize $ \dot M $}
  \put(96,-9)      {\scriptsize $ \dot M $}
  \put(284,182)    {\scriptsize $ A $}
  \put(311,182)    {\scriptsize $ A $}
  \put(342,182)    {\scriptsize $ \dot M $}
  \put(340,-9)     {\scriptsize $ \dot M $}
  \put(143.3,96.6) {\small fusion}
  \put(146.6,85.1) {$\longleftrightarrow$}
  \epicture24 \labl{M-fusion-move}
and to
  \bea \begin{picture}(300,140)(0,49)
  \put(0,0)        {\Includeourbeautifulpicture 45a }
  \put(243,0)      {\Includeourbeautifulpicture 45b }
  \put(123.5,182)  {\scriptsize $ \dot M $}
  \put(123,-9)     {\scriptsize $ \dot M $}
  \put(239.4,182)  {\scriptsize $ \dot M $}
  \put(238,-9)     {\scriptsize $ \dot M $}
  \put(170,95.6)   {\small bubble}
  \put(175,84.1)   {$\longleftrightarrow$}
  \epicture29 \labl{M-bubble-move}
respectively.

\dtlLemma{lem:bdy-fusion}
The two sides of the fusion move \erf{M-fusion-move} coincide.
\ENDL

\Proof
Invoking the module property, the \lhs\ of the fusion move \erf{M-fusion-move}
becomes the \lhs\ of the following sequence of equalities:
  \bea \begin{picture}(380,146)(0,42)
  \put(3,0)        {\Includeourbeautifulpicture 46a }
  \put(156,0)      {\Includeourbeautifulpicture 46b }
  \put(265,0)      {\Includeourbeautifulpicture 46c }
  \put(-0.5,182)   {\scriptsize $ A $}
  \put(28,182)     {\scriptsize $ A $}
  \put(97,182)     {\scriptsize $ \dot M $}
  \put(96,-9)      {\scriptsize $ \dot M $}
  \put(303,182)    {\scriptsize $ A $}
  \put(330,182)    {\scriptsize $ A $}
  \put(362,182)    {\scriptsize $ \dot M $}
  \put(360,-9)     {\scriptsize $ \dot M $}
  \put(153,182)    {\scriptsize $ A $}
  \put(181,182)    {\scriptsize $ A $}
  \put(212,182)    {\scriptsize $ \dot M $}
  \put(210,-9)     {\scriptsize $ \dot M $}
  \put(128,87)     {$ = $}
  \put(239,87)     {$ = $}
  \epicture26  \labp 46
The first equality uses just the Frobenius, unit and coassociativity 
properties, and the second follows by the Frobenius, unit and counit
properties. The \rhs\ of \erp 46, in turn, is just a deformation of the
ribbon graph on the \rhs\ of the fusion move \erf{M-fusion-move}.
\qed

\dtlLemma{lem:bdy-bubble}
The two sides of the bubble move \erf{M-bubble-move} coincide.
\ENDL

\Proof
A deformation of the \lhs\ of the bubble move \erf{M-bubble-move}
results in the \lhs\ of
  \bea \begin{picture}(330,142)(0,42)
  \put(0,0)        {\Includeourbeautifulpicture 47a }
  \put(140,0)      {\Includeourbeautifulpicture 47b }
  \put(268,0)      {\Includeourbeautifulpicture 47c }
  \put(78.5,182)   {\scriptsize $ \dot M $}
  \put(77,-9)      {\scriptsize $ \dot M $}
  \put(194,182)    {\scriptsize $ \dot M $}
  \put(193,-9)     {\scriptsize $ \dot M $}
  \put(288.4,182)  {\scriptsize $ \dot M $}
  \put(287.8,-9)   {\scriptsize $ \dot M $}
  \put(108,87)     {$ = $}
  \put(231,87)     {$ = $}
  \epicture26  \labp 47
The first equality employs the Frobenius, unit and counit properties, while
the second uses the module property and specialness. That the \rhs\ of
\erp 47 is equal to the \rhs\ of the bubble move \erf{A-bubble-move} is
a direct consequence of the (unit) module property.
\qed

This completes the proof of proposition \ref{prop:triangulation}.
Proving theorem~\ref{thm:iso-or} is now straightforward.  

\proofof[theorem~{\ref{thm:iso-or}}]

We are given two oriented world sheets $\X$ and ${\rm Y}$
and an orientation preserving homeomorphism $f{:}\ \X \To {\rm Y}$.
To obtain the correlator $C_\AA(\X)$ via the construction of appendix 
\ref{The ass} we need to choose a triangulation $T_{\X}$ of the world sheet 
\X, i.e.\ an embedding $T_{\X}{:}\ K\To\X$ for some simplicial complex $K$. 
Similarly, to obtain $C_\AA({\rm Y})$ we have to choose a triangulation 
$T_{\rm Y}$ of the world sheet ${\rm Y}$. By definition, we have $C_\AA(\X) 
\eq Z(\MX[R_{T_{\X}}])$ and $C_\AA({\rm Y}) \eq Z(\rmM_{\rm Y}[R_{T_{\rm Y}}])$,
and by proposition \ref{prop:triangulation} this definition is independent of 
the choice of $T_{\X}$ and $T_{\rm Y}$. On ${\rm Y}$ we can define a second 
triangulation in terms of the triangulation on $\X$ and the map $f$ as 
$T_{\rm Y}' \,{:=}\, f\cir T_\X$. By construction, we have the relation
  \be
  \rmM_{\hat f} \cir \MX[R_{T_{\X}}] \,\cong\, \rmM_{\rm Y}[R_{T_{\rm Y}'}] \,,
  \labl{eq:M-hom-M}
where the notation `$\cong$' indicates that there exists a homeomorphism 
between the two cobordisms that acts as the identity on the boundary, and
where $\hat f$ is the map that was defined in section \ref{sec:inv-map}.
We can then compute
  \be\bearll
  \hat f_\sharp \cir C_\AA(\X) \!\!&
  = Z(\rmM_{\hat f}) \cir Z(\MX[R_{T_{\X}}]) \\{}\\[-.5em]
  &= Z(\rmM_{\rm Y}[R_{T_{\rm Y}'}]) = Z(\rmM_{\rm Y}[R_{T_{\rm Y}}])
  =  C_\AA({\rm Y}) \,,
  \eear\ee
where the first equality is the definition of $f_\sharp$ and $C_\AA(\X)$,
in the second step we use functoriality\,\footnote{~%
  By construction, the push-forward $\hat{f}_*$ of the Lagrangian subspace
  $\lambda(\Xh)\iN H_1(\Xh,\reals)$ for $\Xh$ gives the Lagrangian subspace
  for $\widehat{\rm Y}$. It is not hard to convince oneself that, as a
  consequence, in the present situation the Maslov indices in the anomaly
  factor of the functoriality relation for $Z$ vanish.\label{foot:lag}}
of $Z$ and \erf{eq:M-hom-M}, the third step is triangulation invariance as 
established in proposition \ref{prop:triangulation}, and the last step is 
just the definition of $C({\rm Y})$.
\qed

\dtRemark
Given a world sheet $\X$ and homeomorphisms $f,\, g{:}\ \X \To \X$, by the 
argument of footnote \ref{foot:lag} we have
$Z(\rmM_{\hat f}\cir\rmM_{\hat g}) \eq Z(\rmM_{\hat f}) \cir Z(\rmM_{\hat g})$, 
i.e.\ the anomaly factor in the functoriality relation vanishes in this case. 
Thus we obtain a genuine (non-projective) representation of ${\rm Map}_{\rm or}
(\X)$ (and of ${\rm Map}(\X)$ in the unoriented case) on $\mathcal{H}(\Xh)$. 
\\
This situation for the subgroup ${\rm Map}(X)\,{\subset}\,{\rm Map}(\Xh)$ 
should be contrasted to the situation for the whole mapping class group
${\rm Map}(\Xh)$. For that group, one only obtains (see formula 
\erf{eq:rep-Map}) a projective representation on $\mathcal{H}(\Xh)$.
\ENDR

\subsection{Unoriented world sheets}

$~$

\noindent
For the construction of correlators on unoriented world sheets
the following statement holds.

\dtlProposition{prop:triangulation-unor}
Let $\X$ be an unoriented world sheet. The correlator $C_{\!\tilde A}(\X)$ 
given by the prescription in appendix \ref{The ass}
does not depend on the particular choice of triangulation of the embedded
world sheet $\imath_{\X}(\X)\,{\subset}\,\MX$.
\ENDP

The proof of proposition \ref{prop:triangulation-unor} proceeds analogously to 
that of proposition \ref{prop:triangulation}.
The main difference is that one has to choose a triangulation $T$ of $\X$
together with a local orientation of $\X$ at every vertex of $T$.
(And there is the obvious modification that in all pictures
the algebra ribbons are now labelled by $\tilde A$.) 
Changing these local orientations does not affect $C_{\!\tilde A}(\X)$, as has 
been demonstrated in section II:3.1. To show that $C_{\!\tilde A}(\X)$ does
not depend on the triangulation, one shows independence under the fusion and
bubble move by choosing the local orientations at the two vertices
involved to be equal and copying the argument of the previous section.

\medskip

The proof of theorem \ref{thm:iso-unor} is analogous to the oriented case, too, 
the main difference being that $\X$ and ${\rm Y}$ are not oriented, and as a
consequence the homeomorphism $f{:}\ \X \To {\rm Y}$ can in general not be 
required to be orientation preserving. However, it can still
be used to obtain a triangulation of ${\rm Y}$ with local orientations at its 
vertices from a triangulation of $\X$ with local orientations at its vertices. 
The proof then proceeds as  the proof of theorem \ref{thm:iso-or} above.


\sect{Proof of boundary factorisation} \label{sec:bdryfac}

In this section we present the proof of theorems~\ref{thm:bnd-or} and
\ref{thm:bnd-unor}, which describe boundary factorisation.
We first develop some ingredients to be used;
there are two main aspects, a geometric and an algebraic one.
A crucial feature of the geometric aspect of factorisation, indeed the very
feature that allows for a general proof of this property, is that it is of
local nature. Here the qualification `local' is used to indicate that 
factorisation only involves a neighbourhood of the interval along which the 
world sheet is cut, i.e.\ the image of the set $\Reps$ in the notation of 
section \ref{sec:facbdry}.

\subsection{Geometric setup of boundary factorisation}\label{sec:bnd-geom}

$~$

\noindent
The construction of this subsection applies both to oriented and to unoriented 
world sheets. Accordingly, below $C$ stands for $C_\AA$ or $C_{\!\tilde A}$. 
The reason that at this point no distinction between the two cases is 
necessary is that the part of the world sheet that is of interest to us
-- the image of $\Reps$ -- inherits an orientation from $\Reps$. It is 
this local orientation of the world sheet that is used in the construction.

Let \X\ and $f$ be as in theorem \ref{thm:bnd-or}.
The first step is to make explicit the geometry of both sides of formulas 
\erf{eq:bnd-or} and \erf{eq:bnd-unor}, i.e.\ of $C(\X)$ and of the summands
  \be \dim(U_k)\,
  {(c^{\rm bnd~~~~-1}_{M_l,M_r,k})}^{}_{\beta\alpha}\,
  G^{\rm bnd}_{f,k}\big(\, C( \cbnd_{f,k,\alpha,\beta}(\X) ) \,\big) \,.
  \ee
Consider a strip-shaped region in
{\X} in the neighbourhood of which the `cutting twice' operation \erf{eq:pic-i3}
is going to be performed. The image of that neighbouring region in {\MX} can
be drawn, as a subset of $\mathbb{R}^3$, as a solid cylinder {\rm Z} containing
two straight ribbons labelled $M_r$ and $M_l$, as shown in the following figure
(the shaded region shows the embedding of the strip-shaped region of $\X$ in 
$\MX$):
  \bea \begin{picture}(220,95)(0,54)
  \put(60,0)        {\Includeourbeautifulpicture 50c }
  \put(80.3,98)     {\scriptsize $M_l$}
  \put(118.5,42)    {\scriptsize $M_r$}
  \put(188.8,132.5) {\small $\MX$}
  \epicture31  \labl{fig:MX}
The cylindrical part of $\partial{\rm Z}$ is part of the double
$\Xh\eq\partial\MX$, and the rest of {\MX}, which is not shown here, is
attached to the top and bottom bounding disks that make up the remaining
part of $\partial{\rm Z}$. With suitable choice of coordinates, the
former part of $\partial{\rm Z}$ is given by
  \be
  \{(x,y,z)\iN\mathbb{R}^3 \,|\, x^2{+}z^2\eq2,\, y\iN[-3,3]\}
  \subset \Xh \subset \MX \,,
  \ee
with orientation induced by the inward pointing normal, and the corresponding
part of the embedded world sheet is the strip
  \be  \{[-1,1]\Times[-3,3]\Times\{0\}\}\subset \imath(\X) \subset \MX
  \,.  \ee
The two ribbons, which touch $\imath(\partial\X)$, are then the strips
  \be
  \{[-1,-\Frac9{10}]\Times[-3,3]\Times\{0\}\} \qquad{\rm and}\qquad
  \{[\Frac9{10},1]\Times[-3,3]\Times\{0\}\}
  \ee
in the $z{=}0$-plane, with orientation opposite to that inherited from the
local orientation of \iX, and core-orientation opposite to the orientation of
$\imath(\partial\X)$. The left ribbon is labelled by the $A$-module $M_l$, and
the right one is labelled by $M_r$.

\smallskip

Next consider the gluing homomorphism $G_{f,k}^{\rm bnd}$ introduced in
\erf{eq:Gbnd-def}. This is obtained using the gluing homomorphism $\hat g_{f,k}$
defined in equation~\erf{eq:glue2}, with $E\eq\Xh$. We identify part of 
$\widehat{\rm M}_{f,k}(\Xh)$ (as defined above \erf{eq:picV-41})
with the following region embedded in $\mathbb{R}^3$:
  \bea \begin{picture}(200,105)(0,64)
  \put(60,0)        {\includeourbeautifulpicture 51 }
  \put(110,113)     {\scriptsize $\bar{k}$}
  \put(110,47)      {\scriptsize ${k}$}
  \put(74,122)      {\scriptsize $(U_{\bar{k}},-)$}
  \put(117,105)     {\scriptsize $(U_{\bar{k}},+)$}
  \put(74,53)       {\scriptsize $(U_{k},+)$}
  \put(117,36)      {\scriptsize $(U_{k},-)$}
  \put(185,155)     {\small $\widehat{\rm M}_{f,k}(\Xh)$}
  \epicture39  \labl{fig:Mfk}
This region is a solid cylinder with some parts cut out, and with two
vertical ribbons labelled $k$ and $\bar{k}$, respectively. The outer boundary,
which is part of \Xh, is as in figure \erf{fig:MX} above, while the inner
boundary has three disconnected components, two of them being hemispheres and
the third a sphere $s$. The two hemispheres are part of the boundary component
isomorphic to the double of $\cbnd_{\!f,k,\alpha,\beta}(\X)$ with opposite
orientation, which we denote by $-\cbndd_{\!f,k,\alpha,\beta}(\X)$.
The top and bottom bounding annuli of the solid
cylinder are attached to the rest of $\widehat{\rm M}_{f,k}$.

The sphere $s$ is a unit sphere centered at $(0,0,0)$, while the hemispheres
are centered at $(0,3,0)$ and at $(0,-3,0)$, respectively, and have unit radius,
too. The two ribbons lie both in the $z{=}0$-plane, and are centered about the
line $x\eq 0$, with width $1/10$. The upper ribbon, labelled by $\bar{k}$,
ends on marked arcs at $y\eq 2$ and $y\eq 1$, labelled by $(U_{\bar{k}},-)$
and by $(U_{\bar{k}},+)$, respectively, while the lower ribbon is labelled by
$k$ and ends on marked arcs at $y\eq{-}1$ and $y\eq{-}2$, labelled $(U_k,+)$
and $(U_k,-)$. The orientation of the arcs is in the $+x$-direction at $y\eq1$
and at $y\eq{-}2$, while it is in the $-x$-direction at $y\eq{-}1$ and at
$y\eq2$; the core of the upper and lower ribbon is oriented in the $+y$- and 
$-y$-direction, respectively.

\smallskip

{}From $\widehat{\rm M}_{f,k}$ we now construct a new cobordism 
$\widetilde{\rm M}_{f,k}$ by gluing the cobordism $B^-_{\bar k k}$ to $s$. 
There is a canonical isomorphism between the extended surfaces $\partial 
B^-_{\bar k k}$ and $-s$, and we use this isomorphism for identification.
The result, shown in figure~\erf{fig:MB}, differs from $\widehat{\rm M}_{f,k}$ 
in the central region, where now the two ribbons are joined by a coupon that 
is labelled by the morphism $\bar\lambda^{k \bar k}$:
  \bea \begin{picture}(210,106)(0,56)
  \put(60,0)        {\includeourbeautifulpicture 52 }
  \put(193,143)     {\small $\widetilde{\rm M}_{f,k}$}
  \put(108,108)     {\tiny $\bar{k}$}
  \put(107,54)      {\tiny ${k}$}
  \put(109,122)     {\scriptsize $(U_{\bar{k}},-)$}
  \put(109,39)      {\scriptsize $(U_{k},-)$}
  \put(73.3,88.8)   {\tiny \begin{turn}{270}$\bar\lambda^{k\bar{k}}$\end{turn}}
  \epicture34  \labl{fig:MB}
Now we have the equality
  \be
  G^{\rm bnd}_{f,k}
  = Z(\widetilde{\rm M}_{f,k},\cbndd_{\!f,k,\alpha,\beta}(\X),\Xh)
  \,,  \ee
and can therefore write
  \be
  G^{\rm bnd}_{f,k}(C(\cbnd_{\!f,k,\alpha,\beta}(\X)))
  = Z(\widetilde{\rm M}_{f,k},\cbndd_{\!f,k,\alpha,\beta} (\X),\Xh)
  \circ Z({\rm M}_{\cbnd_{\!f,k,\alpha,\beta}(\X)},\emptyset,
    \cbndd_{f,k,\alpha,\beta} (\X))1
  \,.  \ee
By functoriality the latter invariant can be replaced by
$Z({\rm M}_k(\X),\emptyset,\Xh)$, where the cobordism ${\rm M}_k(\X)$ is
defined by gluing ${\rm M}_{\cbnd_{\!f,k,\alpha,\beta}(\X)}$ to the inside of
$\widetilde{\rm M}_{f,k}$, using the canonical identification. Furthermore, both
Maslov indices relevant for the functoriality formula vanish 
(by the formula for the Maslov index in section 2.7 of \cite{fffs3}) since, 
by construction, two of the Lagrangian subspaces in their argument coincide.
             
The relevant part of ${\rm M}_k(\X)$ has the form
  \bea \begin{picture}(220,105)(0,59)
  \put(60,0)        {\includeourbeautifulpicture 53 }
  \put(192.2,145)   {\small ${\rm M}_{k}(\X)$}
  \put(71,148)      {\scriptsize $M_l$}
  \put(131,148)     {\scriptsize $M_r$}
  \put(71,12)       {\scriptsize $M_l$}
  \put(131,12)      {\scriptsize $M_r$}
  \put(122.7,110)   {\scriptsize $\bar{k}$}
  \put(85,51)       {\scriptsize ${k}$}
  \put(100.4,120)   {\tiny\begin{turn}{90} $\psi^-_{\beta}$\end{turn}}
  \put(64.2,89)     {\tiny \begin{turn}{270}$\bar\lambda^{k\bar{k}}$\end{turn}}
  \put(103.5,45.5)  {\tiny\begin{turn}{270} $\psi^+_{\alpha}$\end{turn}}
  \epicture39  \labl{fig:Mk(X)}
This contains half-annular ribbons in the $z{=}0$-plane, with unit radius,
piercing the top and bottom boundary disks, respectively, as well as two
additional coupons labelled by the morphisms $\psi_\alpha^+\iN \HomA
(M_l \oti U_{k}, M_r)$ and $\psi_\beta^-\iN\HomA(M_r \oti U_{\bar k}, M_l)$.
The shaded regions in \erf{fig:Mk(X)} show the embedding of the
world sheet $\cbnd_{\!f,k,\alpha,\beta}(\X)$ in ${\rm M}_{k}(\X)$
(compare to the rightmost figure in \erf{eq:pic-i6}).

\subsection{Some algebraic identities}\label{sec:alg-id}

$~$

\noindent
The second ingredient to be used in the proof of boundary factorisation
are some algebraic identities for certain morphism spaces; they are needed
to transform the ribbon graph in \erf{fig:Mk(X)}. Consider a symmetric special 
Frobenius algebra $A$ in a tensor category \C{} together with left and right
$A$-modules, $_AN$ and $M_A$, respectively. We can define a tensor product over
$A$ of the ordered pair consisting of $M_A$ and $_AN$ as
  \be  M_A \,\otA\, {}_AN := {\rm Im}(P_{MN})
  \,,  \ee
with $P_{MN}$ the idempotent $(\rho_{\rm r}\oti\rho_{\rm l}) \cir
(\id_{\dot{M}_A} \oti (\Delta\cir\eta) \oti \id_{_A\dot{N}})
\iN\End_{\cal C}(\dot{M}\oti\dot{N})$. Here $\rho_{{\rm r}/{\rm l}}$ denote
the right and left action of $A$ on the respective modules; graphically,
  \bea \begin{picture}(150,80)(0,34)
  \put(60,0)        {\includeourbeautifulpicture 54 }
  \put(56,-10)      {\scriptsize $\dot M_A$}
  \put(126,-10)     {\scriptsize ${}_A\dot N$}
  \put(56,112)      {\scriptsize $\dot M_A$}
  \put(126,112)     {\scriptsize ${}_A\dot N$}
  \put(0,50)        {$ P_{MN} \;= $}
  \epicture19 \labl{fig:A-tensor}
Next we note that the dual $\dot M^{\vee}$ of the object underlying a left
$A$-module $M$ carries a natural structure of a right $A$-module via
  \bea \begin{picture}(130,85)(0,36)
  \put(0,0)         {\includeourbeautifulpicture 55 }
  \put(-6,-10)      {\scriptsize $\dot M^{\vee}$}
  \put(19,-9)       {\scriptsize $A$}
  \put(-5,115)      {\scriptsize $\dot M^{\vee}$}
  \put(68,-10)      {\scriptsize $\dot M^{\vee}$}
  \put(86,-9)       {\scriptsize $A$}
  \put(126,115)     {\scriptsize $\dot M^{\vee}$}
  \put(42,53)       {$ :\,= $}
  \epicture25 \labl{fig:dualmod}
What we will use is that the intertwiner space $\HomA(M\oti U,N)$ is canonically
isomorphic to $\Hom^{(P)}(U,M^{\vee}\Oti\,N) \,{:=}\,
\{ f\iN\Hom(U,M^{\vee}\Oti\,N)\,|\,P\cir f\eq f\} $ for $P\equiv P_{M^\vee N}$.
The latter space is relevant because $\Hom^{(P)}(U,M^{\vee}\Oti\,N) \,{\cong}\,
\Hom(U,M^{\vee}\OtA\,N)$, see lemma 2.4 in \cite{ffrs}. An isomorphism 
$\HomA(M\oti U,N)\To\Hom^{(P)}(U,M^{\vee}\Oti\,N)$ is given by the map
  \bea \begin{picture}(390,103)(0,31)
  \put(0,0)          {\Includeourbeautifulpicture 56a }
  \put(66,0)         {\Includeourbeautifulpicture 56b }
  \put(203,0)        {\Includeourbeautifulpicture 56c }
  \put(310,0)        {\Includeourbeautifulpicture 56d }
  \put(.3,-10)       {\scriptsize$ \dot M$}
  \put(10,130)       {\scriptsize$ \dot N$}
  \put(12,-10)       {\scriptsize$ U$}
  \put(103,-10)      {\scriptsize$ \dot M$}
  \put(114,-10)      {\scriptsize$ U$}
  \put(109,130)      {\scriptsize$ \dot N$}
  \put(226,130)      {\scriptsize$ \dot M^{\vee}$}
  \put(335,130)      {\scriptsize$ \dot M^{\vee}$}
  \put(255,-10)      {\scriptsize$ U$}
  \put(362,-10)      {\scriptsize$ U$}
  \put(252,130)      {\scriptsize$ \dot N$}
  \put(358,130)      {\scriptsize$ \dot N$}
  \put(40,55)        {$ = $}
  \put(155,55)       {$ \longmapsto $}
  \put(282,55)       {$ = $}
  \epicture18 \labl{fig:modisom1}
(for the first equality, see lemma I:4.4(ii)), with inverse
  \bea \begin{picture}(390,103)(0,30)
  \put(0,0)          {\Includeourbeautifulpicture 57a }
  \put(105,0)        {\Includeourbeautifulpicture 57b }
  \put(207,0)        {\Includeourbeautifulpicture 57c }
  \put(315,0)        {\Includeourbeautifulpicture 57d }
  \put(-3,130)       {\scriptsize$ \dot M^{\vee}$}
  \put(7,-11)        {\scriptsize$ U$}
  \put(13,130)       {\scriptsize$ \dot N$}
  \put(118,-10)      {\scriptsize$ \dot M$}
  \put(133,-10)      {\scriptsize$ U$}
  \put(138,130)      {\scriptsize$ \dot N$}
  \put(219,-10)      {\scriptsize$ \dot M$}
  \put(342,-10)      {\scriptsize$ \dot M$}
  \put(235,-10)      {\scriptsize$ U$}
  \put(359,-10)      {\scriptsize$ U$}
  \put(240,130)      {\scriptsize$ \dot N$}
  \put(363,130)      {\scriptsize$ \dot N$}
  \put(55,55)        {$ \longmapsto $}
  \put(173,55)       {$ = $}
  \put(282,55)       {$ = $}
  \epicture18 \labl{fig:modisom2}
That these two mappings are indeed each others' inverses follows directly
from the properties of the duality in \C. For a simple object $U_k$ 
we define an isomorphism $\HomA(N\oti U_k,M)  
     $\linebreak[0]$
{\to}\, \Hom^{(P)}
(M^\vee\oti N,U_{\bar k})$, $\psi\,{\mapsto}\, e_\psi$, by 
  \bea \begin{picture}(140,91)(0,33)
  \put(0,0)          {\Includeourbeautifulpicture 43a }
  \put(90,0)         {\Includeourbeautifulpicture 43c }
  \put(-3,-10)       {\scriptsize$ \dot M^{\vee}$}
  \put(13,-10)       {\scriptsize$ \dot N$}
  \put(6.7,121.5)    {\scriptsize$ U_{\bar{k}}$}
  \put(7,56)         {\scriptsize $e_{\psi}$}
  \put(97,-10)       {\scriptsize$ \dot M^{\vee}$}
  \put(113,-10)      {\scriptsize$ \dot N$}
  \put(131.7,121.5)  {\scriptsize$ U_{\bar{k}}$}
  \put(115,54.5)     {\scriptsize $\psi$}
  \put(129,14)       {\scriptsize $\bar\lambda^{k\bar k}$}
  \put(50,55)        {$ := $}
  \epicture20 \labl{fig:fdef}
where $\bar\lambda^{k\bar k} \iN \Hom(\one,U_k\oti U_{\bar k})$ is the basis 
vector introduced in \erf{fig:llbar}. This mapping is indeed an isomorphism; 
its inverse is a similar composition, using a coevaluation instead of the 
evaluation morphism, and $\lambda_{k\bar k}$ instead of $\bar\lambda^{k\bar k}$.

Given a basis $\{\psi_\alpha\}$ of $\HomA(N\oti U_k,M)$, the isomorphism 
\erf{fig:fdef} maps it to a basis $\{e_\alpha\}$ of 
$\Hom^{(P)}(M^{\vee}\Oti\, N,U_{\bar k})$. We also use two bases of 
$\Hom^{(P)}(U_{\bar{k}},M^{\vee}\Oti\, N)$: the basis $\{\bar e_\alpha\}$ 
dual to $\{e_\alpha\}$, and the basis $\{f_\beta\}$ induced from 
$\psi_{\beta}\iN\HomA(M\oti U_{\bar k},N)$ by \erf{fig:modisom1}.

Now let $M\eq M_r$ and $N\eq M_l$. As shown in~(\ref{fig:Dstrconst}), the matrix
of structure constants for the boundary two-point function is given by
  \be
  \big(c^{\rm bnd}_{M_l,M_r,k}\big)_{\alpha\beta} = \dim(U_k)\,
  L_{\alpha\beta} \,,
  \labl{eq:c=dim-L}
where $L \,{\equiv}\, L(M_l,M_r,k)$ is the invertible matrix affording the 
basis transformation, according to $f_\beta \eq \sum_{\gamma} 
L_{\gamma\beta}\,\bar{e}_{\gamma}$ (see \erf{eq:bastr}).

We also use the following

\dtlLemma{lem::mbrel}
For $\{e_{\alpha}\}$ a basis of $\,\Hom^{(P)}(M^{\vee}\Oti\, N,U_{\bar k})$, and
$\{\bar{e}_{\alpha}\}$ a dual basis of $\,\Hom^{(P)}
     $\linebreak[0]$
(U_{\bar k},M^{\vee}\Oti\,N)$,
the equality
  \be
  \sum_{k\in I}\sum_{\alpha}\bar e_{\alpha}\cir e_{\alpha}
  = P_{M^{\vee}N}
  \labl{eq:mbrel}
holds.
\\
Graphically,
  \bea \begin{picture}(240,90)(0,30)
  \put(64,0)         {\Includeourbeautifulpicture 58a }
  \put(150,0)        {\Includeourbeautifulpicture 58b }
  \put(61,-10)       {\scriptsize$ \dot M^{\vee}$}
  \put(61,114)       {\scriptsize$ \dot M^{\vee}$}
  \put(77,53)        {\scriptsize $U_{\bar k}$}
  \put(77,-10)       {\scriptsize$ \dot N$}
  \put(78,114)       {\scriptsize$ \dot N$}
  \put(145,-10)      {\scriptsize$ \dot M^{\vee}$}
  \put(145,114)      {\scriptsize$ \dot M^{\vee}$}
  \put(219,-10)      {\scriptsize$ \dot N$}
  \put(220,114)      {\scriptsize$ \dot N$}
  \put(71,71)        {\footnotesize $\bar e_{\alpha}$}
  \put(71,35.5)        {\footnotesize $e_{\alpha}$}
  \put(0,52)         {$ \dsty \sum_{k\in\I}\,\sum_\alpha $}
  \put(113,52)       {$ = $}
  \epicture19 \labl{fig:modbasrel}
\ENDL

\Proof
The statement is a consequence of dominance in \C, and of $A$ being special 
and Frobenius. Starting from the right hand side, we proceed as follows:
  \begin{eqnarray} \begin{picture}(420,145)(0,0)
  \put(0,0)          {\Includeourbeautifulpicture 42a }
  \put(120,0)        {\Includeourbeautifulpicture 42b }
  \put(285,0)        {\Includeourbeautifulpicture 42c }
  \put(415,0)        {\Includeourbeautifulpicture 42d }
  \put(-5,-10)       {\scriptsize$ \dot M^{\vee}$}
  \put(60,-11)       {\scriptsize$ \dot N$}
  \put(60.8,140.7)   {\scriptsize$ \dot N$}
  \put(-4,140)       {\scriptsize$ \dot M^{\vee}$}
  \put(116,-10)      {\scriptsize$ \dot M^{\vee}$}
  \put(180,-10)      {\scriptsize$ \dot N$}
  \put(180,140.7)    {\scriptsize$ \dot N$}
  \put(115,140.7)    {\scriptsize$ \dot M^{\vee}$}
  \put(285,-10)      {\scriptsize$ \dot M^{\vee}$}
  \put(287,140.7)    {\scriptsize$ \dot M^{\vee}$}
  \put(303,-10)      {\scriptsize$ \dot N$}
  \put(303.8,140.7)  {\scriptsize$ \dot N$}
  \put(413,-10)      {\scriptsize$ \dot M^{\vee}$}
  \put(413,135.7)    {\scriptsize$ \dot M^{\vee}$}
  \put(427,-10)      {\scriptsize$ \dot N$}
  \put(428,135.7)    {\scriptsize$ \dot N$}
  \put(288,64)       {\scriptsize$ U_{\bar k}$}
  \put(428.2,64)     {\scriptsize$ U_{\bar k}$}
  \put(296,51)       {\footnotesize $\tau^a$}
  \put(296,76)       {\footnotesize $\bar{\tau}_a$}
  \put(422,88.1)     {\footnotesize $\bar e_{\alpha}$}
  \put(422,42.5)       {\footnotesize $ e_{\alpha}$}
  \put(88,67)        {$ = $}
  \put(210,67)       {$ =~~ \dsty \sum_{k\in\I}\,\sum_a $}
  \put(340,67)       {$ =~~ \dsty \sum_{k\in\I}\,\sum_\alpha $}
  \end{picture} \nonumber\\[11pt]
  {} \label{fig:mbrelproof} \end{eqnarray}
where $\tau^a \,{\equiv}\, \tau^a_{(M^{\vee}\dot{N})\bar k}$
and $\bar\tau_a \,{\equiv}\, \bar\tau_a^{(M^{\vee}\dot{N})\bar k}$.
In the first step we use the fact that $P_{M^{\vee}N}$ is an idempotent.
Next, by dominance the identity morphism is decomposed as
  \be
  \id_{\dot{M}^{\vee}\otimes\dot{N}}
  = \sum_{k\in\I}\sum_{a}\bar\tau_{a}^{(\dot{M}^{\vee},\dot{N})\bar k}
  \cir \tau^{a}_{(\dot{M}^{\vee},\dot{N})\bar k}
  \labl{dom}
with dual bases $\{ \tau^a_{(\dot M^\vee,\dot N)\bar k} \}$ and
$\{ \bar\tau_{a}^{(\dot M^\vee,\dot N)\bar k} \}$ of $\Hom(\dot M^\vee\oti
 \dot N,U_{\bar k})$ and $\Hom(U_{\bar k},\dot M^\vee\oti\dot N)$,
respectively. Now we may choose the former basis in such a way that each basis 
vector is an eigenvector of $P_{M^{\vee}N}$. Denote the index subset labelling 
the basis vectors with eigenvalue $1$ as $\{\alpha\}$. By definition, the basis 
vectors indexed by $\{\alpha\}$ provide a basis for $\Hom^{(P)}(U_{\bar k}, 
M^\vee\Oti\,N)$, which we can choose to coincide with the basis $e_{\alpha}$ 
defined above. The basis dual to $\{ \tau^a \}$ then has the same decomposition,
so that the effect of the idempotents $P_{M^{\vee}N}$ in \erf{fig:mbrelproof} 
is to restrict the sum from the index set $\{a\}$ to the subset $\{\alpha\}$.
\qed

\subsection{Oriented world sheets}\label{sec:fac-bnd-or}

$~$

\noindent
We are now ready to prove theorem \ref{thm:bnd-or}, that is,
boundary factorisation for oriented world sheets.

\proofof[theorem {\ref{thm:bnd-or}}]

Our task is to establish the equality \erf{eq:bnd-or}. To this end we express 
both sides of \erf{eq:bnd-or} through cobordisms and prove that these two 
cobordisms give the same vector in $\mathcal{H}(\Xh)$. The relevant part of 
the cobordism for the \lhs\ has been displayed in \erf{fig:MX}, while the one 
for the \rhs\ is shown in \erf{fig:Mk(X)}. Note that as three-manifolds,
the two cobordisms coincide, they only differ in the embedded ribbon graph. 
\\
In fact, instead of using the fragment of $\MX[R_T]$ shown in \erf{fig:MX},
which is obtained by applying the construction of appendix \ref{The ass}
for the triangulation $T$ of $\X$, we use the ribbon graph $R_{T'}$ obtained
from a different triangulation $T'$ that is given by adding one edge to $T$.
The additional edge covers the interval along which $\X$ is to be cut.
The ribbon graph $R_{T'}$ differs from $R_T$ by an idempotent $P_{M^\vee N}$
inserted in \erf{fig:MX}. By proposition \ref{prop:triangulation}, changing
the triangulation does not affect the vector in $\mathcal{H}(\Xh)$, 
i.e.\ $Z(\MX[R_T]) \eq Z(\MX[R_{T'}])$.
\\
Projecting the ribbon graphs to the plane (and looking upon them from the 
`white side') we see that in order to establish the theorem it is sufficient
to prove the following equality for morphisms in
$\End({\dot M}^\vee_r \oti {\dot M}_l)$:
  \bea \begin{picture}(220,90)(0,54)
  \put(-260,13){
     \put(180,0)        {\Includeourbeautifulpicture 58b }
     \put(175,-10)      {\scriptsize$ \dot M_r^{\vee}$}
     \put(248,-10)      {\scriptsize$ \dot M_l$}
     \put(175,115)      {\scriptsize$ \dot M_r^{\vee}$}
     \put(248,115)      {\scriptsize$ \dot M_l$}
     }
  \put(140,0){   
     \put(60,0)        {\includeourbeautifulpicture 59 }
     \put(55,-9)       {\scriptsize$ \dot M_r^\vee$}
     \put(111,-9)      {\scriptsize$ \dot M_l^{}$}
     \put(55,146)      {\scriptsize$ \dot M_r^\vee$}
     \put(111,146)     {\scriptsize$ \dot M_l^{}$}
     \put(78.2,92)     {\scriptsize$ \bar{k}$}
     \put(107,51.5)    {\scriptsize$ k$}
     \put(120.5,66)    {\tiny \begin{turn}{90}$\bar\lambda^{k\bar k}$\end{turn}}
     \put(81.4,28)     {\tiny ${\psi^+_{\alpha}}$}
     \put(87,113 )     {\tiny ${\psi^-_{\beta}}$}
     }
  \put(20,70)        {$\dsty =~\sum_{k\in\I} \sum_{\alpha,\beta} \dim(U_k) \,
                         {(c^{\rm bnd~~~~-1}_{M_l,M_r,k})}^{}_{\beta\alpha}$}
  \epicture40  \labl{fig:bdpr1}
To see that this equality holds, start from the \rhs, apply the isomorphisms 
\erf{fig:fdef} and \erf{fig:modisom1} to $\psi^+_{\alpha}$ and $\psi^-_{\beta}$, 
respectively, and expand the latter in terms of the basis $\{\bar e_{\gamma}\}$; 
this gives
  \bea \begin{picture}(70,87)(0,30)
  \put(50,0){
     \put(50,0)        {\Includeourbeautifulpicture 58a }
     \put(46,-10)      {\scriptsize$ \dot M_r^{\vee}$}
     \put(63,-10)      {\scriptsize$ \dot M_l$}
     \put(46,114)      {\scriptsize$ \dot M_r^{\vee}$}
     \put(63,114)      {\scriptsize$ \dot M_l$}
     \put(63,52)       {\scriptsize$ U_{\bar{k}}$}
     \put(57,71.6)     {\footnotesize $\bar e_{\gamma}$}
     \put(57,36)       {\footnotesize $e_{\alpha}$}
     }
  \put(-100,53)        {$ \dsty \sum_{k\in\I} \sum_{\alpha,\beta} \dim(U_k) \,
                        {(c^{\rm bnd~~~~-1}_{M_l,M_r,k})}^{}_{\beta\alpha}
                        \sum_{\gamma}\, L_{\gamma\beta} $}
  \epicture22 \labl{fig:bdpr2}
Using the inverse of the relation \erf{eq:c=dim-L}, carrying out the sum over 
the bases of intertwiner spaces and isomorphism classes of simple objects, 
and applying \erf{fig:modbasrel}, we can rewrite \erf{fig:bdpr2} as the \lhs\ of
  \bea \begin{picture}(290,88)(0,32)
  \put(100,0)        {\Includeourbeautifulpicture 58a }
  \put(180,0)        {\Includeourbeautifulpicture 58b }
  \put(97,-10)       {\scriptsize$ \dot M_r^{\vee}$}
  \put(113,-10)      {\scriptsize$ \dot M_l$}
  \put(97,115)       {\scriptsize$ \dot M_r^{\vee}$}
  \put(113,115)      {\scriptsize$ \dot M_l$}
  \put(113,52)       {\scriptsize$ U_{\bar{k}}$}
  \put(107,71.6)     {\footnotesize $\bar e_{\gamma}$}
  \put(107,36)       {\footnotesize $e_{\alpha}$}
  \put(175,-10)      {\scriptsize$ \dot M_r^{\vee}$}
  \put(248,-10)      {\scriptsize$ \dot M_l$}
  \put(175,115)      {\scriptsize$ \dot M_r^{\vee}$}
  \put(248,115)      {\scriptsize$ \dot M_l$}
  \put(0,53)         {$ \dsty \sum_{k\in\I}\,\sum_{\alpha,\beta,\gamma}
                        L^{}_{\gamma\beta}\, L^{-1}_{\beta\alpha} $}
  \put(145,53)       {$ = $}
  \epicture24 \labl{fig:bdpr3}
which is equal to the \lhs\ of \erf{fig:bdpr1}.
This establishes the equality \erf{fig:bdpr1}.
\qed

\subsection{Unoriented world sheets}\label{sec:bnd-proof-unor}
					
$~$

\noindent
The proof of theorem \ref{thm:bnd-unor} for the case that the map $f$ satisfies
condition (i) of the theorem is identical to the argument given above for 
oriented world sheets. Before we can outline the proof for $f$ satisfying 
condition (ii) instead, we need to present the details on how to obtain the 
world sheet $\cbnd_{\!f,k,\alpha,\beta}(\X)$ in that case.

\subsubsection*{Construction of $\cbnd_{\!f,k,\alpha,\beta}(\X)$ for
$f{:}\ \Reps' \To \X$}

The reason to present these details only here is a technical complication
in the labelling of boundary conditions which makes the description
somewhat lengthy. To construct $\cbnd_{\!f,k,\alpha,\beta}(\X)$ for 
a given map $f{:}\ \Reps' \To \X$ we distinguish two cases:
\Itemize
\item[\Nxt] 
$f(1)$ and $f(-1)$ lie on different connected components of
$\partial\X$. Then we choose an equivalent labelling of the boundary component
containing $f(1)$ with opposite orientation (as explained in section
\ref{app:equiv-lab}). The map $f$ is then orientation preserving as
a map from $\Reps$ to \X, and the procedure of section \ref{sec:facbdry}
can be applied to obtain $\cbnd_{\!f,k,\alpha,\beta}(\X)$. 
\\
In fact, this procedure reduces the proof of theorem \ref{thm:bnd-unor} in
case (ii) to the proof of case (i), so that the only situation requiring a 
special treatment is the next one.
\item[\Nxt] 
$f(1)$ and $f(-1)$ lie on the same connected component of $\partial\X$.
This happens for example when cutting the M\"obius strip along an interval.
\end{itemize}

\medskip

The rest of this section deals with the case that 
$f(1)$ and $f(-1)$ lie on the same connected component of $\partial\X$.

Let us again start from a situation analogous to that displayed in figure 
\erf{eq:pic-i6}. Let as in section \ref{app:equiv-lab}
$(M_1,...\,,M_{n'};\varPsi_1,...\,,\varPsi_n;\oro)$ be the boundary data 
labelling the boundary component shown in the leftmost picture of 
\erf{eq:pic-i6}, such that
\oro{} assigns to both boundary intervals an orientation downwards, and the left
component is labelled by $M_1$ while the right is labelled by $M_m$ for some
$1\,{\leq}\, m\,{\leq}\, n'$. We then glue the half disks $D^\pm_\eps$ as in 
\erf{bndglue}. Next we must provide labels
$(M'_1,...\,,M'_{n'+2};\varPsi'_1,...\,,\varPsi'_{n+2};\oro')$ for the resulting
boundary component. Choose $\oro'$ to be the orientation induced by \oro{} on 
the left boundary component above the cut. Start on the segment to the left, 
above the cut, and move along the boundary in the direction opposite to $\oro'$,
and label the segments between boundary insertions with $M'_j\,{:=}\,M_j$ for
$1\,{\leq}\, j\,{\leq}\,m$. We then end up on the right segment, below
the cut. Switch to the left segment
below the cut and again move along the boundary in the direction opposite to
$\oro'$, labelling the segments with $M'_{m+1}\,{:=}\,M_1^\sigma$
(with $M^\sigma$ the $\tilde A$-module defined in \erf{Msigma}), and
$M'_{j}\,{:=}\,M_{n'+m+2-j}^\sigma$ for $m{+}2\,{\leq}\,j\,{\leq}\,n'{+}2$.
For the field insertions we perform the same manipulations again, starting
on the left segment above the cut and defining $\varPsi'_j\,{:=}\,\varPsi_j$
for $1\,{\leq}\,j\,{\leq}\,m{-}1$. 

Next, define
  \be
  \varPsi'_m := (M'_m,M'_{m+1},U_k,s(\psi^+_{\alpha})
             \cir (\id_{M'_{m+1}}{\otimes}\,\theta^{-1}_{U_k}),p^+,[\gamma^+])
  \ee
and, for $m{+}1\,{\leq}\,j\,{\leq}\,n'{+}1$,
  \be
  \varPsi'_j := (M'_j,M'_{j+1},V_{n'+m+2-j},\psi_j',[\gamma_{n'+m+2-j}])
  \ee
with
  \be
  \psi_j' = \left\{ \bearll s(\psi_{n'+m+2-j})
  & \mbox{ if }~\oro'\eq {\rm or}(\gamma_j')\,, \\{}\\[-.4em]
  s(\psi_{n'+m+2-j}) \cir (\id_{M'_j}\oti \theta^{-1}_{V_{n'+m+2-j}})
  & \mbox{ if }~\oro'\eq{-}{\rm or}(\gamma_j')\,,  \eear \right. 
  \ee
and finally
  \be
  \varPsi'_{n+2}:=(M'_{n'+2},M'_1,U_{\bar k},\psi^-_{\beta},[\gamma^-]).
  \ee
Adopting all other data from \X{},
this defines the new world sheet $\cbnd_{\!f,k,\alpha,\beta}(\X)$.

\subsubsection*{Boundary factorisation for $f{:}\ \Reps' \To \X$}

For the proof in case (ii) of theorem \ref{thm:bnd-unor},
let us proceed as in case (i) and first consider the geometric setup.

The part of the world sheet $\X$ containing the image of $\Reps'$ under $f$ 
is given by the leftmost figure in \erf{eq:pic-i6}, with both boundary 
fragments oriented downward, and with the right fragment labelled by 
$M_r^\sigma$ instead of $M_r$. The corresponding part of the
world sheet $\cbnd_{\!f,k,\alpha,\beta}(\X)$ obtained after
cutting is given by the rightmost figure in \erf{eq:pic-i6},
with the lower boundary fragment having opposite orientation and the labels 
being $M_l^\sigma$, $M_r^\sigma$ and $s(\psi^+) \cir (\id_{M_r^\vee} \oti 
\theta_{U_k}^{-1})$ instead of $M_l$, $M_r$ and $\psi^+$.

The \lhs\ of \erf{eq:bnd-unor} is given by the invariant of the cobordism 
$\MX$, of which the relevant part is, similar to figure \erf{fig:MX}
  \bea \begin{picture}(350,95)(0,54)
  \put(-60,0){
    \put(60,0)        {\Includeourbeautifulpicture 29a }
    \put(80.3,42)     {\scriptsize $M_l$}
    \put(118.5,42)    {\scriptsize $M_r^\sigma$}
    \put(188.8,132.5) {\small $\MX$}
    }
  \put(130,0){
    \put(60,0)        {\Includeourbeautifulpicture 29b }
    \put(80.3,42)     {\scriptsize $M_l$}
    \put(118.5,42)    {\scriptsize $M_r$}
    \put(188.8,132.5) {\small $\MX$}
    }
  \put(140,72){$=$}
  \epicture36  \labl{eq:fac-unor-lhs}
Here we used a triangulation of $\X$ with an edge lying on the interval along 
which $\X$ is to be cut. In the second representation we inserted definitions 
(II:3.4) and (II:2.26) and slightly deformed the resulting ribbon graph.
(Also, unlike in figure \erf{fig:MX}, here we refrain from
explicitly indicating the location of the world sheet by a shading.)

Expressing the \rhs\ of \erf{eq:bnd-unor} via cobordisms leads to
  \begin{eqnarray} \begin{picture}(420,165)(0,0)
  \put(80,0){
    \put(60,0)        {\Includeourbeautifulpicture 30a }
    \put(71,148)      {\scriptsize $M_l$}
    \put(131,148)     {\scriptsize $M_r$}
    \put(68,12)       {\scriptsize $M_l^\sigma$}
    \put(131,12)      {\scriptsize $M_r^\sigma$}
    \put(122.8,111)   {\scriptsize $\bar{k}$}
    \put(87,73)       {\scriptsize ${k}$}
    \put(100.4,120)   {\tiny\begin{turn}{90} $\psi^-_{\beta}$\end{turn}}
    \put(64.2,89)     {\tiny \begin{turn}{270}$\bar\lambda^{k\bar{k}}$\end{turn}}
    \put(104,37.5)    {\tiny\begin{turn}{90} $\psi$\end{turn}}
    }
  \put(260,0){
    \put(60,0)        {\Includeourbeautifulpicture 30b }
    \put(71,148)      {\scriptsize $M_l$}
    \put(131,148)     {\scriptsize $M_r$}
    \put(71,12)       {\scriptsize $M_l$}
    \put(131,12)      {\scriptsize $M_r$}
    \put(122.7,110)   {\scriptsize $\bar{k}$}
    \put(80,65)       {\scriptsize ${k}$}
    \put(100.4,120)   {\tiny\begin{turn}{90} $\psi^-_{\beta}$\end{turn}}
    \put(64.2,89)     {\tiny \begin{turn}{270}$\bar\lambda^{k\bar{k}}$\end{turn}}
    \put(103.5,63.5)    {\tiny\begin{turn}{270} $\psi^+_{\alpha}$\end{turn}}
    }
  \put( 0,80)        {$\dsty \sum_{k,\alpha,\beta} \dim(U_k) \,
                         {(c^{\rm bnd~~~~-1}_{M_l,M_r,k})}^{}_{\beta\alpha}$}
  \put(250,80)        {$\dsty =~\sum_{k,\alpha,\beta} L_{\beta\alpha}^{-1}$}
  \end{picture} \nonumber\\~ \label{eq:fac-unor-rhs}
  \end{eqnarray} 
where, like in the corresponding figure \erf{fig:Mk(X)} in the oriented
case, only the relevant fragment of the cobordism for
$G^{\rm bnd}_{f,k}\big( C_{\!\tilde A}( \cbnd_{f,k,\alpha,\beta}(\X) ) \big)$
is shown. Also, in the first figure we have set $\psi \eq 
s(\psi_\alpha^+) \cir (\id_{M_r^\sigma}\oti \theta^{-1}_{U_k})$, while in the 
second figure the definition \erf{fig:sdef} of $s$ has been inserted.  

One can now follow the argument in section \ref{sec:fac-bnd-or}
to show that \erf{eq:fac-unor-lhs} and \erf{eq:fac-unor-rhs} describe
the same vector in $\mathcal{H}(\Xh)$. Note that to this end one
needs to push the half-twist on the top-right module ribbon in
\erf{eq:fac-unor-lhs} further along the boundary component 
(not shown in the figure) until it emerges at the bottom left
module ribbon. That this is the place where it reappears when
following the boundary leaving figure \erf{eq:fac-unor-lhs} at the
top right corner follows from the choice of orientations and the
fact that by assumption both module ribbons in \erf{eq:fac-unor-lhs} 
lie on the same connected component of $\iota(\partial\X)$.

   
\sect{Proof of bulk factorisation} \label{sec:bulkfac}

In this section we present the proof of factorisation for bulk correlators,
theorem~\ref{thm:bulk-or}. Similarly to the case of boundary factorisation,
we must first clarify the geometry underlying the two sides of the 
equalities \erf{eq:bulk-or} and \erf{eq:bulk-unor}, 
and then establish some algebraic identities to be used in the proof.

\subsection{Geometric setup of bulk factorisation}

$~$

\noindent
The construction in this section applies to both oriented
and unoriented world sheets. Like in section \ref{sec:bnd-geom}, the reason
is that the embedding $f{:}\ \Aeps \To \X$ induces a local
orientation on the relevant part of the world sheet \X.

We are interested in the geometry of the relevant surfaces and
three-manifolds in a neighbourhood of the region to which the `cutting twice'
procedure \erf{eq:pic-i3} is applied.
Throughout this section we will make use of a ``wedge presentation" of surfaces
and three-manifolds to illustrate the geometry. By this we mean that a
horizontal disk is drawn as a disk {\em sector\/}, with identification of the
legs of the sector implied (compare also e.g.\ picture (II:A.81)).
Further, curly lines indicate interfaces at which the displayed
piece is connected to other parts of the manifold. To give an example, in this
presentation the pictures
  \bea \begin{picture}(400,245)(0,44)
  \put(0,148)        {\Includeourbeautifulpicture 60a }
  \put(130,75)       {\Includeourbeautifulpicture 60b }
  \put(280,0)        {\Includeourbeautifulpicture 60c }
  \epicture22 \labl{fig:pocrep}
stand for a cylindrical part of a
two-manifold, for a solid three-ball, and for a solid torus, respectively;
in the latter case, the additional identification of top and bottom faces
is indicated by drawing the corresponding legs as dashed lines.
Also, it is implied that whenever top and bottom or two vertical faces
of a wedge are identified, this is to be performed without any additional 
rotation or reflection. (For instance, in the middle picture the boundary of 
the three-manifold is a sphere rather than a cross cap.)  

The left hand side of formula \erf{eq:bulk-or} concerns the connecting manifold
\MX. Let $\NX\,{\subset}\,\MX$ consist of all points that are mapped to 
$f(\Aeps)$ via the canonical projection $\pi{:}\ \MX \To \X$.
$\NX$ has the topology of a solid cylinder with a solid cylinder cut out,
and it contains $A$-ribbons which cover the part of a triangulation of 
$\imath(\X)$ that lies in $\NX$. Accordingly, the boundary $\partial\NX$ has 
two annular connected components $\partial\NX^{(1)}$ and $\partial\NX^{(2)}$. 
Let $\partial\NX^{(1)}$ be the component for which (the restriction to 
$\partial\NX$ of) the projection $\pi$ is orientation preserving.
With suitable choice of coordinates, $\partial\NX^{(2)}$ is a cylinder of
radius $2$ and $\partial\NX^{(1)}$ a cylinder of radius $1$, and
the common axis of $\partial\NX^{(2)}$ and $\partial\NX^{(1)}$ is the $z$-axis. 
In this description, the rest of \MX\ is connected to the annular top and bottom
parts of $\partial\NX$. This situation is illustrated in the following picture:
  \begin{eqnarray}   \begin{picture}(420,229)(0,0)
  \put(0,0)             {\Includeourbeautifulpicture 90a }
  \put(283,35)          {\Includeourbeautifulpicture 90c }
  \end{picture} \nonumber\\
                     \begin{picture}(130,183)(0,0)
  \put(0,0)             {\Includeourbeautifulpicture 90e }
  \put(-35,70)          {=}
  \end{picture} \nonumber\\
  ~\label{V-90} \end{eqnarray}
The two collars forming the subset $\partial\MX\,{\cap}\,\NX$ of the boundary 
$\partial\MX$ are indicated in colours. Note that by definition they coincide 
with the images $f_i(\Aeps)$ and $f_j(\Aeps)$ for $f_i$ and $f_j$ as introduced
in \erf{eq:fifj-def}. Zooming in on \NX\ results in the upper figure on the 
\rhs, which is then redrawn in wedge presentation in the lower figure.
\\ 
It will be convenient, however, to slightly deform \NX\ in a region close to its
boundary.  Including also the ribbon graph, the resulting manifold, which we 
still denote by \NX, then looks as follows (in this picture, now the part 
$\partial\NX^{(1)}$ of the boundary extends over both the inner cylindrical
piece and the top and bottom parts):
        \pagebreak[1]
  \bea \begin{picture}(230,258)(0,55)
  \put(0,0)          {\includeourbeautifulpicture 61 }
  \put(-50,166)      {$ \NX \;= $}
  \put(34,226.5)     {\tiny $1$}
  \put(20.9,218.5)   {\tiny $2$}
  \put(235,93)       {\tiny $1$}
  \put(222.4,115)    {\tiny $2$}
  \put(83,105)       {\small $\Delta$}
  \put(131.8,107.3)  {\scriptsize \begin{turn}{115}$\eta$\end{turn}}
  \put(123,28)       {\small $m$}
  \put(153,26)       {\small $m$}
  \epicture29 \labl{fig:bulkLHS}

The relevant region of the extended surface ${\rm Y}_{ij}$ that was defined in 
\erf{Yij} is embedded in two copies of $\reals^3$ in the following manner:\,%
  \footnote{~In each copy of $\reals^3$ there are two calottes and one sphere.
  Thus, in non-wedge representation, each of these surface fragments is a 
  copy of the piece of two-manifold displayed in the rightmost figure of
  \erf{eq:pic-i3}.}
  \bea \begin{picture}(300,149)(0,60)
  \put(0,0)        {\Includeourbeautifulpicture 62b }
  \put(180,0)      {\Includeourbeautifulpicture 62a }
  \put(41,183)     {\tiny $2$}
  \put(54,170)     {\tiny $1$}
  \put(17,96)      {\tiny $2$}
  \put(30,110)     {\tiny $1$}
  \put(20,23)      {\tiny $2$}
  \put(33,36)      {\tiny $1$}
  \put(221,182)    {\tiny $2$}
  \put(234,170)    {\tiny $1$}
  \put(197,96)     {\tiny $2$}
  \put(210,110)    {\tiny $1$}
  \put(200,23)     {\tiny $2$}
  \put(214,36)     {\tiny $1$}
  \put(62,204)     {\scriptsize $(U_\ib,+)$}
  \put(63,137)     {\scriptsize $(U_\ib,-)$}
  \put(58,68)      {\scriptsize $(U_i,-)$}
  \put(56,25)      {\scriptsize $(U_i,+)$}
  \put(243,204)    {\scriptsize $(U_j,+)$}
  \put(245,137)    {\scriptsize $(U_j,-)$}
  \put(240,68)     {\scriptsize $(U_\jb,-)$}
  \put(239,25)     {\scriptsize $(U_\jb,+)$}
  \epicture37 \labl{fig:cut2}
Next, we consider part of the cobordism\,%
  \footnote{~In non-wedge representation, this amounts to two copies of
  the three-manifold \erf{eq:picV-41}. The annular parts at the
  top and bottom, at which \erf{eq:picV-41} is joined to the rest
  of the cobordism, correspond to the curly lines in \erf{fig:glcob1}.}
from ${\rm Y}_{ij}$ to \Xh, used in defining
$\hat{g}_{f_i,i}\cir\hat{g}_{f_j,j}$:
  \begin{eqnarray} \begin{picture}(420,325)(22,0)
  \put(0,0)          {\Includeourbeautifulpicture 63b }
  \put(266,0)        {\Includeourbeautifulpicture 63a }
  \put(142,316)      {\scriptsize $(U_{\ib,-})$}   
  \put(157,206)      {\scriptsize $(U_{\ib,+})$}	
  \put(149,131)      {\scriptsize $(U_{i,+})$}
  \put(159,47)       {\scriptsize $(U_{i,-})$}
  \put(417,317)      {\scriptsize $(U_{j,-})$}
  \put(416,204.5)    {\scriptsize $(U_{j,+})$}
  \put(415,130.2)    {\scriptsize $(U_{\jb,+})$} 
  \put(414,45)       {\scriptsize $(U_{\jb,-})$}
  \put(40,290)       {\tiny $1$}
  \put(38,279)       {\tiny $2$}
  \put(34,213)       {\tiny $1$}
  \put(21,207)       {\tiny $2$}
  \put(100,23)       {\tiny $1$}
  \put(74,27)        {\tiny $2$}
  \put(117,159)      {\tiny $1$}
  \put(104,174)      {\tiny $2$}
  \put(307,290)      {\tiny $1$}
  \put(303,279)      {\tiny $2$}
  \put(301,213)      {\tiny $1$}
  \put(287,207)      {\tiny $2$}
  \put(338,25)       {\tiny $2$}
  \put(365,23)       {\tiny $1$}
  \put(384,160)      {\tiny $1$}
  \put(370,175)      {\tiny $2$}
  \end{picture} \nonumber\\{} \label{fig:glcob1}
  \end{eqnarray}
There are now ribbons running inside the cobordism, ending on the marked arcs
on $-{\rm Y}_{ij}$, with orientations as shown in the picture.
The core orientation of each ribbon points away from the spherical boundary
components. The two (`outgoing') cylindrical boundary components, which
correspond to $\partial\NX^{(2)}$ and $\partial\NX^{(1)}$ in \erf{fig:bulkLHS},
respectively, have different orientation.

The gluing homomorphism $G^{\rm bulk}_{f,ij}$ is the invariant of the cobordism
that is obtained by gluing $B^-_{i\ib}$ and $B^-_{j\jb}$ to the (`incoming')
spherical boundary components of \erf{fig:glcob1}, using the canonical
identification of the spherical boundary components. The result is the following
cobordism from $\Xh'_{ij}$ to \Xh\ (recall that ${\rm X}'_{ij}$ is an 
abbreviation for $\cblk_{f,i,j,\alpha,\beta}(\X)$):
  \begin{eqnarray} \begin{picture}(420,316)(10,0)
  \put(0,0)          {\Includeourbeautifulpicture 64a }
  \put(255,0)        {\Includeourbeautifulpicture 64b }
  \put(107,302)      {\scriptsize $(U_{\ib ,-})$}
  \put(122,33.5)     {\scriptsize $(U_{i,-})$}
  \put(369,303)      {\scriptsize $(U_{j,-})$}
  \put(365,31)       {\scriptsize$(U_{\jb ,-})$}
  \put(133,264)      {\scriptsize $\ib $}
  \put(142,73)       {\scriptsize $i$}
  \put(396,73)       {\scriptsize $\jb $}
  \put(387,265)      {\scriptsize $j$}
  \put(47,291)       {\tiny $1$}
  \put(45,278)       {\tiny $2$}
  \put(34,213)       {\tiny $1$}
  \put(21,205.5)     {\tiny $2$}
  \put(98,19)        {\tiny $1$}
  \put(72,18)        {\tiny $2$}
  \put(295,288)      {\tiny $1$}
  \put(294,279)      {\tiny $2$}
  \put(289,212)      {\tiny $1$}
  \put(277,204.5)    {\tiny $2$}
  \put(328,24)       {\tiny $2$}
  \put(352,22)       {\tiny $1$}
  \put(98,152)	   {\small\begin{turn}{310}$\bar\lambda^{i\ib}$\end{turn}}
  \put(352,149)	   {\small\begin{turn}{310}$\bar\lambda^{j\jb}$\end{turn}}
  \end{picture} \nonumber\\{} \label{fig:glcob2}
  \end{eqnarray}
Of the connecting manifold ${\mathrm M}_{\X'_{ij}}$, the part of interest to us
is a neighbourhood of the region where $\X'_{ij}$ is cut and pasted, or more
precisely, small disks containing the bulk field insertions. We embed these 
disks in two copies of $\mathbb{R}^3$ in the plane $z\eq 3/2$, centered at the 
point $(0,0,3/2)$, at which the insertion points $p^{\pm}\iN\X$ are placed. The 
orientation of the first disk is the one inherited from the $x$-$y$-plane 
passing through $z$, and a representative arc $\gamma^-$ is aligned along the 
$x$-axis with the same orientation as the axis. The orientation and 
arc-orientation 
(for $\gamma^+$) on the second disk are opposite to those of the first.
On $\Xh'_{ij}$ each of the disks has two pre-images, at $z\eq 1$ and $z\eq 2$,
respectively. The marked arcs are labelled by $(U_{\ib},+)$, $(U_{\jb},+)$,
$(U_i,+)$ and $(U_j,+)$, respectively, and the surface- and arc-orientations are
as indicated in the following figure:
  \bea \begin{picture}(300,80)(0,40)
  \put(0,0)          {\Includeourbeautifulpicture 65a }
  \put(180,0)        {\Includeourbeautifulpicture 65b }
  \put(32,99)      {\tiny $1$}
  \put(18,86)      {\tiny $2$}
  \put(21,38)      {\tiny $2$}
  \put(31,29)      {\tiny $1$}
  \put(212,99)     {\tiny $1$}
  \put(198,86)     {\tiny $2$}
  \put(200.8,38)   {\tiny $2$}
  \put(211,29)     {\tiny $1$}              
  \put(67,44)      {\scriptsize $(U_\ib,+)$}
  \put(67,106)     {\scriptsize $(U_\jb,+)$}
  \put(248,44)     {\scriptsize $(U_j,+)$}  
  \put(248,106)    {\scriptsize $(U_i,+)$}  
  \epicture25 \labl{fig:cbulk}

The relevant part of the connecting manifold ${\rm M}_{\X'_{ij}}$ consists
of two full cylinders, with the rest of ${\rm M}_{\X'_{ij}}$ connected to
the cylindrical parts of their boundaries:
  \begin{eqnarray} \begin{picture}(420,270)(15,0)
  \put(0,0)          {\Includeournicemediumpicture 66a }
  \put(260,0)        {\Includeournicemediumpicture 66b }
  \put(147,263)      {\scriptsize $(U_\jb,+)$}
  \put(160,46)       {\scriptsize $(U_\ib,+)$}
  \put(148,158)      {\scriptsize \begin{turn}{305}$\phi_\beta^-$\end{turn}}
  \put(183,162)      {\scriptsize \begin{turn}{305}$\eps$\end{turn}} 
  \put(432,155)      {\scriptsize \begin{turn}{125}$\eta$\end{turn}}
  \put(371,253)      {\scriptsize $(U_i,+)$}
  \put(377,35)       {\scriptsize $(U_j,+)$}
  \put(389.5,145.5)      {\scriptsize \begin{turn}{125}$\phi_\alpha^+$\end{turn}}
  \put(35,249)	     {\tiny $1$}
  \put(32,239)	     {\tiny $2$}
  \put(79,23)	     {\tiny $1$}
  \put(58,24)	     {\tiny $2$}
  \put(295,249)	     {\tiny $1$}
  \put(292,239)	     {\tiny $2$}
  \put(339,23)	     {\tiny $1$}
  \put(318,24)	     {\tiny $2$}
  \end{picture}  \nonumber\\{} \label{fig:Mcbulk}
  \end{eqnarray}
The coupon in the piece of three-manifold in the left part of the figure is
labelled by a bimodule morphism
$\phi^-_{\beta}\iN\HomAA(U_{\ib}\oT+A\ot-U_{\jb},A)$, and the one in the
right part by a bimodule morphism $\phi^+_{\alpha}\iN\HomAA(U_i\oT+A\ot-U_j,A)$.

When using the canonical identification to glue ${\rm M}_{\X'_{ij}}$ to the
cobordism~\erf{fig:glcob2}, the resulting cobordism from $\emptyset$ to $\Xh$ is
  \bea \begin{picture}(200,488)(0,20)
  \put(0,0)          {\includeournicemediumpicture 67 }
  \put(148.5,319.5)	   {\scriptsize \begin{turn}{305}$\phi_\beta^-$\end{turn}}
  \put(130.5,118.5)	   {\scriptsize \begin{turn}{125}$\phi_\alpha^+$\end{turn}}
  \put(123,406)	   {\scriptsize \begin{turn}{305}$\bar\lambda^{j\jb}$\end{turn}}
  \put(111,223)	   {\scriptsize \begin{turn}{305}$\bar\lambda^{i\ib}$\end{turn}}
  \put(184,323)	   {\scriptsize \begin{turn}{305}$\eps$\end{turn}}
  \put(174,127.5)	   {\scriptsize \begin{turn}{125}$\eta$\end{turn}}
  \put(197,440)	   {\scriptsize $j$}
  \put(181,390)	   {\scriptsize $\jb$}
  \put(163,268)	   {\scriptsize $\ib$}
  \put(191,195)	   {\scriptsize $i$}
  \put(29,415)	   {\tiny $1$}
  \put(18,408)	   {\tiny $2$}
  \put(28,225.5)   {\tiny $1$}
  \put(17,218)	   {\tiny $2$}
  \epicture03 \labl{fig:RHScob}
Here top and bottom are identified; for the coupons labelled by $\phi_\alpha^+$ 
and $\bar\lambda^{j\jb}$ it is the white side that is facing upwards, while for 
the coupons labelled by $\phi_\beta^-$ and $\bar\lambda^{i\ib}$ it is the black 
side. Again two of the Lagrangian subspaces coincide in each of the two Maslov 
indices relevant for the functoriality formula, so the latter cobordism is 
precisely the cobordism on the right hand side of the desired equality
\erf{eq:bulk-or}.

\subsection{A surgery identity}

$~$

\noindent
Having sorted out the situation of our interest at the geometrical level, we
proceed to establish a few intermediate results in preparation of the proof. 
First, consider a tubular neighbourhood ${\rm M}_{ij,\alpha\beta}$ of all the 
vertical ribbons in figure (\ref{fig:RHScob}), obtained by cutting along surface
that in the presentation used in figure (\ref{fig:RHScob}) is a vertical 
cylinder close to the displayed parts of the boundary. Orient its boundary 
$\partial{\rm M}_{ij,\alpha\beta}$ by the inward-pointing normal. This boundary 
is pierced at two arcs by $A$-ribbons; the directions of these arcs are given 
by the orientation induced from the ribbon orientation via the inward-pointing 
normal. The upper arc is labelled by $(A,+)$, and the lower one by $(A,-)$. 
With this prescription, and with the canonical choice of Lagrangian subspace 
(see \erf{def:Lagr}), the boundary becomes an extended surface, to be denoted by
$\TAA$, and ${\rm M}_{ij,\alpha\beta}$ a cobordism from $\emptyset$
to $\TAA$. 

In blackboard framing we have
  \bea \begin{picture}(105,165)(0,68)
  \put(0,0)          {\includeourbeautifulpicture 68 }
  \put(18,200)     {\tiny $1$}
  \put(5,194)      {\tiny $2$}
  \put(138,213)    {\scriptsize $j$}
  \put(132,188)    {\scriptsize $\jb$}
  \put(113,128)    {\scriptsize $\ib$}
  \put(119.6,108)  {\scriptsize $i$}
  \put(89,93)      {\small $\alpha$}
  \put(129,145)    {\small $\beta$}
  \put(4,136)      {\scriptsize $(A,+)$}
  \put(4,84)       {\scriptsize $(A,-)$}
  \put(-66,121)    {$ {\rm M}_{ij,\alpha\beta} \;= $}
  \epicture44 \labl{fig:Mij}
Next, define another cobordism ${\rm M}_{\rm T}$ from $\emptyset$ to
$\TAA$ by taking ${\rm M}_{\rm T}$ to be a solid torus with boundary $\TAA$. 
However, referring to the graphical presentation in figure \erf{fig:Mij},
while in ${\rm M}_{ij,\alpha\beta}$ a horizontal cycle is contractible, we 
choose ${\rm M}_{\rm T}$ in such a manner that a vertical cycle is contractible.
Note that with respect to ${\rm M}_{\rm T}$ it is then not the canonical 
Lagrangian subspace that appears in $\TAA$.

Place $A$-ribbons in ${\rm M}_{\rm T}$ as indicated in the following figure:
  \bea \begin{picture}(150,180)(0,62)
  \put(30,0)          {\includeourbeautifulpicture 69 }
  \put(48,197)        {\tiny $1$}
  \put(34,192)        {\tiny $2$}
  \put(34,147)        {\scriptsize $(A,+)$}
  \put(38,58)         {\scriptsize $(A,-)$}
  \put(-40,120)       {${\rm M}_{\rm T}$\; := }
  \epicture42 \labl{fig:MT}
(The $A$-ribbons entering on the front face and leaving the back face
are connected, due to the identification of those two faces.)

\dtlProposition{prop:bulksurgery}
The cobordisms ${\rm M}_{ij,\alpha\beta}$ and ${\rm M}_{\rm T}$ are related by
  \be
  Z({\rm M}_{\rm T},\emptyset, \TAA)
  = \sum_{i,j,\alpha,\beta}\dim(U_i)\dim(U_j) 
  ({\Cbulk_{i,j}}^{\,\scriptstyle -1})^{}_{\beta\alpha} \, 
  Z({\rm M}_{ij,\alpha\beta},\emptyset,\TAA) \,.
  \labl{eq:bulksurg}
\ENDP

Before proving this statement, we will need to establish two lemmas: 
\Itemize
\item[\Nxt]
The left side of \erf{eq:bulksurg} can be written as an expansion with respect 
to a basis of $\mathcal{H}(\TAA)$, and it is shown that the resulting summation 
can be restricted to the vectors $Z({\rm M}_{ij,\alpha\beta},\emptyset,\TAA)$.
\item[\Nxt]
By evaluating both sides on a basis of $\Hom(\mathcal{H}(\TAA),\mathbb{C})$,
the expansion coefficients are shown to be those given in \erf{eq:bulksurg}.
\end{itemize}

We first select a convenient basis of $\Hom(\mathbb{C},\mathcal{H}(\TAA))$
and introduce a certain cobordism ${\rm P}{:}\;\TAA\To\TAA$. The basis of 
$\Hom(\mathbb{C},\mathcal{H}(\TAA))$ is provided by the invariants of the 
cobordisms
  \bea \begin{picture}(250,169)(0,61)
  \put(80,0)          {\includeourbeautifulpicture 70 }
  \put(99,199)        {\tiny $1$}
  \put(85,194)        {\tiny $2$}
  \put(218,213)       {\scriptsize $j$}
  \put(211,188)       {\scriptsize $\jb$}
  \put(198,80)        {\scriptsize $j$}
  \put(189.7,128)     {\scriptsize $\ib$}
  \put(199,108)       {\scriptsize $i$}
  \put(169,91)        {\small $\alpha$}
  \put(213,145)       {\small $\beta$}
  \put(84,137)        {\scriptsize $(A,+)$}
  \put(84,84)         {\scriptsize $(A,-)$}
  \put(190,188)       {\tiny $\bar{\lambda}^{j\jb}$}
  \put(165,128)       {\tiny \begin{turn}{270}$\bar{\lambda}^{i\ib}$\end{turn}}
  \put(0,116)         {$ L_{ij,\alpha\beta} \;:= $}
  \epicture39 \labl{fig:TAAbasis}
with $\alpha$ and $\beta$ running over a basis of $\Hom(U_i\oti U_j,A)$ and of 
$\Hom(U_{\ib}\oti A\oti U_{\jb},\one)$, respectively. It is not difficult to 
see that the collection of invariants of the cobordisms $L_{ij,\alpha\beta}$, 
for all values of $i,j,\alpha,\beta$, indeed forms a basis. By the general 
construction given in \cite{TUra} (see lemma 2.1.3 in chapter IV), the space 
$\mathcal{H}(\TAA)$ is isomorphic to
  \be \bearl\dsty
  \bigoplus_{i\in\I}\Hom(\one,A^\vee\oti U_i\oti U_i^\vee\oti A)
  \\{}\\[-1.2em]\dsty \hsp6 \cong\,
  \bigoplus_{i,j\in\I}\Hom(\one,A^\vee\oti U_i\oti U_j)\otic
  \Hom(\one,U_j^\vee\oti U_i^\vee\oti A) \,.  \eear
  \labl{eq:2ptT2morph}
In the case of \erf{fig:TAAbasis}, the isomorphism is provided by first choosing
bijections $\Hom(U_i\oti U_j,
     $\linebreak[0]$
A) \,{\cong}\, \Hom(\one,A^\vee\oti U_i\oti U_j)$ 
and $\Hom(U_\ib\oti A \oti U_\jb,\one) \,{\cong}\, \Hom(\one,U_j^\vee\oti 
U_i^\vee\oti A)$. Then one uses these bijections to assign labels to the two 
larger coupons in \erf{fig:TAAbasis} for each element of \erf{eq:2ptT2morph}.
\\
The cobordism ${\rm P}$ is obtained 
as follows. In the cylinder $\TAA\Times[0,1]$ over $\TAA$, insert two straight
$A$-ribbons that connect the marked arcs on $\TAA\Times\{0\}$ to those at
$\TAA\Times\{1\}$, as well as two $A$-ribbons in the interior running along
one of the non-contractible cycles, joined to the straight $A$-ribbons by
a coproduct on one side and a product on the other. Thus 
  \begin{eqnarray} \begin{picture}(415,282)(0,0)
  \put(80,0)          {\includeourbeautifulpicture 71 }
  \put(98,254)        {\tiny $1$}
  \put(83,249)        {\tiny $2$}
  \put(315.6,243)     {\tiny $1$}
  \put(302.8,265)     {\tiny $2$}
  \put(84,165)        {\scriptsize $(A,+)$}
  \put(84,66)         {\scriptsize $(A,-)$}
  \put(332,183)       {\scriptsize $(A,-)$}
  \put(337,82)        {\scriptsize $(A,+)$}
  \put(23,136)        {$ {\rm P} \;= $}	  
  \put(158,192)	      {\small $\Delta$}
  \put(244,192)	      {\small $m$}
  \put(158,94)	      {\small $m$}
  \put(244,94)	      {\small $\Delta$}
  \end{picture} \nonumber\\~ \label{fig:Proj}
  \end{eqnarray} 
\dtlLemma{1lem} $~$\\[.5em]
{\rm(i)}~~$Z({\rm P}) \iN \End(\mathcal{H}(\TAA))$ is a projector.
\\[.5em]
{\rm(ii)}~\,$Z({\rm P})$ projects onto the
subspace spanned by $\{Z({\rm M}_{ij,\alpha\beta},\emptyset,\TAA)1\}_{i,j,
\alpha,\beta}$, with ${\rm M}_{ij,\alpha\beta}$ the cobordisms introduced in
\erf{fig:Mij}.
\\[.5em]
{\rm(iii)}~The invariants of the cobordisms ${\rm P} \cir {\rm M}_{\rm T}$ and 
${\rm M}_{\rm T}$ are equal, $Z(P\cir{\rm M}_{\rm T})\eq Z({\rm M}_{\rm T})$.
\ENDL

\Proof
(i)~~That $Z({\rm P})\cir Z({\rm P}) \eq Z({\rm P})$ follows easily by using 
associativity, coassociativity, specialness and the Frobenius property;
compare the similar calculation in (I:5.36). 
\\[.3em]
(ii)~\,Consider the action of $Z(P)$ on one of the basis elements
$Z(L_{ij,\alpha\beta})1 \iN \mathcal{H}(\TAA)$. The ribbon graph
for $Z(P \cir L_{ij,\alpha\beta})$ can be deformed as follows.
  \begin{eqnarray} \begin{picture}(420,267)(5,0)
  \put(0,0)       {\Includeourbeautifulpicture 72a }
  \end{picture} \nonumber \\ \begin{picture}(270,270)(0,0)
  \put(130,400)	  {\scriptsize $\beta$}
  \put(86,334)	  {\scriptsize $\alpha$}
  \put(130,458)	  {\scriptsize $j$}
  \put(115,430)	  {\scriptsize $\jb$}
  \put(110,387)	  {\scriptsize $\ib$}
  \put(115,340)	  {\scriptsize $i$}
  \put(-135,390)  {\scriptsize $(A,+)$}
  \put(-135,325)  {\scriptsize $(A,-)$}
 \put(-13,0) {
  \put(0,0)       {\Includeourbeautifulpicture 72b }
  \put(-45,125)   {$ = $}
  \put(221,157)	  {\scriptsize $\beta$}
  \put(164,90)	  {\scriptsize $\alpha$}
  \put(282,212)	  {\scriptsize $j$}
  \put(266,182)	  {\scriptsize $\jb$}
  \put(207,145)	  {\scriptsize $\ib$}
  \put(209,100)	  {\scriptsize $i$}
  \put(20,146)	  {\scriptsize $(A,+)$}
  \put(20,80)	  {\scriptsize $(A,-)$}
  \begin{picture}(0,0)(-76,-12)
  \put(0,0)		{\includegraphics[scale=0.3]{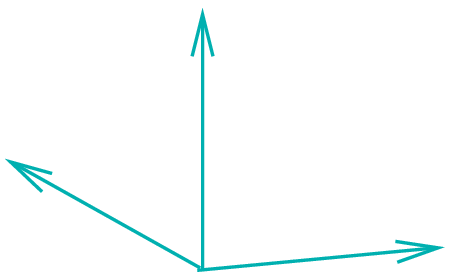}}
  \put(39,1.5)	{\tiny $x$}
  \put(-3,12)	{\tiny $y$}
  \put(16,25)	{\tiny $z$}
  \end{picture}
 }
  \end{picture}
  \begin{picture}(0,0)(350,-285)
  \put(0,0)		{\includegraphics[scale=0.3]{3fr.eps}}
  \put(39,1.5)	{\tiny $x$}
  \put(-3,12)	{\tiny $y$}
  \put(16,25)	{\tiny $z$}
  \end{picture} \nonumber\\~
  \label{fig:locbim} \end{eqnarray}
To see that this equality holds, notice that the figure on the \rhs\ is obtained
from the left by a deformation: the $A$-ribbons parallel to the $y$-axis are 
moved to the right until the ends join at the trivial cycle at the far right.
The morphisms together with the encircling $A$-ribbons can now be recognised 
as local morphisms according to lemma \ref{lem:A3-2}. Further, by combining 
proposition I:5.7 and lemma IV:2.2 one sees that bases for the corresponding 
spaces of local morphisms are provided by 
$\phi_{(ij)\alpha} \iN \HomAA(U_i\oT+A\ot-U_j,A)$ and 
$\phi_{(\ib\jb)\beta}\iN\HomAA(U_{\ib}\oT+A\ot-U_{\jb},A)$. It follows that 
$Z(P \cir L_{ij,\alpha\beta})$ can be expanded in terms of 
$Z({\rm M}_{ij,\alpha\beta})$, establishing that the image of $Z(P)$ is 
contained in the subspace spanned by the vectors $Z({\rm M}_{ij,\alpha\beta})$.
\\[.3em]
(iii)~The Frobenius property and specialness guarantee that the cobordism 
resulting from gluing ${\rm P}$ to ${\rm M}_{\rm T}$ has the same invariant 
as ${\rm M}_{\rm T}$. In more detail, this is demonstrated by the following 
sequence of equalities:
  \begin{eqnarray} \begin{picture}(420,215)(0,0)
  \put(0,0)         {\Includeournicemediumpicture 73a }
  \put(12,121)	    {\scriptsize $(A,+)$}
  \put(12,55)	    {\scriptsize $(A,-)$}
  \end{picture} \nonumber 
  \end{eqnarray}
  \begin{eqnarray} 
  \begin{picture}(135,196)(0,0)
  \put(0,0)         {\Includeournicemediumpicture 73b }
  \put(12,121)	    {\scriptsize $(A,+)$}
  \put(12,55)	    {\scriptsize $(A,-)$}
  \put(-45,97)      {$ = $}
  \end{picture} \nonumber
  \end{eqnarray}
  \begin{eqnarray} \begin{picture}(135,195)(0,0)
  \put(0,0)         {\Includeournicemediumpicture 73c }
  \put(-45,97)      {$ = $}
  \put(12,121)	    {\scriptsize $(A,+)$}
  \put(12,55)	    {\scriptsize $(A,-)$}
  \end{picture} \nonumber
  \end{eqnarray}
  \begin{eqnarray} \begin{picture}(135,197)(0,0)
  \put(0,0)         {\Includeournicemediumpicture 73d }
  \put(-45,97)      {$ = $}
  \put(12,121)	    {\scriptsize $(A,+)$}
  \put(12,55)	    {\scriptsize $(A,-)$}
  \end{picture} \nonumber 
  \end{eqnarray}
  \begin{eqnarray}  \begin{picture}(135,197)(0,0)
  \put(0,0)         {\Includeournicemediumpicture 73e }
  \put(-45,97)      {$ = $}
  \put(12,121)	    {\scriptsize $(A,+)$}
  \put(12,55)       {\scriptsize $(A,-)$}
  \end{picture} \nonumber \\[-1.9em]~
  \label{fig:ProjL} \end{eqnarray}
To see these equalities, one proceeds as follows. In the first step the 
Frobenius and associativity properties are employed to attach the `left-most' 
$A$-ribbon -- i.e., the lower one coming from ${\rm P}$ that is piercing the 
front and back faces of the wedge (which are to be identified) --
to the ribbons to its right (which come from ${\rm M}_{\rm T}$).
In the second step the loop that results from the first step is removed
with the help of the Frobenius property, associativity, and specialness, and
in addition the `upper' $A$-ribbon (the one coming from ${\rm P}$ that is 
entering and leaving the part of the manifold drawn in the upper part 
of the picture) is moved towards the region in which the previous manipulations 
took place, thereby now showing its white rather than black side.
The third step consists in deforming the latter ribbon and using
associativity and the Frobenius property. The last
step is similar to the first two, first using associativity and Frobenius 
property, and then associativity and specialness to remove the resulting loop.
\qed

By lemma \ref{1lem}(ii), $Z(\rmM_{\rm T},\emptyset,\TAA)$ can be expanded in 
terms of the vectors $\{ Z({\rm M}_{ij,\alpha\beta},\emptyset,
     $\linebreak[0]$
\TAA)\}_{i,j,\alpha,\beta}$. Accordingly we write
  \be
  Z({\rm M}_{\rm T},\emptyset,\TAA)
  = \sum_{i,j,\alpha,\beta} K_{ij,\alpha\beta}\,
  Z({\rm M}_{ij,\alpha\beta},\emptyset,\TAA)
  \labl{eq:Saction}
with suitable coefficients $K_{ij,\alpha\beta}\iN\complex$.
It remains to show that the expansion coefficients $K_{ij,\alpha\beta}$ in a
given basis coincide with the inverses of the structure constants of the 
two-point functions on the sphere in that same basis. We first observe 

\dtlLemma{2lem}
The coefficients $K_{ij,\alpha\beta}$ satisfy
  \bea \begin{picture}(80,75)(0,43)
  \put(89,0)         {\includeourbeautifulpicture 83 }
  \put(113,35)       {\scriptsize $i$}
  \put(136,36)       {\scriptsize $j$}
  \put(125,21)       {\footnotesize $\bar{\alpha}$}
  \put(125,67.5)     {\footnotesize $\bar{\beta}$}
  \put(-108,58)      {$\displaystyle  K_{ij,\alpha\beta} ~=~ S_{0,0}\,
                       \Big(\frac{\dim(U_i)\dim(U_j)}{\dim(A)}\Big)^{\!2} $}
  \epicture17 \labl{fig:expcoef1}
\ENDL

\Proof
We start by verifying that the vector dual to $Z({\rm M}_{ij,\alpha\beta},
\emptyset,\TAA)$ is given by the invariant of the cobordism
  \bea \begin{picture}(245,166)(0,66)
  \put(80,0)       {\includeourbeautifulpicture 74 
    \setlength{\unitlength}{.38pt}
     \put(10,496)        {\scriptsize $\lambda_{j\jb}$} 
     \put(176,314)       {\scriptsize $\lambda_{i\ib}$}
    \setlength{\unitlength}{1pt}
    }
  \put(-10,115)    {$\widetilde{{\rm M}}_{ij,\alpha\beta} 
                     \;:=~\mathcal N  $}
  \put(216,199)    {\tiny $1$}
  \put(229,194)    {\tiny $2$}
  \put(104,221)    {\scriptsize $j$}
  \put(134,165)    {\scriptsize $\jb$}
  \put(122,72)     {\scriptsize $j$}
  \put(132,130)    {\scriptsize $\ib$}
  \put(124,107)    {\scriptsize $i$}
  \put(101.8,143.5)  {\small $\bar{\beta}$}
  \put(140.8,91)     {\small $\bar{\alpha}$}
  \put(208,84)     {\scriptsize $(A,+)$}
  \put(209,139)    {\scriptsize $(A,-)$}
  \epicture41 \labl{fig:Mijdual}
with $\mathcal N$  a normalisation constant to be determined.
\\[.3em]
Thus $\widetilde{{\rm M}}_{ij,\alpha\beta}$ is the same topological
three-manifold as ${\rm M}_{ij,\alpha\beta}$, but with opposite bulk- and
boundary orientation; the ribbons are labelled by the same objects, but their
cores are oriented differently. The coupons are now labelled by morphisms in
$\HomAA(A,U_i\oT+A\ot-U_j)$ and in $\HomAA(A,U_\ib\oT+A\ot-U_\jb)$,
respectively, and we choose bases $\{{\bar{\phi}}^{\pm}_{\alpha}\}$ that are
dual to $\{\phi^{\pm}_{\alpha}\}$ in the sense that
  \bea \begin{picture}(70,73)(0,28)
  \put(0,3)          {\Includeourbeautifulpicture 75a }
  \put(95,3)         {\Includeourbeautifulpicture 75b }
  \put(46,48)        {$ =~~\delta_{\alpha,\beta} $}
  \put(-6,50)        {\scriptsize$ U_i$}
  \put(20,50)        {\scriptsize$ U_j$}
  \put(6,50)         {\scriptsize$ A$}
  \put(7.5,28)       {\footnotesize $\bar{\phi}{}_{\beta}$}
  \put(7.5,69.5)     {\footnotesize $\phi_{\alpha}$}
  \epicture10 \labl{fig:dualbimorph}
That this is possible follows from the presence of a non-degenerate pairing 
between the morphism spaces
$\Hom(A,U_i\oti A\oti U_j)$ and $\Hom(U_i\oti A\oti U_j,A)$ (defined by 
the trace, see \cite{TUra}), as is shown in appendix \ref{sec:2ptS2}.
Gluing ${\rm M}_{ij,\alpha\beta}$ to $\widetilde{\rm M}_{kl,\gamma\delta}$
via the canonical identification of the boundaries results in the following
ribbon graph:
  \bea \begin{picture}(260,202)(0,40)
  \put(0,0)          {\includeourbeautifulpicture 76 }
  \put(180,112)      {\scriptsize $\ib $}
  \put(181,93)       {\scriptsize $i$}
  \put(179,148)      {\scriptsize $\jb $}
  \put(199,195)      {\scriptsize $j$}
  \put(197,43.5)     {\scriptsize $j$}
  \put(54,112)       {\scriptsize $\bar k$}
  \put(55,93)        {\scriptsize $k$}
  \put(57,148)       {\scriptsize $\bar l$}
  \put(36,192)       {\scriptsize $l$}
  \put(53,43)        {\scriptsize $l$}
  \put(159.5,76.2)   {\small $\alpha$}
  \put(72,76)        {\small $\bar{\gamma}$}
  \put(198.8,128.5)  {\small $\beta$}
  \put(33,128)       {\small $\bar{\delta}$}
  \put(254,215)      {\StSebox}
  \put(-30,111)      {$\mathcal N$}
  \epicture26  \labp 76
Here $S^2\Times S^1$ is drawn embedded in $\mathbb{R}^3$, with the $S^1$ factor
in vertical, direction, i.e.\ top and bottom are identified, and each
horizontal plane is the stereographic image of an $S^2$. By using the identity
  \bea \begin{picture}(335,88)(0,13)
  \put(0,0)          {\Includeourbeautifulpicture 77a }
  \put(205,0)        {\Includeourbeautifulpicture 77b }
  \put(130,46)       {$ =~ \dsty\frac{\delta_{i,j}}{\dim(U_i)} $}
  \put(13,66)        {\scriptsize $U_{i}$}
  \put(47.7,29)      {\scriptsize $U_{j}$}
  \put(214.8,29)     {\scriptsize $U_{i}$}
  \put(256.4,66)     {\scriptsize $U_{i}$}
  \put(74,85)        {\StIbox}
  \put(279,85)       {\StIbox}
  \epicture07  \labp 77
(which follows from dominance and $\mathcal H(S^2,U_k)\eq\delta_{k,0}\complex$)
at four places in \erp 76, we end up with the following ribbon graph:
  \bea \begin{picture}(400,330)(0,0)
  \put(120,0)        {\Includeourbeautifulpicture 78a }
  \put(166,185)      {\Includeourbeautifulpicture 78b }
  \put(166,115)      {\Includeourbeautifulpicture 78c }
  \put(166,45)       {\Includeourbeautifulpicture 78d }
  \put(0,150)        {$ \dsty\frac{\mathcal N\,\delta_{i,k}\,\delta_{j,l}}
                        {{\big(\dim(U_i)\,\dim(U_j)\big)}^2} $}
  \put(170,150)      {\scriptsize $\ib $}
  \put(170,126)      {\scriptsize $i$}
  \put(180,252)      {\scriptsize $\jb $}
  \put(180,275)      {\scriptsize $j$}
  \put(219,206.5)    {\scriptsize $\ib $}
  \put(219,182.8)    {\scriptsize $\jb $}
  \put(238,91)       {\scriptsize $j$}
  \put(238,51)       {\scriptsize $i$}
  \put(254,70.5)     {\small $\alpha$}
  \put(204.5,70.5)   {\small $\bar{\gamma}$}
  \put(258.5,193.5)  {\small $\beta$}
  \put(198.5,192)    {\small $\bar{\delta}$}
  \put(380,315)      {\StSebox}
  \epicture-9  \labp 78
Here two components of the ribbon graph have been rotated such that the white 
side of the ribbons faces up. Next, using the identities \erf{fig:cid2a}, 
\erf{fig:cid2c} and \erf{fig:prm7}, one concludes that the value of the 
invariant corresponding to this cobordism evaluated on $1\iN\complex$ is 
given by the product of the invariants of the following two graphs in\,%
  \footnote{~The notation $\Stn$ used in the picture indicates that what is
  displayed is the graphical representation of the {\em normalised\/} invariant
  of a ribbon graph in the three-sphere. That is, for a ribbon graph $R$ in
  $S^3$ we define $Z(\Stn[R]) \,{:=}\, Z(S^3[R]) / S_{0,0}$. Denoting the empty
  ribbon graph by $\emptyset$, by the identity $S_{0,0}\eq Z(S^3[\emptyset])$
  this amounts to the normalisation $Z(\Stn[\emptyset])\eq1$. This convention is
  also used explicitly in \cite{fuRs10}, while in \cite{fuRs4,fuRs8,fuRs9} it
  is understood implicitly that $\Stn$ should be used 
  instead of $S^3$ (see section 5.1 of \cite{fuRs4}).\label{Stnfootnote}}
$S^3$:
  \bea \begin{picture}(315,88)(0,28)
  \put(-20,0)        {\Includeourbeautifulpicture 81a }
  \put(150,0)        {\Includeourbeautifulpicture 81b }
  \put(260,0)        {\Includeourbeautifulpicture 81a }
  \put(-12,75)       {\footnotesize $\beta$}
  \put(-11,24)       {\footnotesize $\bar{\delta}$}
  \put(175.5,26)     {\footnotesize $\alpha$}
  \put(175.5,76)       {\footnotesize $\bar{\gamma}$}
  \put(268.6,75.5)  {\footnotesize $\alpha$}
  \put(268.6,26.5)   {\footnotesize $\bar{\gamma}$}
  \put(-19.8,47)     {\scriptsize$ \ib $}
  \put(.8,47)        {\scriptsize$ \jb $}
  \put(149,85)       {\scriptsize$ i $}
  \put(199,86)       {\scriptsize$ j $}
  \put(260,47)       {\scriptsize$ i $}
  \put(281,47)       {\scriptsize$ j $}
  \put(74,46)        {and}
  \put(227,48)       {$ = $}
  \put(12,91)        {\Stnbox}
  \put(292,91)       {\Stnbox}
  \epicture18 \labl{fig:dualgraphs}
Since the respective basis elements are dual to each other,
it follows that these numbers equal $\dim(A)\,\delta_{\delta,\beta}$ and
$\dim(A)\,\delta_{\alpha,\gamma}$, respectively. Hence the vectors 
$\widetilde{{\rm M}}_{ij,\alpha\beta}$ are indeed dual to 
${\rm M}_{ij,\alpha\beta}$, and the value of the normalisation constant 
$\mathcal N$ is $\big( \dim(U_i)\,\dim(U_j) / \dim(A) \big)^2.$
\\[.2em]
On the other hand, gluing $\widetilde{{\rm M}}_{ij,\alpha\beta}$ to the
cobordism ${\rm M}_{\rm T}$ results in the ribbon graph
  \bea \begin{picture}(270,178)(0,10)
  \put(0,0)          {\includeourbeautifulpicture 82 }
  \put(56.5,102)     {\scriptsize $\ib $}
  \put(45.2,80.8)    {\scriptsize $i$}
  \put(59,134)       {\scriptsize $\jb $}
  \put(125,165)      {\scriptsize $j$}
  \put(48.2,28)      {\scriptsize $j$}
  \put(66,57)        {\small $\bar{\gamma}$}
  \put(27,115)       {\small $\bar{\delta}$}
  \put(272,170)      {\Stnbox}
  \put(-60,100)      {$\mathcal N \cdot S_{0,0}$}
  \epicture02 \labp 82
(the factor of $S_{0,0}$ appears when passing from $S^3$ to $\Stn$).
This can be simplified by deformations and using~(\ref{fig:prm7}), leading to
  \bea \begin{picture}(55,95)(0,35)
  \put(0,0)          {\includeourbeautifulpicture 83 }
  \put(24,81)        {\scriptsize $\ib $}
  \put(24,34)        {\scriptsize $i$}
  \put(54.7,87)      {\scriptsize $\jb $}
  \put(47,37)        {\scriptsize $j$}
  \put(35.5,21)      {\small $\bar{\gamma}$}
  \put(36,66.5)      {\small $\bar{\delta}$}
  \put(-60,54)       {$\mathcal N \cdot S_{0,0}$}
  \epicture17 \labp 83
With $\mathcal N \eq \big(\dim(U_i)\dim(U_j)/\dim(A)\big)^2$,
this is precisely formula \erf{fig:expcoef1}.
\qed

\proofof[proposition {\ref{prop:bulksurgery}}]

By equation \erf{eq:Saction}, in order to establish \erf{eq:bulksurg}
we need to show that
  \be
  \dim(U_i)\,\dim(U_j)\, ({\Cbulk_{i,j}}^{\,\scriptstyle -1})^{}_{\delta\alpha}
  = K_{ij,\alpha\delta} \,.
  \labl{eq:cinv-K1}
Lemma \ref{2lem} provides a ribbon graph representation for 
$K_{ij,\alpha\delta}$. 
\\[.3em]
As an auxiliary calculation, place the ribbon graph in 
\erf{fig:expcoef1} together with the one for $\Cbulk_{ij,\alpha\beta}$
in \erf{fig:S2strc} inside an $\Stn$ and sum over $\alpha$.
In the graph defining $\Cbulk_{ij,\alpha\beta}$, use the function $f$ from 
\erf{pic:bimtr} on the morphism $\alpha$, and similarly use the function 
$g$ from \erf{pic:bimtr} on the morphism $\bar{\delta}$ in the graph defining 
$K_{ij,\alpha\delta}$. Using the matrix $\Delta$ and its inverse, defined 
just after lemma \ref{lem:A3-4}, we find
  \bea \begin{picture}(390,305)(12,0)
                        \put(0,145){\begin{picture}(0,0)
  \put(30,0)         {\Includeourbeautifulpicture 88a }
  \put(105,35)       {\Includeourbeautifulpicture 88b }
  \put(280,20)       {\Includeourbeautifulpicture 88g }
  \put(370,20)       {\Includeourbeautifulpicture 88f }
  \put(0,69)         {$ \dsty\sum_\alpha $}
  \put(218,69)       {$ = $}
  \put(238,69)	     {$\dsty\sum_\alpha $}
                        \end{picture}}
                        \put(40,10){\begin{picture}(0,0)
  \put(69,10)        {\Includeourbeautifulpicture 88f }
  \put(120,10)       {\Includeourbeautifulpicture 88f }
  \put(280,10)       {\Includeourbeautifulpicture 88f }
  \put(330,10)       {\Includeourbeautifulpicture 88f }
  \put(69,60)        {\scriptsize $\ib$}
  \put(90,60)        {\scriptsize $\jb$}
  \put(120,60)       {\scriptsize $i$}
  \put(142,60)       {\scriptsize $j$}
  \put(279,60)       {\scriptsize $\ib$}
  \put(301,60)       {\scriptsize $\jb$}
  \put(330,60)       {\scriptsize $i$}
  \put(352,60)       {\scriptsize $j$}
  \put(8,60)         {$ =~\dsty\sum_\alpha$}
  \put(181,60)       {$ =~\dsty\sum_{\alpha,\gamma,\tau}
                        \Delta_{\alpha\gamma}\,\Omega_{\delta\tau}$}
  \put(71,36)        {\footnotesize $f(\alpha)$}
  \put(77,85)        {\footnotesize $\beta$}
  \put(129,35.5)     {\footnotesize $\bar\alpha$}
  \put(124,85)       {\footnotesize $g(\delta)$}
  \put(289,36.5)     {\footnotesize $\bar{\gamma}$}
  \put(288.5,85)     {\footnotesize $\beta$}
  \put(339,35.5)     {\footnotesize $\bar{\alpha}$}
  \put(339.5,85)     {\footnotesize $\tau$}
                        \end{picture}}
  \put(61,230.5)     {\footnotesize $\alpha$}
  \put(61,183)       {\footnotesize $\beta$}
  \put(136.5,247.3)  {\footnotesize $\bar{\delta}$}
  \put(136.5,201.5)  {\footnotesize $\bar{\alpha}$}
  \put(296.5,240.5)  {\footnotesize $f(\alpha)$}
  \put(303,191)      {\footnotesize $\beta$}
  \put(374,240.5)    {\footnotesize $g(\delta)$}
  \put(379,191)      {\footnotesize $\bar{\alpha}$}
  \put(52,220)       {\scriptsize $i$}
  \put(73,220)       {\scriptsize $j$}
  \put(52,173)       {\scriptsize $\ib$}
  \put(73,173)       {\scriptsize $\jb$}
  \put(126,259)      {\scriptsize $\ib$}
  \put(148,259)      {\scriptsize $\jb$}
  \put(126,212)      {\scriptsize $i$}
  \put(148,212)      {\scriptsize $j$}
  \put(294,251)      {\scriptsize $\ib$}
  \put(316,251)      {\scriptsize $\jb$}
  \put(294,181)      {\scriptsize $\ib$}
  \put(316,181)      {\scriptsize $\jb$}
  \put(370,216)      {\scriptsize $i$}
  \put(392,216)      {\scriptsize $j$}
  \put(95,-20)	     {$ =\,~\dsty(\dim(A))^2_{}\sum_{\alpha,\gamma,\tau}
                        \Delta_{\alpha\gamma}\,\Omega_{\delta\tau}\,
                        \delta_{\beta,\gamma}\,\delta_{\alpha,\tau} 
                        ~=~\dsty\frac {{(\dim(A))}^2_{}} {\dim(U_i)\,\dim(U_j)}
                        \, \delta^{}_{\beta,\delta}\,. $}
  \epicture20 \labl{fig:Kinv}
Here in the first step the definitions \erf{pic:bimtr} are used. For the left 
graph, this also involves a deformation, moving the thin coupons closer to 
the coupon labelled by $\alpha$. The second step is again a deformation of 
the left graph, followed by using the identity \erf{fig:prm7}. The rest 
follows using \erf{eq:bimmatr} and \erf{eq:bimtrmat}.

This auxiliary calculation shows that
\be
  \sum_\alpha S_{0,0} \, \Cbulk_{ij,\alpha\beta} \cdot
  \frac{1}{S_{0,0}} \Big( \frac{\dim(A)}{\dim(U_i)\dim(U_j)}\Big)^{\!2}
  K_{ij,\alpha\delta} =
  \frac{\dim(A)^2}{\dim(U_i)\dim(U_j)}\,\delta_{\beta,\delta} \,.
  \labl{eq:cinv-K2}
This establishes \erf{eq:cinv-K1}.
\qed


\subsection{Oriented world sheets}

$~$

\noindent
We are now finally in a position to prove the bulk factorisation theorem.

\proofof[theorem {\ref{thm:bulk-or}}]

By proposition~\ref{prop:bulksurgery}, the sum on the right hand side of
\erf{eq:bulk-or} can be replaced by the cobordism obtained from~\erf{fig:RHScob}
by exchanging the tubular neighbourhood of the vertically running ribbons by 
${\rm M}_{\rm T}$. Moreover, as illustrated in the following figure, the 
geometry of the relevant
region of the three-manifold is precisely the geometry of $C(\X)$:
  \begin{eqnarray} \begin{picture}(420,423)(10,0)
  \put(0,0)         {\Includeournicemediumpicture 89a }
  \put(29,182.5)    {\tiny $1$}
  \put(17.5,174)    {\tiny $2$}
  \put(29,335.5)    {\tiny $1$}
  \put(17.5,327)    {\tiny $2$}
  \put(117,184)     {\small $\Delta$}
  \put(152,187)     {\scriptsize \begin{turn}{125}$\eta$\end{turn}}
  \put(142,122)     {\small $m$}
  \put(167,136)     {\small $m$}
  \end{picture} \nonumber\\ \begin{picture}(200,148)(4,-5)
  \put(0,0)         {\Includeournicemediumpicture 89b }
  \put(-40,145)     {$ = $}
  \put(29,189)      {\tiny $1$}
  \put(18,181.6)    {\tiny $2$}
  \put(198,78)      {\tiny $1$}
  \put(187,95.8)    {\tiny $2$}
  \put(108,166)     {\small $\Delta$}
  \put(146,170)     {\scriptsize \begin{turn}{125}$\eta$\end{turn}}
  \put(132,104)     {\small $m$}
  \put(153.5,118)   {\small $m$}
  \end{picture}  \label{fig:proofgeom} \end{eqnarray}
This proves the theorem.
\qed


\subsection{Unoriented world sheets}

$~$

\noindent
The proof of theorem \ref{thm:bulk-unor} is identical to that
of theorem \ref{thm:bulk-or}. 

A property of the unoriented bulk factorisation
not present in the oriented one is the following. With every embedding
$f{:}\ \Aeps\To\X$ we obtain another embedding $\tilde f{:}\ \Aeps \To \X$ by 
setting $\tilde f(x,y,z) \,{:=}\, f(x,-y,z)$, which results in the opposite 
local orientation of \X\ on the image of $\Aeps$. By theorem 
\ref{thm:bulk-unor}, both $f$ and $\tilde f$ give factorisation identities for 
the correlators. These identities can be related by a change of the local 
orientation that enters the definition of a bulk field (see definition IV:3.6
or appendix \ref{app:equiv-lab}). Such a change of local orientation
is immaterial only because $\tilde A$ is a Jandl algebra.

\appendix

\sect{Algebras in modular tensor categories}\label{variousoldstuff}

\subsection{Modular tensor categories and graphical calculus}\label{mtcstuff}

$~$

\noindent
A {\em ribbon category\/} \cite[chap.\,XIV.3]{KAss} is a strict monoidal 
category endowed with duality, braiding, and twist, which are the following 
families of morphisms: A (right) {\em duality\/} on a strict monoidal category 
\C\ associates to every $U\iN\obj(\C)$ an object $U\Vee{\in}\,\obj(\C)$, 
called the (right-)\,dual of $U$, and morphisms
  \be
  b_U \in\Hom(\one,U\Oti U\Vee)\,, \qquad
  d_U \in \Hom(U\Vee\Oti U,\one) \,,
  \ee
and to every morphism $f\iN\Hom(U,Y)$ the morphism
  \be
  f\Vee := (d_Y\oti\id_{U\Vee}) \circ (\id_{Y\Vee}\oti f\oti\id_{U\Vee})
  \circ (\id_{Y\Vee}\oti b_U) \ \in \Hom(Y\Vee\!{,}\,U\Vee) \,.
  \ee
A {\em braiding\/} on \C\ is a family of isomorphisms 
$c_{U,V}\iN\Hom(U\Oti V,V\Oti U)$, one for each pair of objects $U,V\iN\objc$,
and a {\em twist\/} is a family of isomorphisms $\theta_U$, one for each 
$U\iN\objc$. These morphisms are subject to various axioms, which are analogous
to properties of ribbons in $S^3$ and correspondingly can be conveniently
visualized through a graphical calculus as introduced in \cite{joSt5}
(see e.g.\ also \cite{KAss,MAji}).

With the convention that pictures are read from bottom to top, the graphical 
notation for identity morphisms and for general morphisms and 
their composition and tensor product looks as follows:

\vskip .6em

\begin{center}
\begin{tabular}{|cc|cc|cc|cc|} \hline
\mcll{
~~$\id_U^{}=$
\begin{picture}(21,44)(0,22)       \apppicture{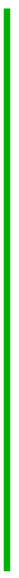}  {7}
\put(4.0,-8.8)   {\sse$U$}
\put(4.5,51.3)   {\sse$U$} \end{picture}     }&
\mcll{
~~$f       =$
\begin{picture}(32,38)(0,22)       \apppicture{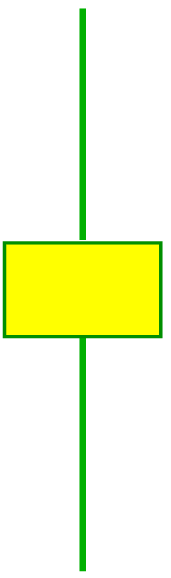}  {7}
\put(10.3,-8.8)  {\sse$U$}
\put(11.2,51.3)  {\sse$V$}
\put(11.6,22.5)  {\tiny$f$} \end{picture}     }&
\mcll{
~~$g\cir f =$
\begin{picture}(36,38)(0,22)       \apppicture{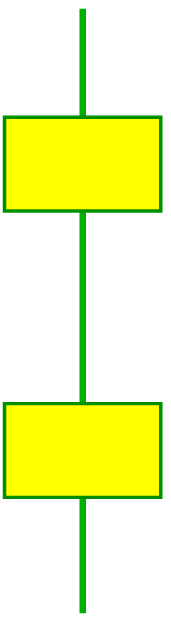}  {7}
\put(10.1,54.3)  {\sse$W$}
\put(10.3,-8.8)  {\sse$U$}
\put(11.6,12.6)  {\tiny$f$}
\put(12.1,37.5)  {\tiny$g$}
\put(14.9,23.9)  {\tiny$V$} \end{picture}     }&
\mcll{
~~$f\Oti f'=$
\begin{picture}(45,38)(0,22)       \apppicture{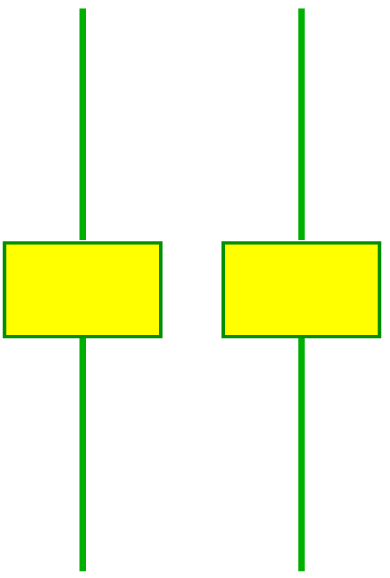}  {7}
\put(10.3,-8.8)  {\sse$U$}
\put(11.2,51.3)  {\sse$V$}
\put(11.6,22.4)  {\tiny$f$}
\put(28.1,-8.8)  {\sse$U'$}
\put(29.0,51.3)  {\sse$V'$}
\put(29.6,22.4)  {\tiny$f'$} \end{picture}     }
\\ \begin{picture}(0,27){}\end{picture} &&&&&&& \\ \hline \end{tabular}
\end{center}

\vskip .3em 

The next pictures show the structural morphisms of a ribbon category:
the braiding, twist, and left and right dualities, 
as well as the definition of the (left and right) dual of a morphism.
(A ribbon category has automatically also a left duality,
with left dual objects $\eev U\,{:=}\,U\Vee$ and left duality morphisms 
$\tilde b_U$ and $\tilde d_U$. It is in fact sovereign,
i.e.\ the left and right dualities coincide both on objects and on morphisms.
Also, lines labelled by the monoidal unit $\one$ are not drawn -- owing to
strictness they are `invisible'.)

\vskip .4em

\begin{center}
\begin{tabular}{|cc|cc|cc|cc|} \hline
\mcll{
~$c_{U,V}^{} =$
\begin{picture}(36,37)(0,18)       \apppicture{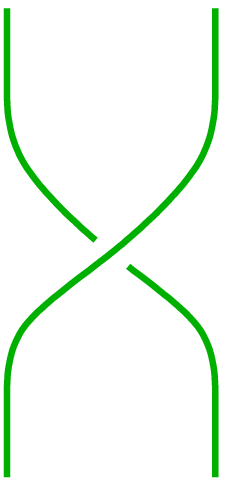}  {7}
\put(4.2,-8.8)   {\sse$U$}
\put(4.8,43.3)   {\sse$V$}
\put(21.2,-8.8)  {\sse$V$}
\put(22.9,43.3)  {\sse$U$} \end{picture}     }&
\mcll{
~$c_{U,V}^{-1} =$
\begin{picture}(36,30)(0,18)       \apppicture{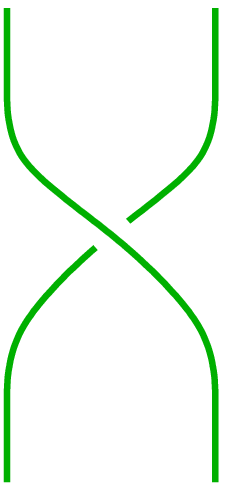}  {7}
\put(4.2,-8.8)   {\sse$V$}
\put(4.8,43.3)   {\sse$U$}
\put(21.2,-8.8)  {\sse$U$}
\put(22.9,43.3)  {\sse$V$} \end{picture}     }&
\mcll{
~~~~$\theta_{U}^{}=$
\begin{picture}(26,30)(0,18)       \apppicture{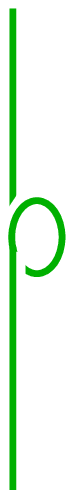}  {7}
\put(5.5,-8.8)   {\sse$U$}
\put(6.4,44.3)   {\sse$U$} \end{picture}     }&
\mcll{
~~~$\theta_{U}^{-1}=$
\begin{picture}(20,30)(0,18)       \apppicture{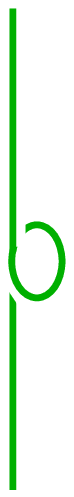}  {7}
\put(5.5,-8.8)   {\sse$U$}
\put(6.4,44.3)   {\sse$U$} \end{picture}     }
\\
\begin{picture}(0,24){}\end{picture} &&&&&&& \\ \cline{1-8}
\mcll{
~$b_{U}^{} =$
\begin{picture}(36,32)(0,18)       \apppicture{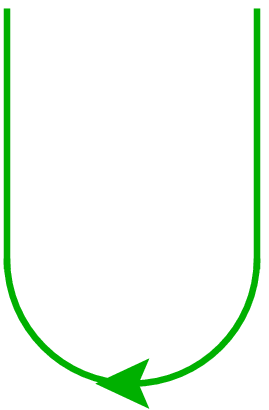}  {7}
\put(4.8,37.3)   {\sse$U$}
\put(24.9,37.3)  {\sse$U^\vee$} \end{picture}    }&
\mcll{
~$d_{U}^{} =$
\begin{picture}(36,30)(0,11)       \apppicture{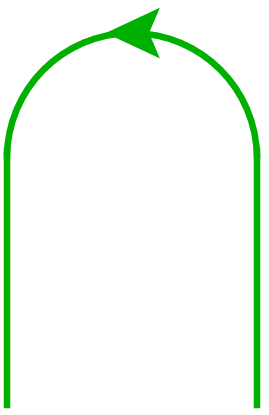}  {7}
\put(3.2,-8.8)   {\sse$U^\vee$}
\put(25.2,-8.8)  {\sse$U$} \end{picture}    }&
\mcll{
~~$\tilde b_{U}^{}=$
\begin{picture}(38,30)(0,18)       \apppicture{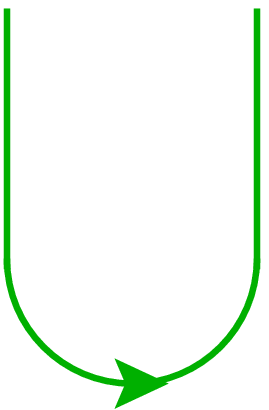}  {7}
\put(-4.4,37.3)  {\sse$ {\phantom U}^\vee_{}\!{U}$}
\put(25.7,37.3)  {\sse$U$} \end{picture}    }&
\mcll{
~~$\tilde d_{U}^{}=$
\begin{picture}(33,30)(0,11)       \apppicture{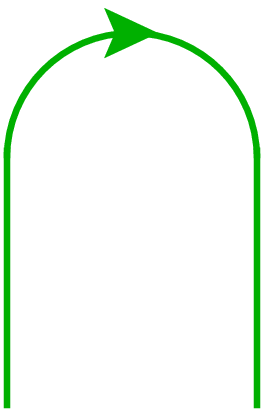}  {7}
\put(4.8,-8.8)   {\sse$U$}
\put(15.9,-8.8)  {\sse$ {\phantom U}^\vee_{}\!{U}$}
                  \end{picture}    }
\\
\begin{picture}(0,18){}\end{picture} &&&&&&& \\ \cline{1-8}
\mclo{\begin{picture}(61,0){}\end{picture}}&\mclll{
~$f^\vee_{} =$
\begin{picture}(36,50)(0,18)       \apppicture{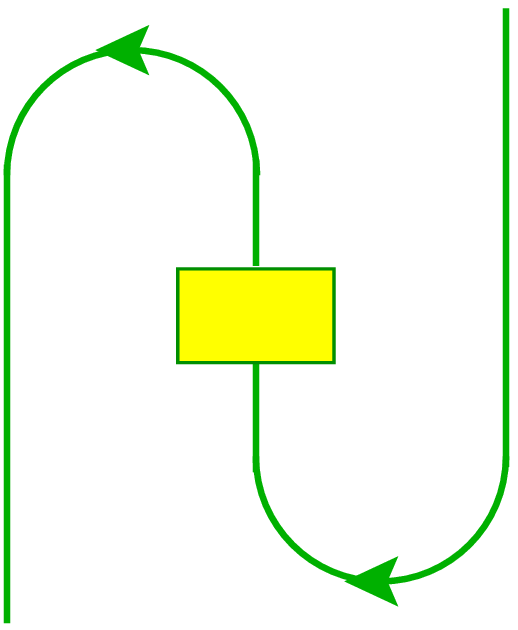}  {7}
\put(3.2,-8.8)   {\sse$V^\vee$}
\put(26.4,25.2)  {\tiny$f$}
\put(44.8,55.7)  {\sse$U^\vee$}
                  \end{picture}    }&
\mclo{\begin{picture}(60,0){}\end{picture}}&\mclll{ 
${\phantom f}^{\vee\!}_{}\!{f} =$
\begin{picture}(60,30)(0,18)       \apppicture{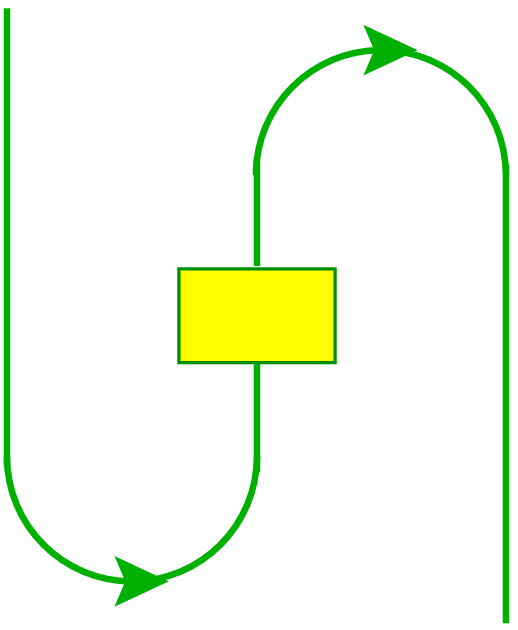}  {7}
\put(-5.8,55.7)  {\sse$ {\phantom U}^\vee_{}\!{U}$}
\put(26.4,25.2)  {\tiny$f$}
\put(35.8,-8.8)  {\sse$ {\phantom U}^\vee_{}\!{V}$}
                 \end{picture}     }
\\ \mclo{}&\mclll{\begin{picture}(0,27){}\end{picture}}&
\mclo{}&\mclll{\begin{picture}(0,25){}\end{picture}} \\ \cline{2-4}\cline{6-8}
\end{tabular}
\end{center}

\vskip .1em 

\noindent
With these conventions, the axioms of a ribbon category -- the properties of
dualities, functoriality and tensoriality of the braiding, functoriality of 
the twist, and compatibility of the twist with duality and with braiding --
look as follows.

\vskip .4em

\begin{center}
\begin{tabular}{|c|c|cc|c|} \hline
\begin{picture}(95,56)(0,23)       \apppicture{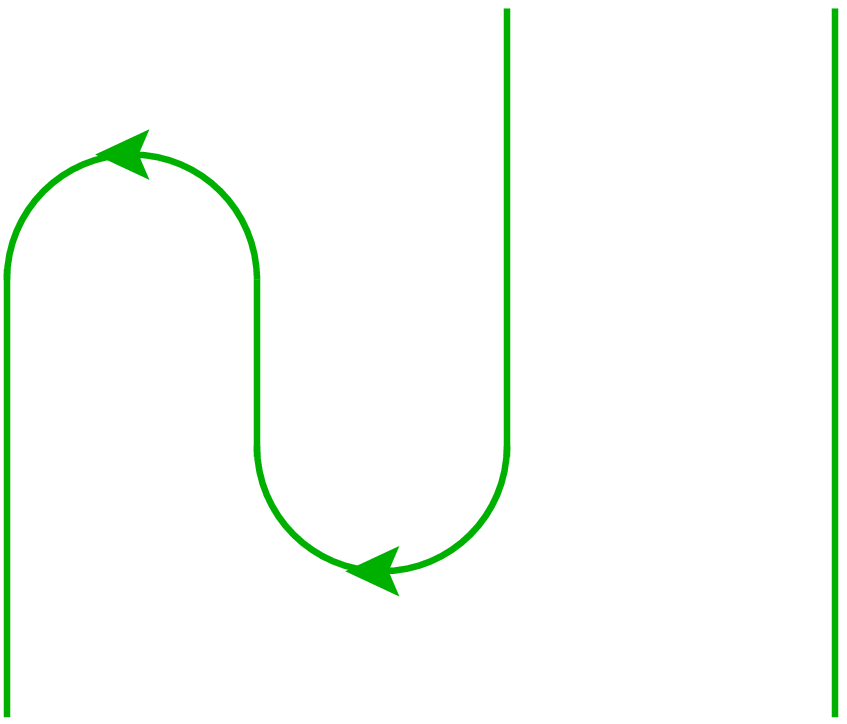}  {14}
\put(10.2,-9.4)  {\sse$U^\vee$}
\put(51.8,64.2)  {\sse$U^\vee$}
\put(65.5,25)    {\small$=$}
\put(79.2,-9.4)  {\sse$U^\vee$}
\put(79.8,64.2)  {\sse$U^\vee$} \end{picture}     &
\begin{picture}(95,49)(0,23)       \apppicture{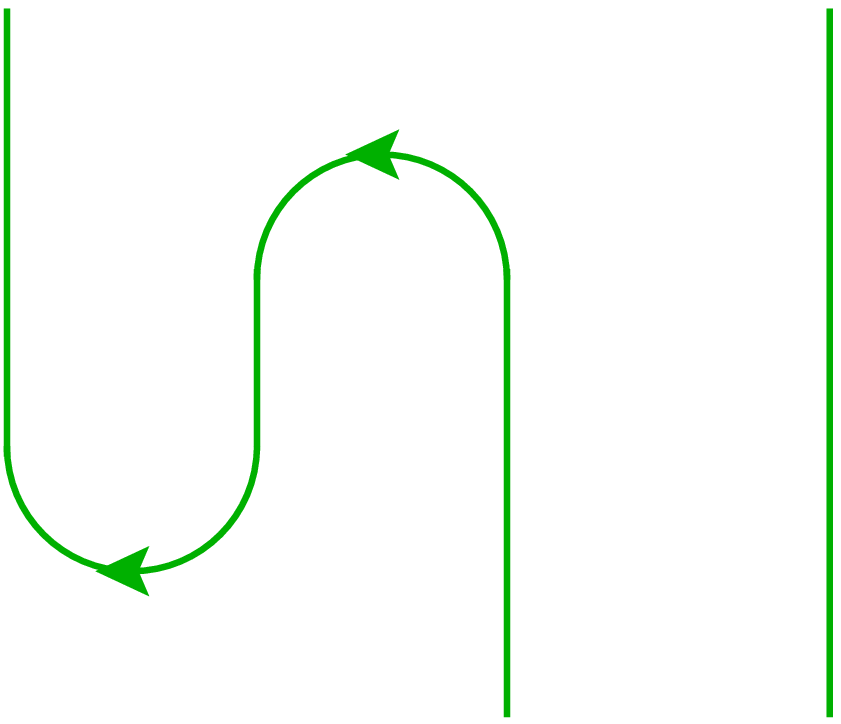}  {14}
\put(12.2,64.2)  {\sse$U$}
\put(52.8,-8.8)  {\sse$U$}
\put(65.5,25)    {\small$=$}
\put(80.4,-8.8)  {\sse$U$}
\put(81.1,64.2)  {\sse$U$} \end{picture}     &
\mcll{
\begin{picture}(95,49)(0,23)       \apppicture{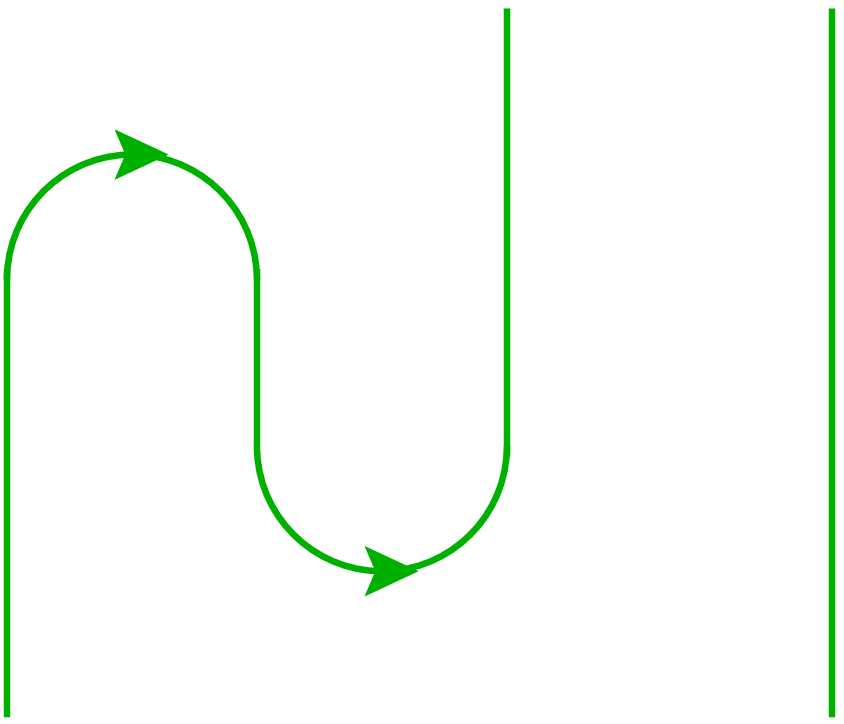}  {14}
\put(10.8,-8.8)  {\sse$U$}
\put(53.6,64.2)  {\sse$U$}
\put(65.5,25)    {\small$=$}
\put(80.4,-8.8)  {\sse$U$}
\put(81.1,64.2)  {\sse$U$} \end{picture}    }&
\begin{picture}(95,49)(0,23)       \apppicture{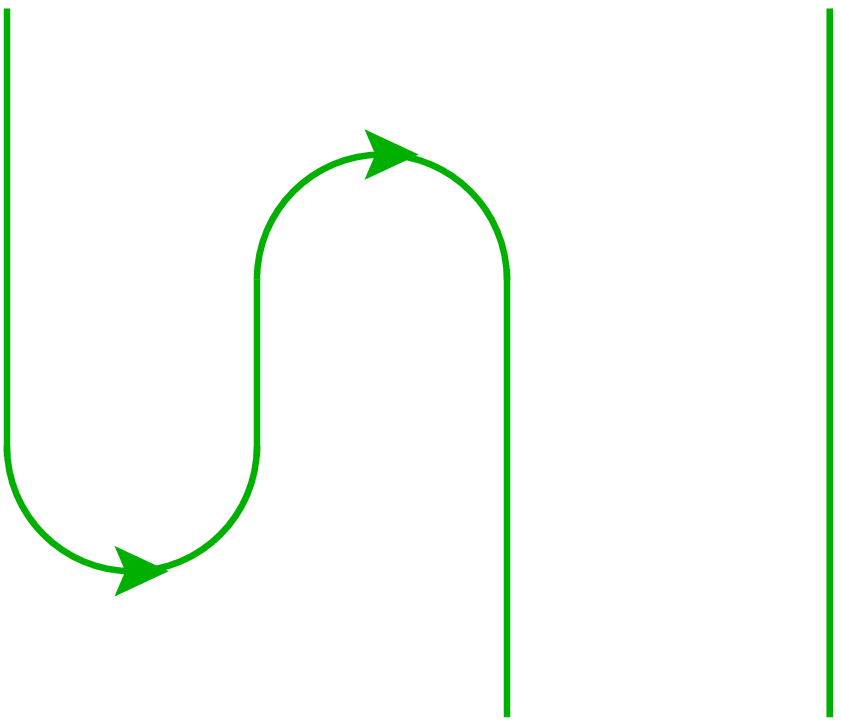}  {14}
\put(3.2,64.2)   {\sse$ {\phantom U}^\vee_{}\!{U}$}
\put(43.8,-9.4)  {\sse$ {\phantom U}^\vee_{}\!{U}$}
\put(65.5,25)    {\small$=$}
\put(70.5,-9.4)  {\sse$ {\phantom U}^\vee_{}\!{U}$}
\put(70.8,64.2)  {\sse$ {\phantom U}^\vee_{}\!{U}$}
                 \end{picture}
\\
\begin{picture}(0,31){}\end{picture} &&&& \\ \cline{1-5}
\begin{picture}(97,49)(0,18)       \apppicture{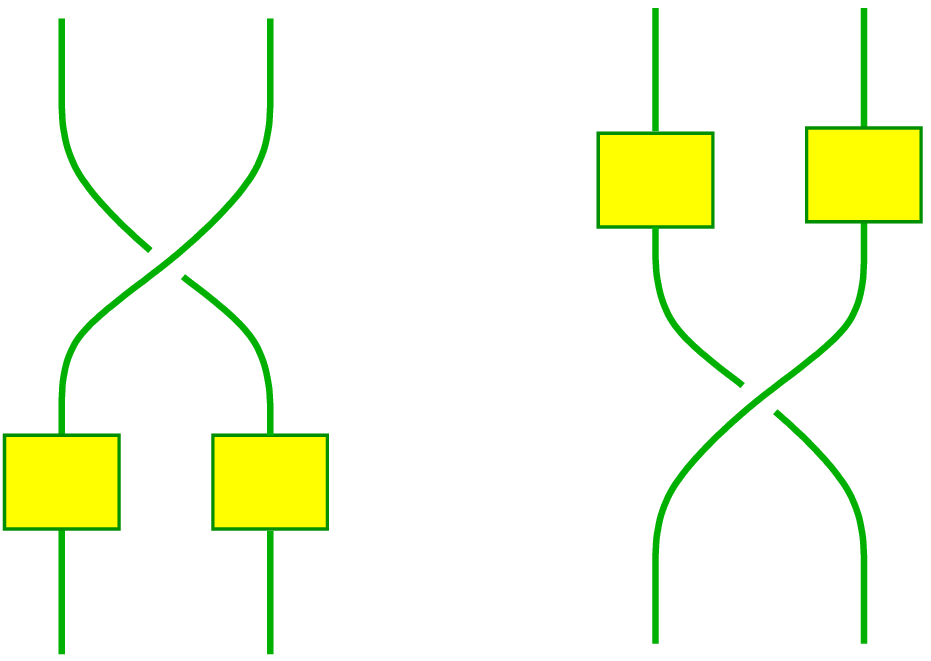}  {10}
\put(11.8,-8.5)  {\sse$U$}
\put(11.8,58.1)  {\sse$W$}
\put(12.5,13.5)  {\tiny$f$}
\put(31,14)      {\tiny$g$}
\put(28.8,-8.5)  {\sse$V$}
\put(29.2,58.1)  {\sse$X$}
\put(45,25)      {\small$=$}
\put(61.3,-8.5)  {\sse$U$}
\put(61.3,58.1)  {\sse$W$}
\put(63,39.7)    {\tiny$g$}
\put(80,39.3)    {\tiny$f$}
\put(78.8,-8.5)  {\sse$V$}
\put(79.2,58.1)  {\sse$X$} \end{picture}     &
\mcll{
\begin{picture}(124,62)(0,28)      \apppicture{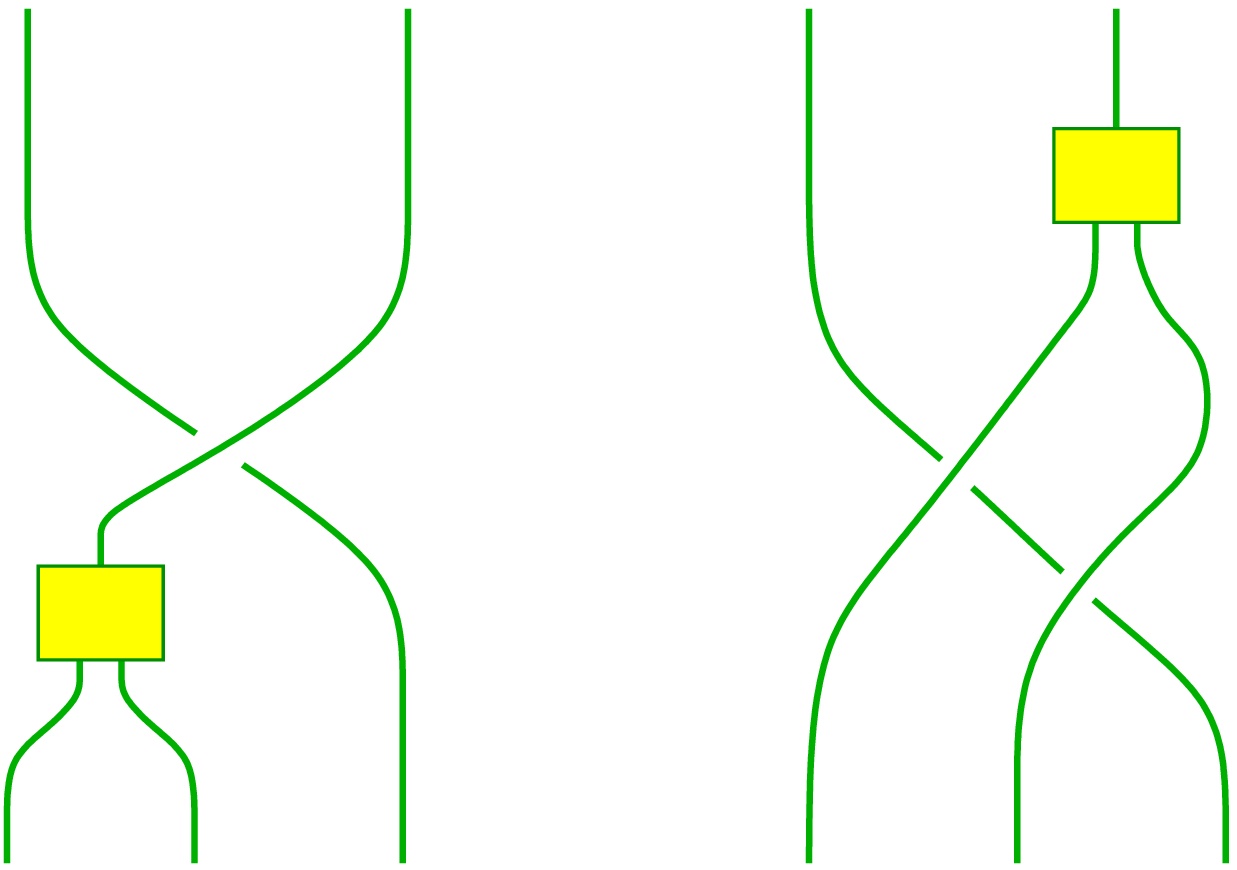}  {10}
\put(7.4,-8.5)   {\sse$U$}
\put(8.4,76.4)   {\sse$W$}
\put(16.3,19.9)  {\tiny$f$}
\put(22.8,-8.5)  {\sse$V$}
\put(34.4,76.4)  {\sse$U\Oti V$}
\put(39.4,-8.5)  {\sse$W$}
\put(58.5,31)    {\small$=$}
\put(74.1,-8.5)  {\sse$U$}
\put(74.4,76.4)  {\sse$W$}
\put(91.5,-8.5)  {\sse$V$}
\put(93.5,76.4)  {\sse$U\Oti V$}
\put(101.3,56.9) {\tiny$f$}
\put(107.7,-8.5) {\sse$W$} \end{picture}     }
\\
\begin{picture}(0,36){}\end{picture} &\mcll{~} \\ \cline{1-4}
\begin{picture}(63,36)(0,-7)       \apppicture{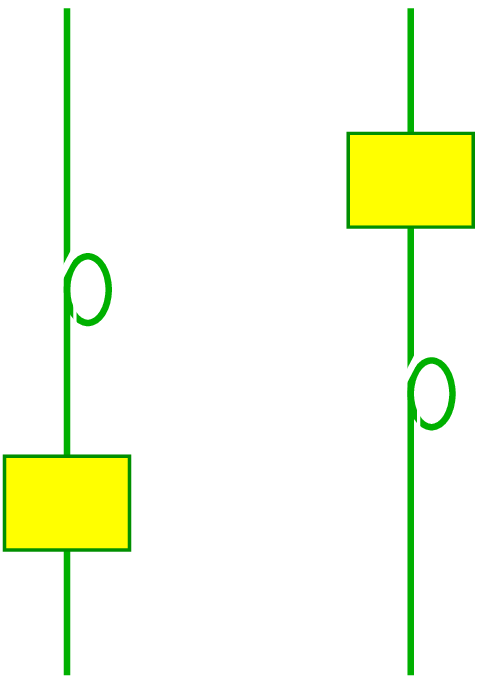}  {10}
\put(12.4,-8.5)  {\sse$U$}
\put(13.2,59.9)  {\sse$V$}
\put(13.5,13.6)  {\tiny$f$}
\put(26,26)      {\small$=$}
\put(41.4,-8.5)  {\sse$U$}
\put(42.2,59.9)  {\sse$V$}
\put(42.5,40.6)  {\tiny$f$} \end{picture}     &
\begin{picture}(63,39)(0,-13)       \apppicture{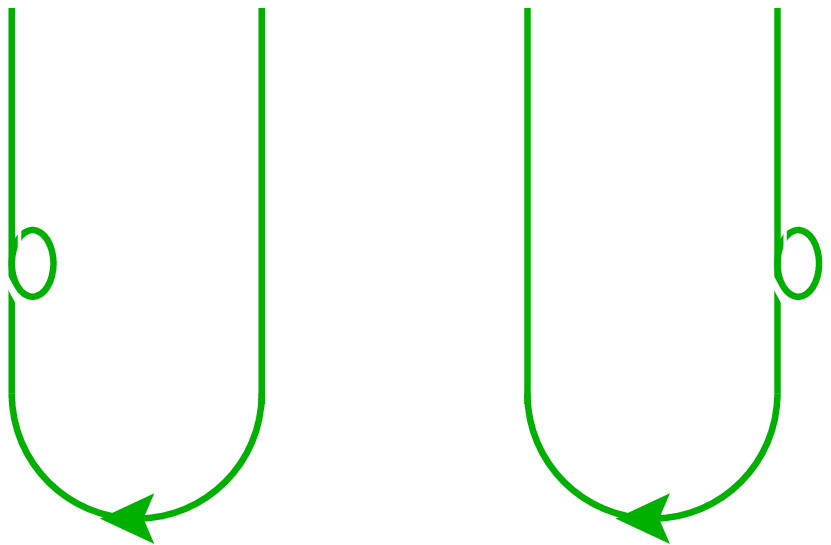} {-2}
\put(-2.8,49.3)  {\sse$U$}
\put(17.5,49.3)  {\sse$U^\vee$}
\put(30.5,18)    {\small$=$}
\put(40.8,49.3)  {\sse$U$}
\put(60.8,49.3)  {\sse$U^\vee$} \end{picture}     &
\mcll{
\begin{picture}(63,98)(0,12)       \apppicture{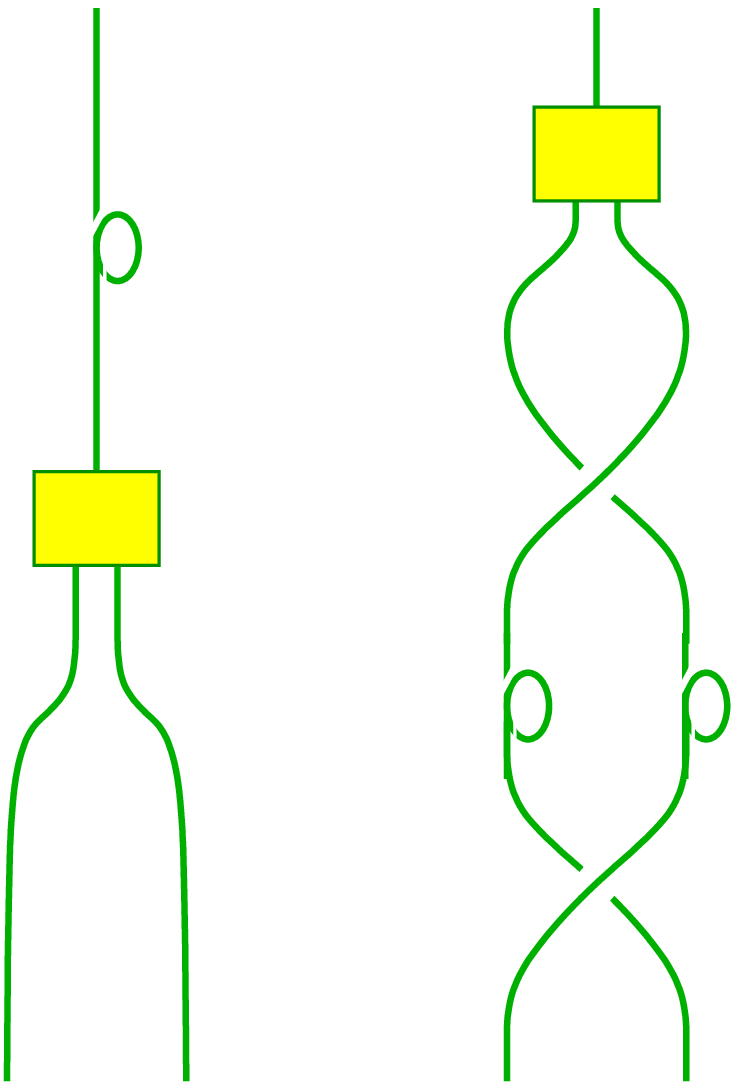}  {18}
\put(15.0,-8.5)  {\sse$U$}
\put(16.2,94.9)  {\sse$U\Oti V$}
\put(23.6,46.1)  {\tiny$f$}
\put(29.7,-8.5)  {\sse$V$}
\put(41,42)      {\small$=$}
\put(57.1,-8.5)  {\sse$U$}
\put(58.2,94.9)  {\sse$U\Oti V$}
\put(65.6,76.6)  {\tiny$f$}
\put(71.9,-8.5)  {\sse$V$} \end{picture}     }
\\ \begin{picture}(0,20){}\end{picture} &&& \\ \cline{1-4} \end{tabular}
\end{center}

A {\em modular tensor category\/} \cite{TUra} is a \complex-linear semisimple 
abelian ribbon category with simple monoidal unit and with a finite number of 
isomorphism classes of simple objects in which the braiding is maximally 
non-symmetric. The latter property means that the numerical
$|\I|\,{\times}\,|\I|$\,-matrix $s\eq(s_{U_i,U_j}^{})^{}_{i,j\in\I}$ with entries
  \be
  s_{U_i,U_j}^{} := {\rm tr}(c_{U_i,U_j}^{}c_{U_j,U_i}^{}) \,\in \complex
  \ee
is non-degenerate. Here we have identified the one-dimensional \complex-vector 
space $\Hom(\one,\one)$ with \complex\ and have chosen a set $\{U_i\,|\,i\iN\I\}$
of representatives of the finitely many isomorphism classes of simple objects 
of \C. Also, as representative of the class of the tensor unit $\one$ we choose 
$\one$ itself and denote the corresponding element of the label set $\I$ by $0$, 
i.e.\ write $\one\eq U_0$.

Instead of $s$, one often also uses the matrix $S \,{:=}\, S_{0,0}\,s$
that is scaled by the number $S_{0,0} \,{:=}\, (\sum_i \dim(U_i)^2 )^{-1/2}_{}$.
One then has $\dim(U_i) \,{\equiv}\, {\rm tr}(\id_{U_i})
\eq s_{U_i,U_0}^{} \eq S_{i,0} / S_{0,0}$.


\subsection{Three-dimensional TFT from modular tensor 
categories}\label{app:3dTFT-MTC}

$~$

\noindent
Next we briefly state our conventions for three-dimensional topological 
field theory; for more details see e.g.\ \cite{TUra,BAki,KArt} or section I:2. 
Given a modular tensor category \C, the construction of \cite{TUra} allows one 
to obtain a three-dimensional TFT, that is, a projective monoidal functor from 
a geometric category $\mathcal{G}_\C$ 
to $\Vect_\complex$, the category of finite-dimensional complex vector spaces.

The geometric category $\mathcal{G}_\C$ has extended surfaces as objects and
homotopy classes of cobordisms between extended surfaces as morphisms.
An {\em extended surface\/} $E$ consists of the following data:
\Itemize
\item[\Nxt] 
A compact oriented two-dimensional manifold with empty boundary, also denoted 
by $E$.
\item[\Nxt]
A finite (unordered) set of marked points -- that is, of quadruples
$(p_a,[\gamma_a],V_a,\eps_a)$, where the $p_a\iN E$ are mutually distinct 
points of the surface $E$ and $[\gamma_a]$ is a germ of arcs\,%
  \footnote{~By a {\em germ of arcs\/} we mean an equivalence class $[\gamma]$
  of continuous embeddings $\gamma$ of intervals 
  $[-\delta,\delta] \,{\subset}\, \reals$ into the extended surface $E$. 
  Two embeddings $\gamma{:}~[-\delta,\delta] \To E$ and
  $\gamma'{:}~[-\delta',\delta'] \To E$ are equivalent iff there is a 
  positive $\eps\,{<}\,\delta,\delta'$ such that the restrictions of $\gamma$
  and $\gamma'$ to the interval $[-\eps,\eps]$ are equal.}
$\gamma_a{:}\; [-\delta,\delta] \To E$ with $\gamma_a(0)\eq p_a$.
Furthermore, $V_a \iN \obj(\C)$, and $\eps_a\iN\{\pm 1\}$ is a sign.
\item[\Nxt]
A Lagrangian subspace $\lambda \,{\subset}\, H_1(E,\reals)$.
\end{itemize}
A morphism $E \To E'$ is an {\em extended cobordism\/} $M$ (or rather
a homotopy class thereof), consisting of the following data:
\Itemize
\item[\Nxt]
A compact oriented three-manifold, also denoted by $M$, such that
$\partial M \eq(-E)\,{\sqcup}\,E'$.
\\
Here $-E$ is obtained from $E$ by reversing its $2$-orientation
and replacing any marked point $(p,[\gamma],U,\eps)$ by
$(p,[\tilde\gamma],U,-\eps)$ with $\tilde\gamma(t)\eq\gamma(-t)$.
The boundary $\partial M$ of a cobordism is 
oriented according to the inward-pointing normal.
\item[\Nxt]
A ribbon graph 
inside $M$ such that for each marked point $(p,[\gamma],U,\eps)$ 
of $(-E)\,{\sqcup}\,E'$ there is a ribbon ending on $(-E)\,{\sqcup}\,E'$.
\\
The notion of a ribbon graph is reviewed in section I:2.3; the allowed ways 
for a ribbon to end on an arc are shown explicitly in (IV:3.1).
\end{itemize}
Composition in $\mathcal{G}_{\C}$ is defined by glueing, the identity 
morphism from $E$ to $E$ is the cylinder $E \Times [0,1]$, and the tensor 
product is given by disjoint union of objects and cobordisms.

Starting from a modular tensor category \C, one can construct a projective 
monoidal functor $(Z,\mathcal{H}){:}\ \mathcal{G}_\C \To \Vect_\complex$
\cite[Theorem\,IV.6.6]{TUra}. Here $\mathcal{H}$ denotes the action
of the functor on objects, i.e.\ $\mathcal{H}(E)$ is a finite-dimensional
\complex-vector space, and $Z$ denotes the action on morphisms, such that 
$Z(M,E,E')$ is a linear map from $\mathcal{H}(E)$ to $\mathcal{H}(E')$. 
The pair $(Z,\mathcal{H})$ is only a projective functor in the sense that
  \be
  Z'(M' \cir M) = \kappa^\mu\, Z'(M') \cir Z'(M) \,,
  \ee
where $M{:}\ E \To E'$, $M'{:}\ E' \To E''$ are extended cobordisms, 
$\kappa \eq S_{0,0} \sum_{j} \theta_j^{-1} \dim(U_j)^2_{}$, and $\mu$ is an 
integer computed from the Lagrangian subspaces in $E$, $E'$ and $E''$ via Maslov
indices; for details see \cite[chap.\,IV.7]{TUra} or \cite[section 2.7]{fffs3}.

\subsection{Algebras}\label{Astuff}

$~$

\noindent
A (unital associative) {\em algebra\/} $A$ in a monoidal \cat\ \C\ is a triple
$(A,m,\eta)$ consisting of an object $A$ of \C\ and morphisms
$m\iN\Hom(A\Oti A,A)$ and $\eta\iN\Hom(\one,A)$ that satisfy
  \be
  m\circ (m\oti\id_\AA) = m \circ (\id_\AA\oti m) \quad\;{\rm and} \quad\;
  m \circ(\eta\oti \id_\AA) = \id_\AA = m \circ (\id_\AA\oti \eta) \,.
  \ee
(What we call an algebra here is often, in particular in \cite{MAcl},
referred to as a monoid.)
Analogously, a (counital coassociative) {\em coalgebra\/} in \C\ is a triple
$(A,\Delta,\eps)$ consisting of an object $A$ and morphisms
$\Delta\iN\Hom(A,A\Oti A)$ and $\eps\iN\Hom(A,\one)$ obeying
coassociativity and counit properties that amount to reversing
the arrows in the associativity and unit properties.
We depict the structural morphisms of a (co)algebra as follows.

\vskip 1.4em

\begin{center}
\begin{tabular}{|c|c|c|c|} \hline
~~$m =$
\begin{picture}(41,32)(0,18)       \apppicture{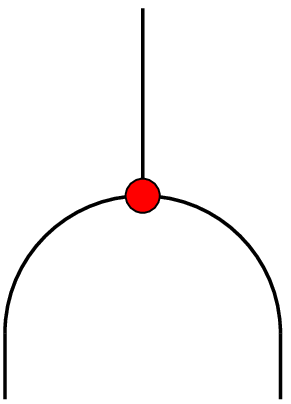}  {7}
\put(3.5,-8.8)   {\sse$A$}
\put(15.5,36.8)  {\sse$A$}
\put(26.5,-8.8)  {\sse$A$} \end{picture}     &
~~$\eta =$
\begin{picture}(24,20)(0,12)        \apppicture{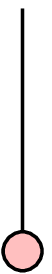}  {7}
\put(5.8,25.8)   {\sse$A$} \end{picture}     &
~~$\Delta =$
\begin{picture}(41,30)(0,18)       \apppicture{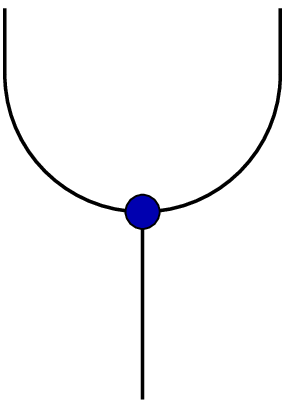}  {7}
\put(3.8,36.8)   {\sse$A$}
\put(15.2,-8.8)  {\sse$A$}
\put(26.8,36.8)  {\sse$A$} \end{picture}     &
~~$\eps =$
\begin{picture}(24,20)(0,6)        \apppicture{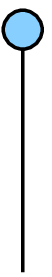}  {7}
\put(5.1,-8.8)   {\sse$A$} \end{picture}
\\ \begin{picture}(0,27){}\end{picture} &&& \\ \hline
\end{tabular}
\end{center}

\vskip .8em 
\noindent
The associativity of the product, unit property, coassociativity of the 
coproduct, and counit property are then given by the following diagrams.

\vskip .2em

\begin{center}
\begin{tabular}{|l|l|l|l|} \hline
\begin{picture}(103,41)(-2,22)       \apppicture{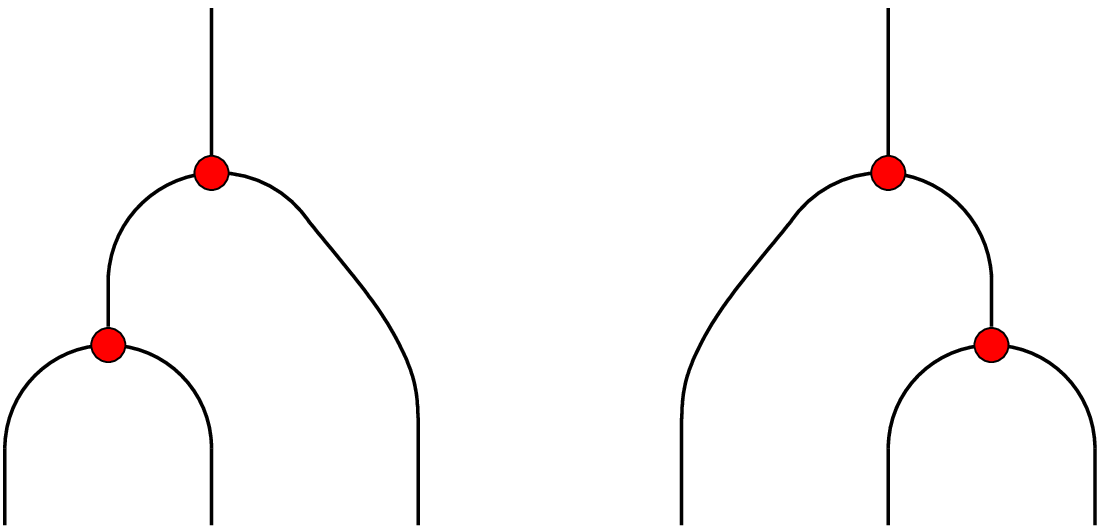}  {3}
\put(-.5,-9.2)   {\sse$A$}
\put(17.1,-9.2)  {\sse$A$}
\put(17.7,47.4)  {\sse$A$}
\put(34.1,-9.2)  {\sse$A$}
\put(44.8,18)    {\small$=$}
\put(56.3,-9.2)  {\sse$A$}
\put(73.9,-9.2)  {\sse$A$}
\put(74.7,47.4)  {\sse$A$}
\put(91.1,-9.2)  {\sse$A$} \end{picture}     &
\begin{picture}(103,29)(-2,22)       \apppicture{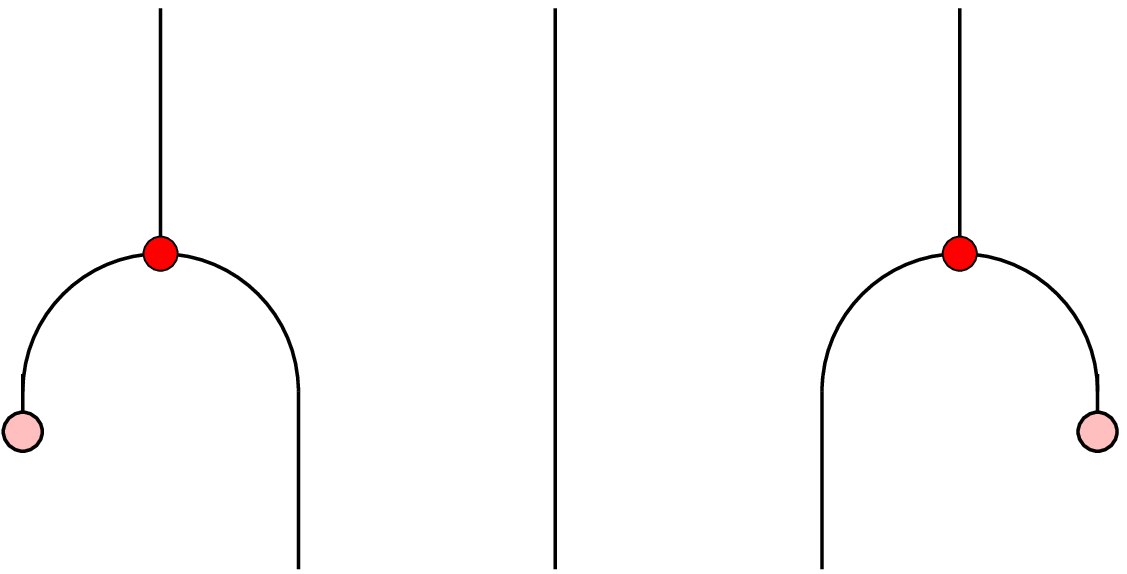}  {3}
\put(13.3,50.9)  {\sse$A$}
\put(23.3,-9.2)  {\sse$A$}
\put(33.5,18)    {\small$=$}
\put(46.1,-9.2)  {\sse$A$}
\put(46.6,50.9)  {\sse$A$}
\put(56.5,18)    {\small$=$}
\put(68.2,-9.2)  {\sse$A$}
\put(80.1,50.9)  {\sse$A$} \end{picture}     &
\begin{picture}(103,29)(-2,22)       \apppicture{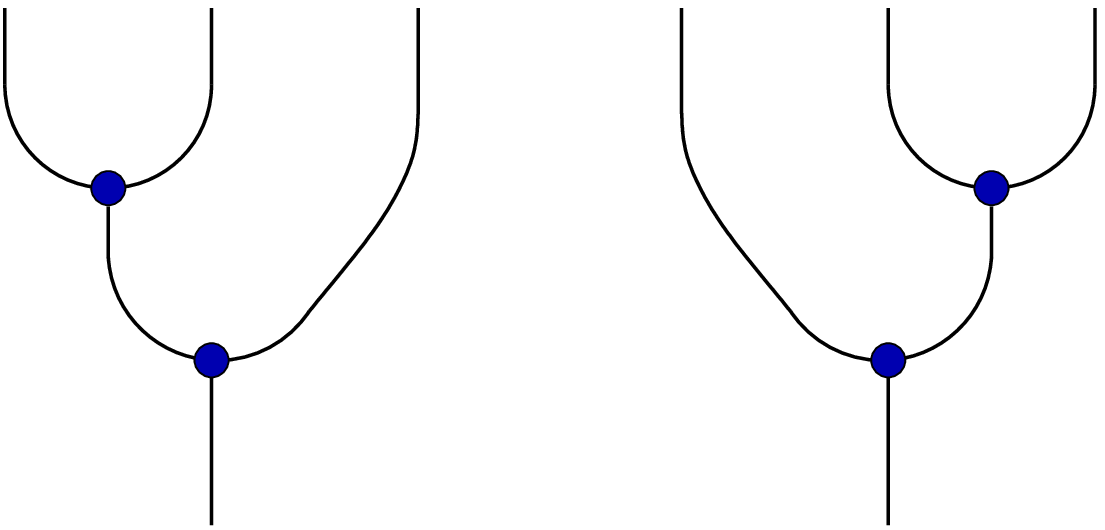}  {3}
\put(-.2,47.7)   {\sse$A$}
\put(16.6,-9.2)  {\sse$A$}
\put(17.3,47.4)  {\sse$A$}
\put(34.5,47.4)  {\sse$A$}
\put(43.5,18)    {\small$=$}
\put(57.0,47.4)  {\sse$A$}
\put(73.4,-9.2)  {\sse$A$}
\put(74.3,47.4)  {\sse$A$}
\put(91.3,47.4)  {\sse$A$} \end{picture}     &
\begin{picture}(103,29)(-2,22)       \apppicture{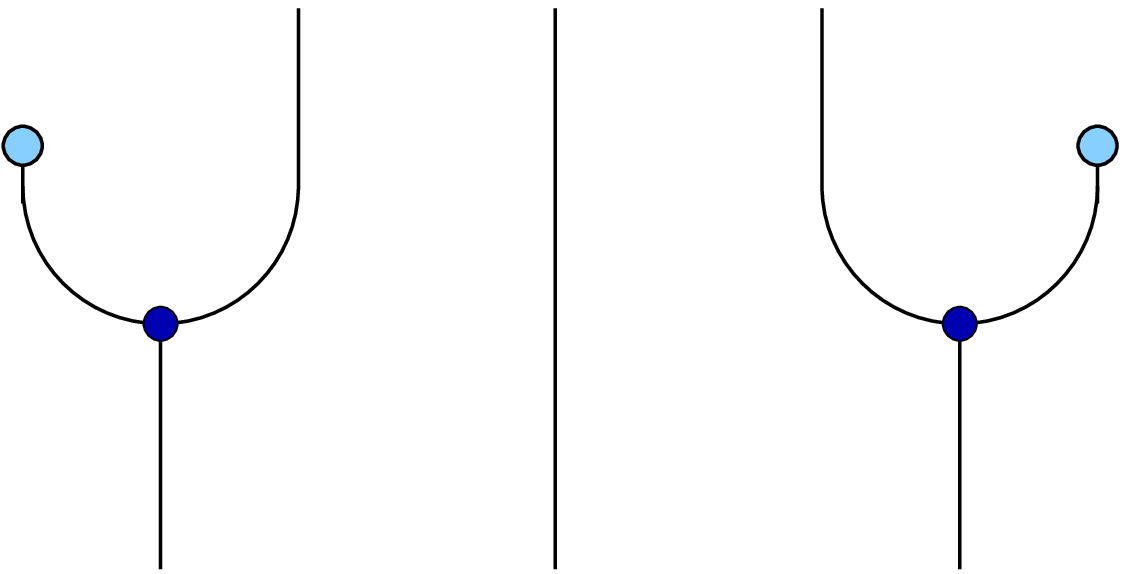}  {3}
\put(12.9,-9.2)  {\sse$A$}
\put(24.1,50.9)  {\sse$A$}
\put(34.5,18)    {\small$=$}
\put(46.1,-9.2)  {\sse$A$}
\put(46.6,50.9)  {\sse$A$}
\put(56.5,18)    {\small$=$}
\put(68.9,50.9)  {\sse$A$}
\put(79.7,-9.2)  {\sse$A$} \end{picture}
\\ \begin{picture}(0,31){}\end{picture} &&& \\ \hline
\end{tabular}
\end{center}

\vskip .8em

A {\em Frobenius \alg\/} in a monoidal \cat\ \C\ is a quintuple
$(A,m,\eta,\Delta,\eps)$ such that $(A,m,\eta)$ is an \alg\ and
$(A,\Delta,\eps)$ is a coalgebra, with the two structures related by
  \be
  (\id_\AA\oti m) \circ (\Delta\oti\id_\AA)
  = \Delta \circ m = (m\oti\id_\AA) \circ (\id_\AA\oti\Delta) \,,
  \ee
which says that the coproduct $\Delta$ is a morphism of $A$-bimodules.
In the special case that \C\ is the \cat\ of complex vector spaces,
this definition of Frobenius \alg\ is equivalent to more familiar ones
as an \alg\ with a nondegenerate invariant bilinear form or as an \alg\
that is isomorphic to its dual as a (left or right) module over itself.

A Frobenius \alg\ in a monoidal \cat\ with left and right dualities
is called {\em symmetric\/} iff the two morphisms
  \be
  [(\eps\cir m)\oti \id_{A^\vee}] \circ (\id_\AA \otimes b_A)
  \qquad{\rm and}\qquad
  [\id_{A^\vee}\oti (\eps\cir m)] \circ (\tilde b_A \oti \id_\AA)
  \ee
in $\Hom(A,A\Vee)$, which by the Frobenius property are in fact isomorphisms,
are equal.
A Frobenius \alg\ in a monoidal \cat\ is called {\em special\/} iff
  \be
  \eps\cir\eta = \beta_\one\, \id_\one \qquad {\rm and}\qquad
  m\cir\Delta = \beta_{\!A}\, \id_\AA
  \ee
for non-zero numbers $\beta_\one$ and $\beta_{\!A}$. For a symmetric special 
Frobenius \alg\ $A$ one has $\beta_\one\beta_{\!A}\eq\dim(A)$; we normalise
$\eps$ and $\Delta$ such that $\beta_\one\eq\dim(A)$ and $\beta_{\!A}\eq1$.
In pictures, the defining properties of a symmetric special Frobenius \alg\ 
look as follows.

\vskip 1.2em

\begin{center}
\begin{tabular}{|llll} \cline{1-2}
\mcll{
\begin{picture}(202,46)(5,28)       \apppicture{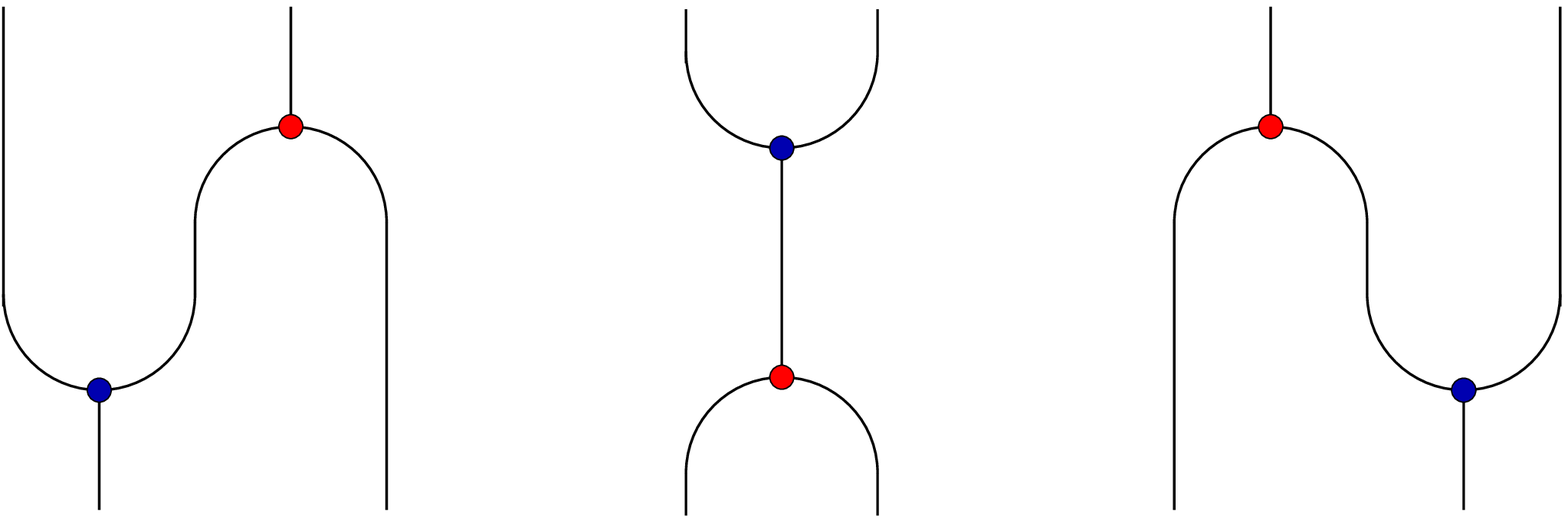}  {22}
\put(19.3,59.9)  {\sse$A$}
\put(28.7,-9.2)  {\sse$A$}
\put(51.2,59.9)  {\sse$A$}
\put(60.6,-9.2)  {\sse$A$}
\put(74.5,25)    {\small$=$}
\put(92.7,-9.2)  {\sse$A$}
\put(93.4,59.9)  {\sse$A$}
\put(113.7,-9.2) {\sse$A$}
\put(114.4,59.9) {\sse$A$}
\put(126.5,25)   {\small$=$}
\put(146.5,-9.2) {\sse$A$}
\put(157.4,59.9) {\sse$A$}
\put(178.5,-9.2) {\sse$A$}
\put(189.4,59.9) {\sse$A$} \end{picture}     }
\\ \mcll{ \begin{picture}(0,38){}\end{picture} } && \\ \hline
\begin{picture}(148,50)(5,28)      \apppicture{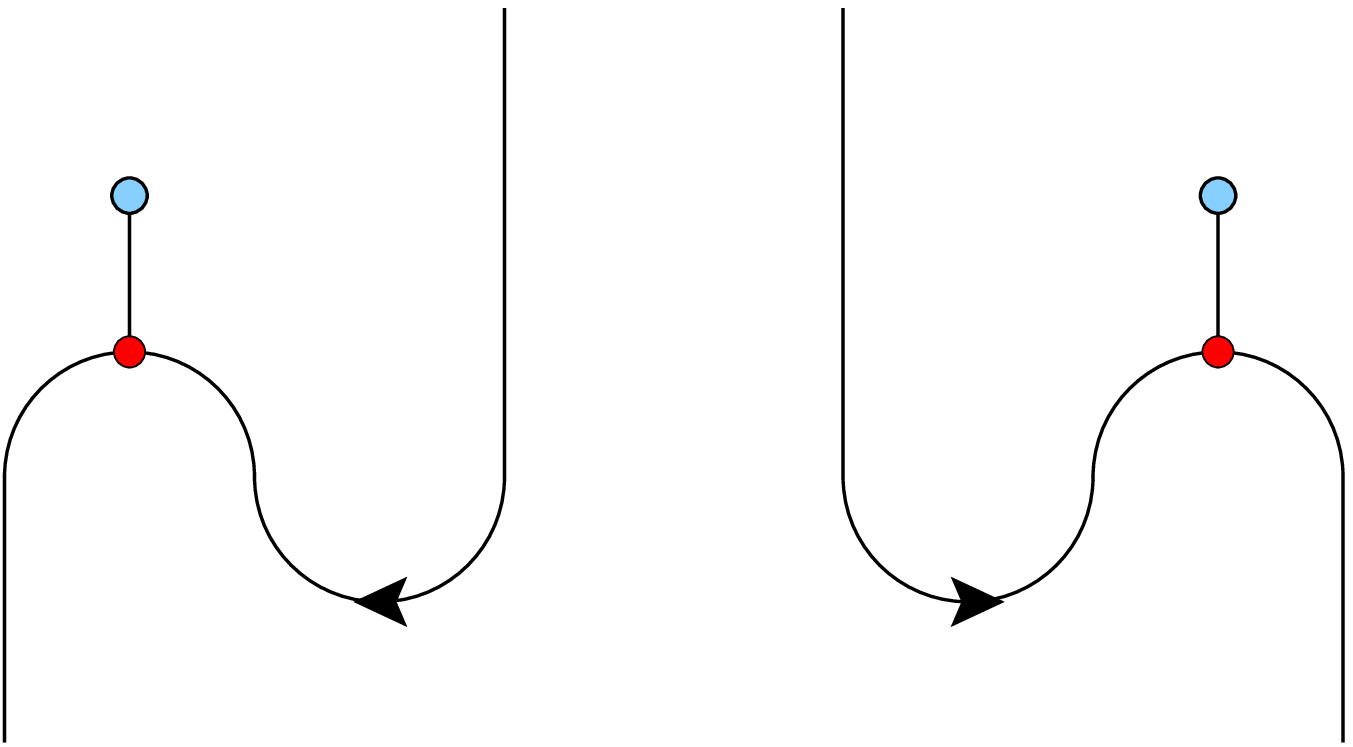}  {22}
\put(18.5,-9.2)  {\sse$A$}
\put(60.7,65.5)  {\sse$A^\vee$}
\put(74.9,29)    {\small$=$}
\put(89.1,65.5)  {\sse$A^\vee$}
\put(130.5,-9.2) {\sse$A$} \end{picture}     &
\multicolumn2{|l}{
\begin{picture}(75,40)(12,20)      \apppicture{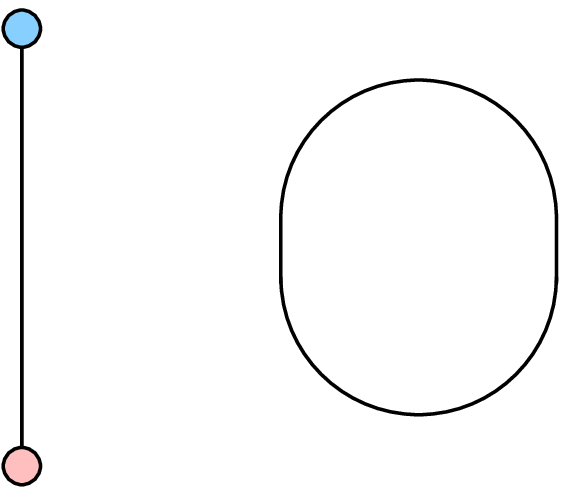}  {31}
\put(39.5,17)    {\small$=$}
\put(76.6,8.6)   {\sse$A$} \end{picture}  }  &
\mclo{
\begin{picture}(88,47)(0,28)       \apppicture{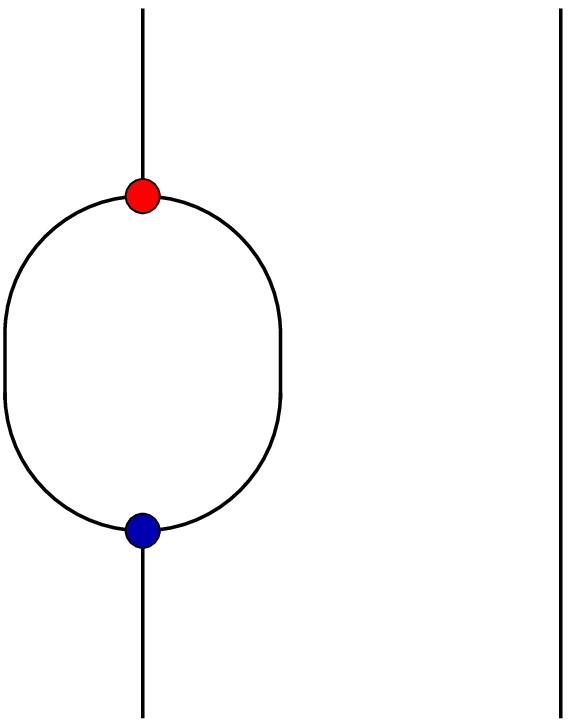}  {22}
\put(30.1,-9.2)  {\sse$A$}
\put(30.6,63.4)  {\sse$A$}
\put(52.5,28)    {\small$=$}
\put(64.9,-9.2)  {\sse$A$}
\put(65.4,63.4)  {\sse$A$} \end{picture}  }
\\ { \begin{picture}(0,36){}\end{picture} }
& \mclll { \begin{picture}(0,36){}\end{picture} } \\ \hline
\end{tabular}
\end{center}

\vskip .8em

A {\em reversion\/} on an algebra $A\eq(A,m,\eta)$ is an
endomorphism $\sigma \iN \Hom(A,A)$ that is an
algebra anti-homomorphism and squares to the twist, i.e.
  \be 
  \sigma \circ \sigma = \theta_\AA
  \,, \qquad
  \sigma \circ m = m \circ c_{A,A} \circ (\sigma\oti\sigma)
  \,, \qquad
  \sigma \circ \eta = \eta \,.
  \labl{eq:sig-def}
If $A$ is also a coalgebra, then in addition
  \be
  \Delta \circ \sigma = (\sigma\oti\sigma) \circ c_{A,A} \circ \Delta
  \qquad {\rm and} \qquad \eps \circ \sigma = \eps
  \labl{eq:sig-coprod}
must hold in order for $\sigma$ to be a reversion. We \cite{fuRs8} call a 
symmetric special Frobenius \alg\ with reversion a {\em Jandl \alg\/}; for a 
Jandl \alg\ $\tilde A \eq (A,m,\eta,\Delta,\eps,\sigma)$ the properties 
\erf{eq:sig-coprod} are actually a consequence of the properties 
\erf{eq:sig-def}.

Given two reversions $\sigma_1$ and $\sigma_2$ on an algebra $A$, the morphism
$\omega\eq\sigma_1^{-1}{\circ}\, \sigma_2$ is an algebra
automorphism of $A$. Conversely, all possible
reversions on an \alg\ $A$ can be obtained by composing a single reversion 
$\sigma_0$ with algebra automorphisms $\omega$ of $A$. More precisely,
for $\sigma$ a reversion on $A$, $\omega \cir\sigma$ is again a reversion 
iff $\omega\cir\sigma\cir\omega\eq\sigma$.

Grapically, the conditions \erf{eq:sig-def} and \erf{eq:sig-coprod} look
as follows.
\vskip 1.2em

\begin{center} \begin{tabular}{|l|l|l|} \hline
\begin{picture}(53,44)(-10,33)
  \put(0,0)     {\scalebox{.38}{\includegraphics{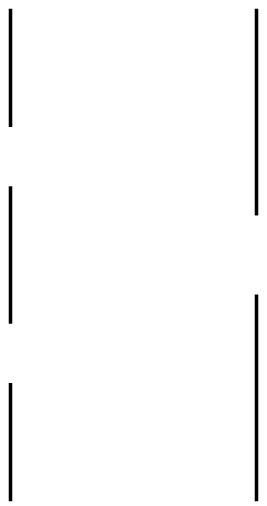}}}
\put(0,-2){
  \put(-.5,-7.5)   {\scriptsize$A$}
  \put(-1,61.5)    {\scriptsize$A$}
  \put(26,-7.5)    {\scriptsize$A$}
  \put(26.5,61.5)  {\scriptsize$A$} }
\setlength{\unitlength}{24810sp}
\put(0,-108){ \put(36,172) {$=$}
    \put(4.5,205.5) {\scriptsize$\sigma$}
    \put(4.5,148.5) {\scriptsize$\sigma$}
    \put(75,174) {\scriptsize$\theta$}}
\end{picture} 
& \begin{picture}(176,44)(-10,37)
  \put(0,0) {\includeourbeautifulpicture{pic4_}2}
\put(-74.5,.2){
  \put(70.5,-9.5)  {\scriptsize$A$}
  \put(83,67.5)    {\scriptsize$A$}
  \put(93.5,-9.5)  {\scriptsize$A$}
  \put(118.5,-9.5) {\scriptsize$A$}
  \put(132,67.5)   {\scriptsize$A$}
  \put(143.5,-9.5) {\scriptsize$A$} }
\put(-187,-.2){
  \put(310,63.5)   {\scriptsize$A$}
  \put(337,63.5)   {\scriptsize$A$} }
\setlength{\unitlength}{24810sp}
\put(-198,-206){ \put(279,276) {$=$}
    \put(226,334) {\scriptsize$\sigma$}
    \put(323,238) {\scriptsize$\sigma$}
    \put(386,238) {\scriptsize$\sigma$}}
\put(-180,-208){ \put(540,294) {$=$}
    \put(509.6,296.5) {\scriptsize$\sigma$}}
\end{picture}
& \begin{picture}(176,44)(-10,37)
  \put(0,0) {\includeourbeautifulpicture{pic4_}3} 
\put(-197,.2){
  \put(194,67.5)   {\scriptsize$A$}
  \put(206.5,-9.5) {\scriptsize$A$}
  \put(216,67.5)   {\scriptsize$A$}
  \put(243,67.5)   {\scriptsize$A$}
  \put(254.5,-9.5) {\scriptsize$A$}
  \put(266,67.5)   {\scriptsize$A$} }
\put(-263,.2){
  \put(385,-4.5)   {\scriptsize$A$}
  \put(413,-4.5)   {\scriptsize$A$} }
\setlength{\unitlength}{24810sp}
\put(-199,0){ \put(278,77) {$=$}
    \put(227.5,35) {\scriptsize$\sigma$}
    \put(324,131) {\scriptsize$\sigma$}
    \put(386,131) {\scriptsize$\sigma$}}
\put(-183,-1){ \put(540,73) {$=$}
    \put(513,79.3) {\scriptsize$\sigma$}}
\end{picture}
\\ { \begin{picture}(0,48){}\end{picture} } &&\\ \hline
\multicolumn3{}{~}
\end{tabular}
\end{center}

\subsection{Modules}\label{Mstuff}

$~$

\noindent
A {\em left module\/} over an \alg\ $A$
is a pair $M\eq(\M,\rho)$ consisting of an object $\M$ of \C\ and a
morphism $\rho\,{\equiv}\,\rho_M^{}\iN\Hom(A\Oti\M,\M)$ that satisfies
  \be
  \rho\circ(m\oti\id_\M) = \rho \circ (\id_\AA\oti\rho)
  \qquad\mbox{and}\qquad \rho\circ(\eta\oti\id_\M) = \id_\M \,.
  \ee
In pictures,

\vskip 1.2em
  
\begin{center}
\begin{tabular}{|l|l|l|}\hline
~~$\rho_M^{} =$
\begin{picture}(39,45)(0,24)       \apppicture{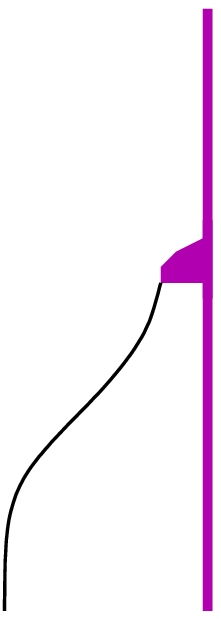}  {7}
\put(3.5,-8.8)   {\sse$A$}
\put(19.5,-9.4)  {\sse$\M$}
\put(20.5,54.4)  {\sse$\M$} \end{picture}    &
\begin{picture}(115,48)(-4,30)       \apppicture{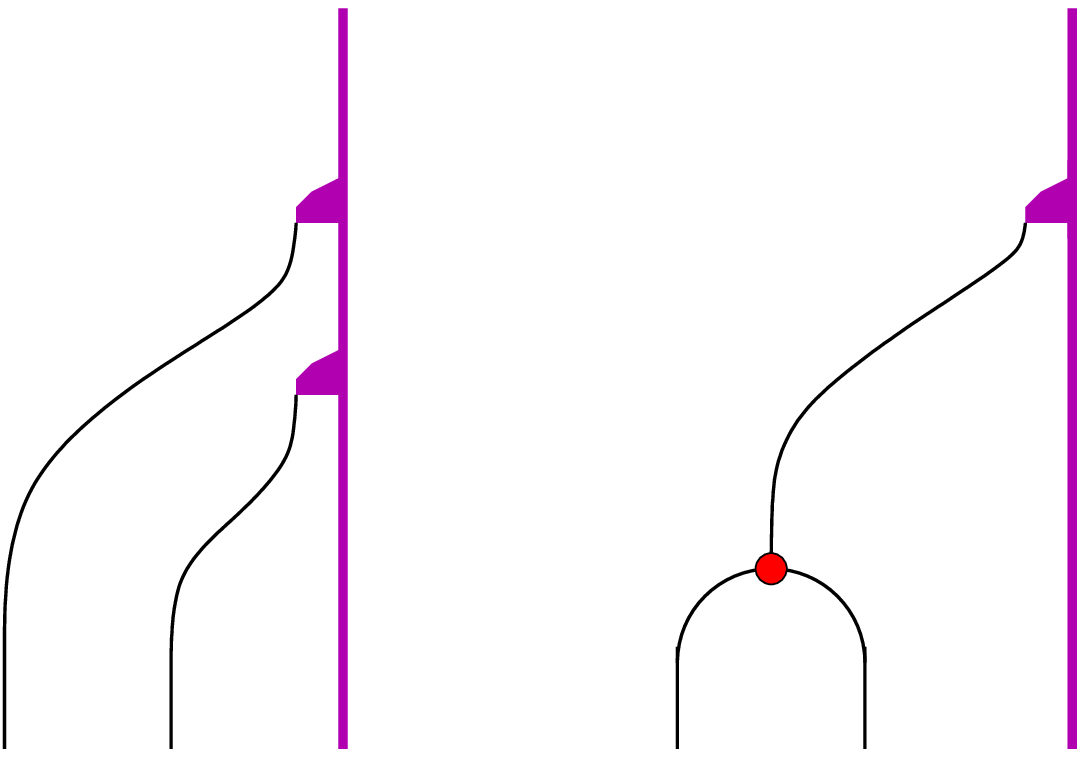}  {6}
\put(2.3,-8.5)   {\sse$A$}
\put(16.3,-8.5)  {\sse$A$}
\put(29.4,-9.8)  {\sse$\M$}
\put(30.5,66.2)  {\sse$\M$}
\put(48.5,28)    {\small$=$}
\put(58.7,-8.5)  {\sse$A$}
\put(74.2,-8.5)  {\sse$A$}
\put(90.9,-9.8)  {\sse$\M$}
\put(92.1,66.2)  {\sse$\M$} \end{picture}     &
\begin{picture}(70,48)(11,30)       \apppicture{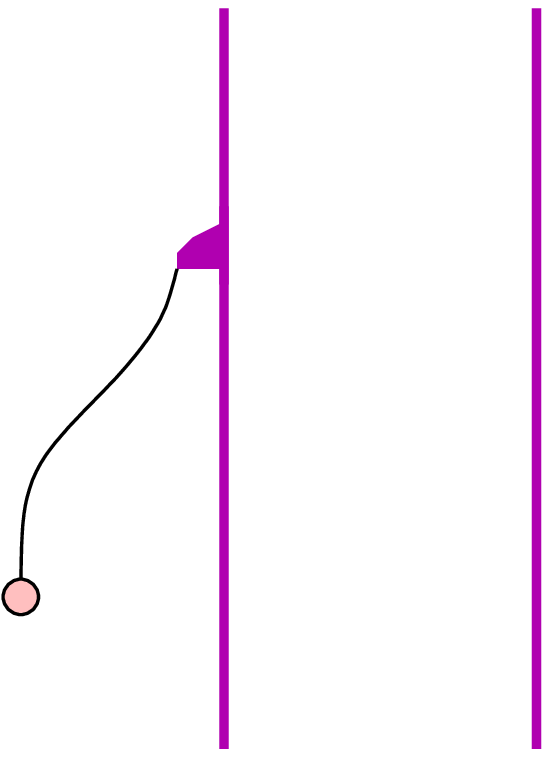}  {22}
\put(35.6,-9.8)  {\sse$\M$}
\put(36.8,66.2)  {\sse$\M$}
\put(49.5,28)    {\small$=$}
\put(61.6,-9.8)  {\sse$\M$}
\put(62.8,66.2)  {\sse$\M$} \end{picture}     
\\ { \begin{picture}(0,38){}\end{picture} } &&\\ \hline
\end{tabular}
\end{center}

\noindent
Analogously one defines a right $A$-module $(\M,\varrho)$. 
An $A$-{\em bimodule\/} is a triple $X \eq(\dot X,\rho,\varrho)$ such that
$(\dot X,\rho)$ is a left $A$-module and $(\dot X,\varrho)$ a right $A$-module
and such that the left and right actions commute,
  \be
  \rho \circ (\id_\AA\oti\varrho) = \varrho \circ (\rho\oti\id_\AA) \,.
  \ee
We denote by
  \begin{eqnarray} &&
  \HomA(N,M) = \{ f\iN \Hom(\dot N,\M) \,|\,
  f \cir\r_N\eq\r_M\cir(\id_\AA\Oti f) \} \qquad \text{and}
  \nonumber \\[.5em]&&
  \HomAA(B,C) = \{f\iN\Hom(\dot B,\dot C) \,|\, 
  f\cir\rho_B \eq \rho_C\cir(\id_A\oti f),~ f\cir\varrho_B
    \eq \varrho_C\cir(f\oti\id_A)\}
  \nonumber\\[.3em]~ \end{eqnarray} 
the subspaces of morphisms in $\C$ that intertwine the left action
of $A$ on the left modules $N$ and $M$, 
and the left and right actions on the bimodules $B$ and $C$, respectively.

Owing to associativity, $A$ is a bimodule over itself.
If \C\ is braided, then for any $A$-bimodule $X$ and any two objects $U,V$
of \C\ one endow the object $U\oti\dot X\oti V$ with several structures of an
$A$-bimodule. In particular there is an $A$-bimodule structure on 
$U\oti A\oti V$, which we denote by $U\oT+A\ot-V$, for which the 
left and right actions of $A$ are defined by
  \be
  \rho := c_{U,A}^{\,-1} \oti \id_V \qquad\mbox{and}\qquad 
  \varrho  :=  \id_U \oti c_{A,V}^{\,-1} \,,
  \ee
respectively.

If $M$ is a module over a Jandl \alg\ $\tilde A$, the presence of the reversion
$\sigma$ allows one to define for any left $A$-module $M$ a left $A$-module 
structure $M^\sigma$ on the object dual to $\M$: 
  \be
  M^\sigma := (\M^\vee , \rho^\sigma)
  \ee
with 
  \bea \begin{picture}(90,95)(0,15)
  \put(22,0)     {\includeourbeautifulpicture M{sigma}}
  \put(-18,42)    {$\rho^\sigma ~:= $}
  \put(23,-8.5)   {\scriptsize$A$}
  \put(25,15.5)   {\scriptsize$\sigma$}
  \put(35,106)    {\scriptsize$\M^\vee{}$}
  \put(61,69.1)   {\scriptsize$\rho$}
  \put(72.3,-9.5) {\scriptsize$\M^\vee{}$}
  \epicture10 \labl{Msigma}
The assignment $M\,{\mapsto}\,M^\sigma$ yields a
($\sigma$-dependent) duality endofunctor on the category left $A$-mo\-dules,
whose square is equivalent to the identity functor.


\sect{The assignment $\X \,{\mapsto}\, C(\X)$}\label{The ass}

The formulation and the proof of modular invariance and of factorisation
require a precise definition of all correlators of a full rational CFT.
In the TFT approach, this is achieved by representing each correlator 
as the invariant of a suitable ribbon graph in a three-manifold,
which specifies the correlator as an element in the relevant space of
conformal blocks. This construction is summarised below; for more details, as 
well as motivations, we refer to the papers \cite{fuRs4,fuRs8,fuRs10} in which 
the construction was developed. To be precise, in our exposition we follow 
closely~\cite{fuRs10}, which introduced a few minor modifications 
(summarised in section IV:3.1) as compared to \cite{fuRs4,fuRs8}.

The basic ingredients of the construction are as follows. As explained in 
section \ref{app:worldsheet}, a world sheet \X{} is a compact two-dimensional 
manifold with a finite number of field insertions, and to each such world sheet
there is associated an extended surface \Xh, a compact oriented surface
with a finite number of marked points.
The CFT correlator $C(\X)$ that corresponds to a world sheet \X{} is defined as
  \be
  C(\X):=Z(\MX,\emptyset,\Xh)1 ~~~ \in \mathcal{H}(\Xh)
  \labl{prescr}
where \MX{}, the {\em connecting manifold\/}, is a cobordism to be defined
in sections \ref{app:worldsheet} and \ref{A2}.

For the rest of this section we select once and for all a modular tensor 
category \C, as well as
\\[.3em]
\nxt a symmetric special Frobenius algebra $A$ in \C, in case we wish to
     describe  a \cft\ on oriented surfaces only,
respectively
\\[.3em]
\nxt a Jandl algebra $\tilde A$ in \C, in case we wish to include also
     unoriented world sheets.
\\[.3em] 
The details of the construction depend on whether \X\ is oriented
or unoriented. Still, in the sequel we treat oriented and unoriented 
world sheets simultaneously. In the parts relevant to both cases
we refer to the algebra $A$, while in the parts applicable to the
unoriented case alone we refer to the algebra $\tilde A$.

\subsection{The world sheet}\label{app:worldsheet}

$~$

\noindent
Oriented and unoriented world sheets are defined as follows.

\dtlDefinition{def:ws}
{\rm(i)}\,~An {\em unoriented world sheet\/} $\X$ is a compact two-dimensional 
topological manifold, also denoted by $\X$ (which may have non-empty boundary 
and may be non-orientable), together with a finite, unordered set of marked 
points and an orientation ${\rm or}(\partial \X)$ of its boundary. 
\\
A marked point is either
\\[.2em]
\nxt a {\em bulk insertion\/}, that is, a tuple 
$\varPhi \eq (i,j,\phi,p,[\gamma],{\rm or}_2(p))$, where $i,j \iN \I$,
$\phi\iN\Hom_{\!\tilde A|\tilde A}(U_i 
     $\linebreak[0]$
\oT+ \tilde A\ot-U_j,
\tilde A)$, $p \iN \X \,{\setminus}\, \partial \X$, $[\gamma]$ is an arc-germ
with $\gamma(0)\eq p$ and ${\rm or}_2(p)$ 
is an orientation of a neighbourhood of $p\iN\X$. 
\\[.2em]
\nxt a {\em boundary insertion\/}, that is, a tuple $\varPsi \eq 
(M,N,U,\psi,p,[\gamma])$, where $M$, $N$ are left $\tilde A$-modules,
$U \iN \objc$, $\psi\iN\Hom_{\!\tilde A}(M\oti U,N)$, $p\iN\partial \X$ and 
$[\gamma]$ is an arc-germ with $\gamma(0)\eq p$. There has to exist a
representative $\gamma$ of $[\gamma]$ that is a subset of $\partial \X$.
\\[.2em]
No two bulk or boundary insertions are allowed to be located at the same point 
of \X. If two boundary insertions $\varPsi_1 \eq (M_1,N_1,U,\psi_1,p_1,
[\gamma_1])$ and $\varPsi_2 \eq (M_2,N_2,B,\psi_2,p_2,[\gamma_2])$ 
are adjacent and $\varPsi_1$ comes ``after'' $\varPsi_2$ 
with respect to $\,{\rm or}(\partial \X)$, i.e.\ pictorially,
  \bea \begin{picture}(100,11)(0,20)
  \put(0,0) {
  \put(0,0)    {\includeournicesmallpicture 87 } 
  \setlength{\unitlength}{0.5pt}
  \put(34,43)  {\scriptsize $M_1$}
  \put(91,28)  {\scriptsize $N_1$}
  \put(147,29) {\scriptsize $M_2$}
  \put(212,42) {\scriptsize $N_2$}
  \put(65,36)  {\scriptsize $\varPsi_1$}
  \put(177,35) {\scriptsize $\varPsi_2$}
  \setlength{\unitlength}{1pt}
  }
  \epicture08 \labl{eq:bnd-ins-labels}
then we require $N_1 \eq M_2$. If there is a connected component of $\partial\X$
without boundary insertions, it gets labelled by a left $\tilde A$-module.
\\[.4em]
{\rm(ii)}~An {\em oriented world sheet\/} $\X$ is an unoriented world sheet $\X$
together with an orientation ${\rm or}_2(\X)$ of $\X$.  For a bulk insertion 
$\varPhi \eq (i,j,\phi,p,[\gamma],{\rm or}_2(p))$, ${\rm or}_2(p)$ is required 
to agree with ${\rm or}_2(\X)$. Also, ${\rm or}_2(\X)$ induces an orientation 
of $\partial \X$ via the inward pointing normal, which is required to agree 
with ${\rm or}(\partial\X)$. 
\ENDD

The machinery of TFT is to be applied to {\em extended surfaces\/}.
An extended surface $E$ is a closed compact two-dimensional manifold with a
finite number of marked points and a choice of Lagrangian subspace
$\lambda\,{\subset}\, H_1(E,\reals)$. A marked point on $E$ is a quadruple
$(p,[\gamma],U,\varepsilon)$ with $p\iN E$, $[\gamma]$ an arc germ centered
at $p$, $U$ an object of \C, and $\varepsilon\iN\{\pm\}$.
The {\em double\/} \Xh\ of a world sheet \X{} is an extended surface; as a
topological surface, \Xh\ is the total space of the orientation bundle over
\X{} modulo an identification of the two points of the fiber over any
boundary point,
  \be
  \Xh := Or(\X){/}{\sim}\quad\mbox{with}\quad (x,\ort)\sim (x,-\ort)
  \quad{\rm for}~~ x\iN\partial\X\,.
  \labl{def:double}
Thus points of \Xh{} are equivalence classes $[x,\ort]$. By construction
\Xh{} is oriented.

The {\em connecting manifold\/} \MX{} of \X{} is defined as a topological
three-manifold by
  \be
  \MX := (\Xh\Times[-1,1]){/}{\sim}\quad\mbox{with}\quad
  ([x,\ort],t)\sim([x,-\ort],-t)\,.
  \labl{def:unorconnmfd}
Points on \MX{} are equivalence classes $[[x,\ort],t]\,{=:}\,[x,\ort,t]$.
There is a natural embedding
  \be
  \imath:\quad \X\rightarrow\MX,\quad~ x\mapsto[x,\pm\ort,0]
  \labl{def:unoremb}
of the world sheet in the connecting manifold, as well as a projection
  \be
  \pi:\quad \MX\rightarrow\X,\quad~ [x,\ort,t]\mapsto x\,.
  \labl{def:unorproj}
We equip \MX{} with the orientation that is induced by the orientation of
\Xh{} together with the standard orientation of the interval. By construction,
$\partial\MX\eq\Xh$, where the orientation of the boundary is induced by the
inward-pointing normal.

The construction of the connecting manifold provides naturally a Lagrangian
subspace of the first homology group with real coefficients of the double:
  \be
  \lambda:={\rm Ker}\,f_*
  \labl{def:Lagr}
with $f{:}~\Xh\eq\partial\MX\To\MX$ the inclusion map and $f_*$ the induced
map $H_1(\Xh,\reals)\To H_1(\MX,\reals)$.
(An alternative definition of Lagrangian subspace is given in lemma 3.5 of
\cite{fffs3}, compare also theorem 5.1.1 of \cite{quin3};
its relation to the present one is explained in remark 3.1 of \cite{fuRs8}.)

\medskip

Field insertions on the world sheet \X{} yield marked points on \Xh. Each
boundary field insertion $\vPsi\eq(M,N,V,\psi,p,[\gamma])$ gives rise to a
single marked point $(\tilde{p},[\tilde{\gamma}],V,+)$ with
$\tilde{p}\eq[p,\pm\ort(p)]$. To define $[\tilde{\gamma}]$, note that any
representative $\gamma$ of the class $[\gamma]$ has a restriction to $\partial\X$,
i.e.\ one has $\gamma([-\delta,\delta])\,{\subset}\,\partial\X$ for a suitable
$\delta\,{>}\,0$. This restriction provides a unique pre-image on \Xh{}, and
then we set $\tilde{\gamma}\,{:=}\,[\gamma,\pm\ort]$ for the restriction of
any representative $\gamma$ of $[\gamma]$. 

Each bulk field insertion $(i,j,\phi,p,[\gamma],\ort)$ gives two marked points
on \Xh, $(\tilde{p}_i,[\tilde{\gamma}_i],U_i,+)$ and
$(\tilde{p}_j,[\tilde{\gamma}_j],U_j,+)$, with $\tilde{p}_i\eq [p,\ort(p)]$,
$\tilde{p}_j\eq [p,-\ort(p)]$, and $\tilde{\gamma}_i\eq[\gamma,\ort(\gamma)]$,
$\tilde{\gamma}_j\eq[\gamma,-\ort
     $\linebreak[0]$
(\gamma)]$.

\subsection{The ribbon graph}\label{A2}

$~$

\noindent
Next, we provide \MX{} with a ribbon graph $R$, turning it into a cobordism 
$\MX[R]$ of extended surfaces. The construction of the ribbon graph involves 
a number of arbitrary choices, in particular that of a triangulation $T$; 
denote the resulting ribbon graph by $R_T$. It turns out that the linear map 
$Z(\MX[R_T]){:}\ \complex \To \mathcal{H}(\hat \X)$ is independent of all 
these choices. In steps (i)\,--\,(ix) below, we will always think of $\X$ 
as embedded in $\MX$ via $\iota_\X$.
 
\medskip\noindent
(i) Choose a triangulation $T$ of $\X$ that has two- or three-valent 
vertices and faces with an arbitrary number of edges\,\footnote{~%
  Thus $T$ is actually the dual of what is commonly referred to 
  as a triangulation.}
-- \underline{Choice \#1}. 
\\[.2em]
The choice of $T$ is subject to the following conditions.
\\[.2em]
\nxt The boundary $\partial\X$ is covered by edges of $T$.
\\[.2em]
\nxt Two-valent vertices in $T$ occur precisely at the marked points of \X.
\\[.2em]
\nxt For a bulk insertion at $p$ with arc germ $[\gamma]$, there has
to be a representative $\gamma$ of $[\gamma]$ such that $\gamma$ is covered
by edges of $T$ (see e.g.\ figure (IV:4.26))

\medskip\noindent
(ii) At each three-valent vertex $v$ of $T$ in the interior of $\X$ choose an 
orientation of a small neighbourhood of $v$. If $\X$ is an oriented world sheet,
choose ${\rm or}_2(\X)$ at each vertex; if $\X$ is an unoriented world sheet,
this is a genuine choice -- \underline{Choice \#2}.
 
\medskip\noindent
(iii) On each three-valent vertex $v$ in the interior
of $\X$ place the following fragment of ribbon graph with
three outgoing $A$-ribbons,
  \bea  \begin{picture}(115,55)(0,42)
  \put(0,0)         {\includeourbeautifulpicture 14 } 
  \put(26,77)       {\scriptsize$A$}
  \put(43.5,25)     {\scriptsize$A$}
  \put(75,76.5)     {\scriptsize$A$}
  \put(91,10)       { \put(0,0) {\tiny $1$}
  \put(20,22) {\tiny $2$} }  
  \epicture18 \labl{fig:vertex-graph}
such that the orientation around $v$ as chosen in (ii) agrees with the one 
indicated in the figure. There are three possibilities to do this (rotating 
the graph) -- \underline{Choice \#3}.

\medskip\noindent
(iv) On each edge of $T$ which does not lie on $\partial\X$, place one of the 
following two fragments of ribbon graph such that the local orientation around 
the two vertices connected by the edge agrees with the one indicated in the 
figure:
  \begin{eqnarray}  \begin{picture}(385,66)(17,0)
  \put(0,0)         {\includeourbeautifulpicture 16 }
  \put(33,21)       {\scriptsize$A$}
  \put(146,21)      {\scriptsize$A$}
  \put(274,21)      {\scriptsize$\tilde A$}
  \put(389,21)      {\scriptsize$\tilde A$}
  \put(-4,27)       { \put(0,0) {\tiny $1$} \put(20,22) {\tiny $2$} }  
  \put(164,27)      { \put(0,0) {\tiny $1$} \put(20,22) {\tiny $2$} }  
  \put(245,27)      { \put(20,0){\tiny $1$} \put(0,22)  {\tiny $2$} }  
  \put(406,27)      { \put(0,0) {\tiny $1$} \put(20,22) {\tiny $2$} }  
  \end{picture} \nonumber \\ \label{edge-graph}
  \end{eqnarray}
In each case this is possible in two ways -- \underline{Choice \#4}.
Note that the situation on the \rhs\ of \erf{edge-graph} can occur only
for unoriented world sheets. Accordingly, the ribbon graph has been
labelled by $\tilde A$ and involves explicitly the reversion (contained
in the definition of the vertical bar in the figure, see figure (II:3.4);
the dashed-line notation refers to the ``black'' 
side of a ribbon, see figure (II:3.3)). 

\medskip\noindent
(v) The edges on the boundary $\partial\X$ get labelled by left $A$-modules
as follows. If an edge $e$ of $T$ lies on a connected component of $\partial\X$
without field insertion, it gets labelled by the $A$-module assigned to that 
boundary component (recall definition \ref{def:ws}\,(i)). Otherwise, $e$ lies 
between two (not necessarily distinct) boundary insertions. In this case it 
gets labelled 
by $N_1\eq M_2$, using the convention in figure \erf{eq:bnd-ins-labels}.

\medskip\noindent
(vi) On each edge on the boundary $\partial\X$ labelled by $M$ place a ribbon 
also labelled by $M$. A small neighbourhood of $\partial \X$ can be oriented
using the orientation of $\partial \X$ and the inward pointing normal. The
orientations of the ribbon core and surface have to be opposite 
to those of $\partial \X$ and $\X$, respectively.

\medskip\noindent
(vii) On each three-valent vertex on the boundary $\partial\X$ 
place ribbon graph fragment
  \bea  \begin{picture}(140,64)(0,36)
  \put(0,-5)        {\includeourbeautifulpicture 15  
  \put(-2,69)       { \put(0,0) {\tiny $1$}  \put(21.7,22) {\tiny $2$} }  
  \put(20,35.5)     {\scriptsize$M$}
  \put(81,78)       {\scriptsize$A$}
  \put(108,35.5)    {\scriptsize$M$}
  }
  \epicture24 \labl{boundary-graph}
such that bulk and boundary orientation indicated in the figure
agree with those of $\X$. Here $M$ is the label of the two edges lying
on the boundary $\partial\X$, as assigned in (v) (they
have the same label by construction).
Shown in \erf{boundary-graph} is a horizontal section of the 
connecting manifold as displayed in figure (IV:3.12).
Correspondingly the lower boundary in \erf{boundary-graph} 
is that of $\MX$ while the ribbons $M$ are placed on the 
boundary of $\X$ as embedded in $\MX$. The arrow on the boundary in 
\erf{boundary-graph} indicates the orientation of $\partial\X$ (transported 
to $\partial\MX$ along the preferred intervals). 

\medskip\noindent
(viii) Let $v \in \partial \X$ be a two-valent vertex of $T$  and
let $\varPsi = (M,N,V,\psi,p,[\gamma])$ be the corresponding boundary
insertion. At $v$ place one of the two ribbon graph fragments
  \bea \begin{picture}(390,59)(0,44)
  \put(0,0)         {\Includeourbeautifulpicture 20a }
  \put(200,0)       {\Includeourbeautifulpicture 20b }
  \put(37,63.5)     {\scriptsize$ M $}
  \put(125,63.5)    {\scriptsize$ N $}
  \put(90,54.5)     {\scriptsize$ \psi $}
  \put(66.4,21)     {\scriptsize$ V $}
  \put(244,63.5)    {\scriptsize$ M $}
  \put(323,63.5)    {\scriptsize$ N $}
  \put(290,54.5)    {\scriptsize$ \psi $}
  \put(267,38)      {\scriptsize$ V $}
  \put(17,96.5)     {\tiny $2$}
  \put(-5,74.3)     {\tiny $1$}
  \put(217,96.5)    {\tiny $2$}
  \put(195,74.3)    {\tiny $1$}
  \epicture25 \labl{fig:bdins}
depending on the relative orientation of the arc-germ $[\gamma]$ and the 
boundary $\partial X$, and such that bulk and boundary orientation 
indicated in the figure agree with those of $\X$.

\medskip\noindent
(ix) Let $v$ be a two-valent vertex of $T$ in the interior of $\X$ and
let $\varPhi \eq (i,j,\phi,p,[\gamma],{\rm or}_2(p))$ be the corresponding
bulk insertion. Place the following fragment of ribbon graph at $v$:
  \bea \begin{picture}(400,206)(0,40)
  \put(30,0)       {\includeournicelargepicture 09 
    \setlength{\unitlength}{.28pt} 
    \put(246,570)       {\scriptsize $\eta$}
    \put(282,495)       {\scriptsize $\Delta$}
    \setlength{\unitlength}{1pt}
    }
  \put(250,227)    {$\partial\MX$}
  \put(243,14)     {$\partial\MX$}
  \put(190.2,234)  {\tiny $1$}
  \put(184,221)    {\tiny $2$}
  \put(216,192)    {\scriptsize$ U_i$}
  \put(210,40)     {\scriptsize$ U_j$}
  \put(83,111.5)   {\scriptsize$A$}
  \put(153,111.5)  {\scriptsize$A$}
  \put(280,111.5)  {\scriptsize$A$}
  \put(195,120)    {$\phi$}
  \put(230,168.1)  {$t\eq0$-plane}
  \put(323,65)     {\tiny $1$}    
  \put(332.5,78.2) {\tiny $2$}   
  \put(289.3,29.6) {\tiny $1$}
  \put(275.1,50)   {\tiny $3$}
  \put(267,45.7)   {\tiny $2$}
  \put(175,17.5)   {\tiny $2$}
  \put(197.8,4)    {\tiny $1$} 
  \epicture23 \labl{fig:bulkins} 
This is done in such a way that the 2-orientation in the $t{=}0$-plane and 
the orientations of the arcs on $\partial\MX$ are as indicated in the figure.
(In the picture, the $A$-ribbons are viewed white side up).

\medskip

Using the various properties of the \cat\ $\C$ and of the algebra $A$
(respectively, $\tilde A$), one
shows that the invariants of ribbon graphs arising from different choices in the
prescription are the same. For choices $\#2$\,--\,$\#4$ this is explained in
detail in \cite{fuRs4,fuRs8}. 
Specifically, for choice $\#2$ see sections II:3.1, IV:3.2 and IV:3.3, for 
choice $\#3$ see section I:5.1, and for choice $\#4$ sections I:5.1 and II:3.1. 
Regarding independence of the triangulation (choice $\#1$), the corresponding 
argument has been indicated in \cite{fuRs4} (see (I:5.11) and (I:5.12)); the 
details are supplied in this paper in propositions \ref{prop:triangulation} and 
\ref{prop:triangulation-unor} as part of the proof of modular invariance.

This completes the prescription for the ribbon graph in \MX. As already stated,
the CFT correlator $C(\X)$ (standing for $C_A$ or $C_{\tilde A}$, depending on 
whether we deal with the oriented or unoriented case) corresponding to the 
world sheet \X{} (and algebra $A$, respectively $\tilde A$, in \C),
is $C(\X) \eq Z(\MX,\emptyset,\Xh)1$.

\subsection{Equivalent labellings of unoriented world sheets}\label{app:equiv-lab}

$~$

\noindent
Not all labellings of boundary conditions, boundary fields and bulk fields
as described in definition \ref{def:ws} lead to distinct correlators
(i.e.\ to different vectors in $\mathcal{H}(\Xh)$). For example, if
two $A$-modules $M$ and $M'$ are isomorphic, then labelling a boundary
condition by $M$ or $M'$ leads to the same correlator.

For unoriented world sheets, there is a less trivial relation between labellings
of fields and boundary conditions which involves the change of the local 
orientation ${\rm or}_1$ of the boundary and the local orientation ${\rm or}_2$ 
around bulk insertions.  
To describe these equivalent labellings we need some additional ingredients. 

For $\tilde A$ a Jandl
algebra and $M$ a left $\tilde A$-module, $M^{\sigma}$ will denote the conjugate
left $\tilde A$-module as defined in (II:2.26). We will need, for any object
$V$ of $\C$ and any two left $\tilde A$-modules $M$ and $N$, a linear map
$s{:}~\Hom_{\tilde A}(M\oti V,N)\To\Hom_{\tilde A}(N^{\sigma}\oti V,M^{\sigma})$;
$s$ is given by
  \bea \begin{picture}(150,89)(0,50)
  \put(56,0)     {\includeourbeautifulpicture 07 }
  \put(0,56)     {$ s(\psi) \;:= $}
  \put(87,130)   {\scriptsize$ \dot{M}^\vee $}
  \put(92.5,-9)  {\scriptsize$ \dot{N}^\vee $}
  \put(122.8,-7) {\scriptsize$ V $}
  \put(93.4,72.5)  {\scriptsize$ \psi $}
  \epicture34 \labl{fig:sdef}
As has been shown in lemma IV:3.2, this is indeed a map of $\tilde A$-module 
morphisms. We also need, for any two objects $U$ and $V$ of \C, the vector space
isomorphism
  \be
  \omega^{\tilde A}_{UV}:\quad 
  \Hom_{\tilde A\tilde A}(U\oti^+\tilde A\oti^-V,\tilde A)
  \rightarrow 
  \Hom_{\tilde A\tilde A}(V\oti^+\tilde A\oti^-U,\tilde A)
  \ee
defined as
  \bea \begin{picture}(170,90)(0,43)
  \put(75,0)     {\includeourbeautifulpicture 08 }
  \put(0,60)     {$ \omega^{\tilde A}_{UV}(\phi) \ := $}
  \put(99.3,46.4) {\scriptsize$ \sigma^{\!-\!1} $}
  \put(102.4,87.9){\scriptsize$ \sigma $}
  \put(101,125.7){\scriptsize$ \tilde A $}
  \put(87,-8)    {\scriptsize$ V $}
  \put(100.5,-8) {\scriptsize$ \tilde A $}
  \put(115,-8)   {\scriptsize$ U $}
  \put(102,68.3) {\scriptsize$ \phi $}
  \epicture29 \labl{fig:omdef}
It is shown in lemma IV:3.5 that $\omega^{\tilde A}_{UV}$ is indeed a map of 
bimodule morphisms and that
$\omega^{\tilde A}_{VU}\cir\omega^{\tilde A}_{UV}\eq\id$.

\medskip

Let now $\X$ be an unoriented world sheet and $B$ be a connected component
of $\partial \X$. Passing along $B$ in the direction opposite to 
${\rm or}(\partial\X)$ we obtain a list of labels for that boundary component
  \be
  (M_1,...\,,M_{n'};\vPsi_1,...\,,\vPsi_n;\oro) \,,
  \labl{def:orbc}
where $M_1,...\,,M_{n'}$ are left $\tilde A$-modules,
$\vPsi_k\eq(M_k,M_{k+1},V_k,\psi_k,p_k,[\gamma_k])$ are boundary fields
and $\oro$ is the orientation of $B$ induced by that of $\partial\X$.
The module $M_k$ labels the
segment of the boundary between the insertion points $p_k$
and $p_{k+1}$. If $n\eq0$, then $n'\eq1$; otherwise $n'\eq n$. 
Denote by $\X'$ the world sheet obtained by labelling the boundary
component $B$ not by \erf{def:orbc} but by a different tuple
$(M'_1,...\,,M'_{n'};\vPsi'_1,...\,,\vPsi'_n;\oro')$. The following
conditions guarantee that $C_{\tilde A}(\X) \eq C_{\tilde A}(\X')$.
\Itemize
\item[(i)~]
$\oro'\eq\oro$, and for every $k\iN\{1,2,...\,,n\}$,
$V'_k\eq V_k$, $[\gamma'_k]\eq[\gamma_k]$, and there exist isomorphisms
$\varphi_k\iN\Hom_{\tilde A}(M'_k,M_k)$ such that

\end{itemize}
  \be
  \psi_k = \varphi_{k+1}\circ\psi'_k\circ(\varphi^{-1}_k\oti \id_{V_k}) \,.
  \ee
\Itemize
\item[(ii)]
$\oro'\eq{-}\oro$, and for every $k\iN\{1,2,...\,,n\}$,
$V'_k\eq V_{n-k}$, $[\gamma'_k]\eq[\gamma_{n-k}]$, and there exist isomorphisms
$\varphi_k\iN\Hom_{\tilde A}(M'_k,M^{\sigma}_{n'-k+1})$ such that
\end{itemize}
  \be
  s(\psi_{n-k}) = \left\{ \bearll
  \varphi_{k+1}\cir\psi_k'\cir(\varphi_k^{-1}\oti \id_{V_k'})
  & \mbox{ if }~\oro'\eq {\rm or}(\gamma_k')\,, \\[4pt]
  \varphi_{k+1}\cir\psi_k'\cir(\varphi_k^{-1}\oti \theta_{V_k'})
  & \mbox{ if }~\oro'\eq{-}{\rm or}(\gamma_k')\,.  \eear \right.
  \ee

\medskip

Similarly, let $\vPhi \eq (i,j,\phi,p,[\gamma],\ort)$ be a bulk insertion
of $\X$ and let $\X'$ the world sheet obtained by replacing the bulk
field  $\vPhi$ with
  \be
  \vPhi' = (j,i,\omega^{\tilde A}_{U_i U_j}(\phi),p,[\gamma],-\ort(p)) \,,
  \labl{def:unorbf}
so that in particular the local orientation around the insertion point $p$ is 
reversed. In this case one also finds $C_{\tilde A}(\X)\eq C_{\tilde A}(\X')$.

\medskip

The equivalences for the bulk and boundary labellings outlined above
are established in sections IV:3.2 and IV:3.3.


\sect{Two-point correlators} \label{labelappB}

In this appendix the two-point correlators on the disk and on the sphere are 
calculated, and it is shown that they are non-degenerate. In both cases the 
world sheet will be endowed with an orientation. 
When considering a disk or sphere without orientation, i.e.\ correlators
$C_{\!\tilde A}$, then of the global orientation we retain just 
the local orientations around the field insertions, as well as 
for the boundary circle in the case of the disk.

\subsection{The two-point correlator on the disk} \label{sec:2ptD}

$~$

\noindent
Here we present some properties of the correlator of two boundary fields
on the disk, which enter the proof of boundary factorisation.
The relevant world sheet is
  \be
  \X = D \equiv D(M_l,M_r,k,\psi^+_{\alpha},\psi^-_{\beta}) \,,
  \ee
as displayed in figure~\erp 95, with $\psi^+_{\alpha}$ and $\psi^-_{\beta}$ 
denoting elements of $\HomA(M_l\oti U_k,M_r)$ and $\HomA(M_r\oti U_{\bar{k}},
M_l)$, respectively. Our aim is to obtain a ribbon graph that determines the 
structure constants $\Cbnd_{M_l,M_r,k,\psi^+_{\alpha},\psi^-_{\beta}}$ that 
are defined by formula \erf{Dscdef}. This is done by composing both sides 
of \erf{Dscdef} with the cobordism 
    $\overline B^+_{k\bar k}{:}\ \mathcal{H}(\widehat D) \To \emptyset$ 
that is defined as being dual to $B^+_{k\bar k}$ 
in the sense that $Z(\overline B^+_{k\bar k}) \cir Z(B^+_{k\bar k})1 \eq 1$.

The connecting manifold ${\rm M}_D$ is a three-ball, with boundary
$\partial{\rm M}_D\,{\cong}\ \partial B^+_{k\bar k}$. It contains an
annular ribbon running along $\imath(\partial D)$, which in the region
$x\,{<}\,0$ is labelled by the $A$-module $M_l$, and in the region $x\,{>}\,0$,
by $M_r$, the two parts being separated by coupons centered at
$p^+\eq(0,1,0)$ and at $p^-\eq{-}p^+$ and labelled by $\psi_{\alpha}^+$ and
$\psi^-_{\beta}$, respectively. In addition, to the coupon centered at $p^+$
there is also attached a ribbon labelled by $U_k$, which ends on the marked
arc $\gamma^+$ on $\partial{\rm M}_D$ labelled by $(U_k,+)$. Similarly, to the
other coupon there is attached a ribbon labelled by $U_{\bar{k}}$, ending on the
other marked arc $\gamma^-$ with label $(U_{\bar{k}},+)$. Orientations and
core-orientations of all ribbons are determined by the general prescription.
We choose a minimal triangulation of $\imath(D)$, consisting of three
edges ending on $\imath(\partial D)$ at the points $(-\cos \frac{2\pi a}{3},
\sin \frac{2\pi a}{3},0)$ with $a\eq0,1,2$, and one vertex at $(0,0,0)$. 
This triangulation is covered by a combination of $A$-ribbons and coupons 
as explained in appendix \ref{A2}. The `white sides' of all ribbons 
and coupons face the negative $z$-direction. Thus ${\rm M}_D$ looks as follows:
  \bea \begin{picture}(180,185)(0,22)
  \put(0,0)       {\Includeourbeautifulpicture 100 
    \setlength{\unitlength}{.38pt}
    \put(543,227.5)     {\scriptsize $x$}
    \put(233,547)       {\scriptsize $y$}
    \put(44,-14)        {\scriptsize $z$}
    \setlength{\unitlength}{1pt}
    }
  \put(152,90)    {\scriptsize $M_r$}
  \put(42,90)     {\scriptsize $M_l$}
  \put(67,37)     {\scriptsize \begin{turn}{90} $\psi^-_\beta$\end{turn}}
  \put(101,145)   {\scriptsize \begin{turn}{270}$\psi^+_\alpha$\end{turn}}
  \put(92,182)    {\scriptsize $(U_k,+)$}
  \put(77,-7)     {\scriptsize $(U_{\bar k},+)$}
  \epicture14 \labl{fig:MD}
The various properties of $A$ and its modules allow for a complete removal
of the $A$-ribbons. This can be done as follows. First, the $A$-ribbon pointing
down is moved upwards along $M_r$, and when approaching the next $A$-ribbon,
the representation property is invoked. Coassociativity and specialness then
remove that ribbon completely. The remaining $A$-ribbon is removed by first
using the intertwining property of $\psi^+_{\alpha}$, then the representation
property, and finally specialness and once more the representation unit property.
Thus we arrive at the graph in (\ref{fig:MD}), but with all $A$-ribbons removed;
gluing $\overline B^+_{k\bar{k}}$ to this three-manifold then gives the graph
(turned upside down with respect to \erf{fig:MD})
  \bea \begin{picture}(280,81)(0,50)
  \put(0,0)        {\INcludeourbeautifulpicture 101a }
  \put(170,0)      {\INcludeourbeautifulpicture 101b }
  \put(179,23)     {\scriptsize $k$}
  \put(255,65)     {\scriptsize $\bar{k}$}
  \put(160,73)     {\scriptsize $M_l$}
  \put(229,73)     {\scriptsize $M_r$}
  \put(199,63)     {\tiny \begin{turn}{270}$\psi_{\alpha}^+$ \end{turn}}
  \put(193,108)    {\tiny \begin{turn}{90}$\psi_{\beta}^-$\end{turn}}
  \put(73,29)      {\scriptsize $k$}
  \put(-5,65)      {\scriptsize $\bar{k}$}
  \put(84.5,73)    {\scriptsize $M_l$}
  \put(14.5,73)    {\scriptsize $M_r$}
  \put(50,55)      {\tiny \begin{turn}{90}$\psi_{\alpha}^+$ \end{turn}}
  \put(53,118)     {\tiny \begin{turn}{270}$\psi_{\beta}^-$\end{turn}}
  \put(134,63)     {$ = $}
  \put(96,115)     {\Stnbox}
  \put(266,115)    {\Stnbox}
  \epicture29 \labl{fig:lhsD}
    (Recall from footnote \ref{Stnfootnote} that the symbol $\Stn$ indicates
    the normalised invariant of a ribbon graph in $S^3$.)
On the other hand, gluing 
$\overline B^+_{k\bar k}$ to the right hand side of (\ref{Dscdef}) we simply get
  \bea \begin{picture}(285,45)(0,18)
  \put(90,0)       {\Includeourbeautifulpicture 102 }
  \put(0,23)       {$ \Cbnd_{M_l,M_r,k,\psi^+_{\alpha},\psi^-_{\beta}} $}
  \put(168,23)     {$ =\; \Cbnd_{M_l,M_r,k,\psi^+_{\alpha},\psi^-_{\beta}} $}
  \put(78,20)      {\scriptsize $U_{\bar{k}}$}
  \put(117,20)     {\scriptsize $U_k$}
  \put(120,51)     {$\Stnbox$}
  \epicture03 \labl{fig:rhsD}

To proceed, it is convenient to rewrite the result in terms of elements of
the morphism spaces $\Hom^{(P)}(U_{\bar k},M_r^{\vee}\Oti\,M_l)$ and 
$\Hom^{(P)}(M_r^{\vee}\Oti\,M_l,U_{\bar k})$, as defined in section 
\ref{sec:alg-id}. The isomorphisms 
\erf{fig:modisom1} and \erf{fig:fdef} allow for an interpretation of the 
graph~\erf{fig:lhsD} as $\Tr{(e_{\alpha}\cir f_{\beta})}$, or graphically,
  \bea \begin{picture}(60,66)(0,40)
  \put(0,0)       {\Includeourbeautifulpicture 103 
    \setlength{\unitlength}{.38pt}
    \put(-8,133)       {\scriptsize $M_r^\vee$}
    \put(58,135)       {\scriptsize $M_l$}
    \setlength{\unitlength}{1pt}
    }
  \put(11.5,77.5) {\footnotesize $e_{\alpha}$}
  \put(11,30)     {\footnotesize $f_{\beta}$}
  \put(45,53)     {\scriptsize $U_{\bar{k}}$}
  \epicture17 \labl{fig:lhsD2}
The morphisms $e_\alpha$ and $f_\beta$ are related to $\psi^+_\alpha$ and
$\psi^-_\beta$ in the manner described in section~\ref{sec:alg-id}.
					    
Whenever the spaces $\Hom(U,V\oti W)$ and $\Hom(V\oti W,U)$ are non-zero
for a simple object $U$, then the bilinear form $\Lambda$ on $\Hom(U,V\oti W)
\,{\times}\, \Hom(V\oti W,U)$ defined by
  \bea \begin{picture}(135,77)(0,35)
  \put(0,0)       {\INcludeourbeautifulpicture 104a }
  \put(130,0)     {\INcludeourbeautifulpicture 104b }
  \put(8,69.9)    {\footnotesize $\varphi$}
  \put(7,34)      {\footnotesize $\psi$}
  \put(8,110)     {\scriptsize $U$}
  \put(7.5,-10)   {\scriptsize $U$}
  \put(127,-10)   {\scriptsize $U$}
  \put(127,110)   {\scriptsize $U$}
  \put(-2,50)     {\scriptsize $V$}
  \put(18,50)     {\scriptsize $W$}
  \put(53,50)     {$ =:\; \Lambda(\varphi,\psi) $}
  \epicture18 \labl{fig:linform}
is non-degenerate (see e.g.\ \cite{TUra}). We apply this result to the
case $U\eq U_{\bar{k}}$, $V\eq\dot M_r^{\vee}$ and $W\eq\dot M_l$.
Then the restriction of $\Lambda$ to the subspace $\Hom^{(P)}(U_{\bar{k}},
M_r^{\vee} \Oti\,M_l)\Times\Hom^{(P)}(M_r^{\vee}\Oti 
     $\linebreak[0]$
M_l,U_{\bar{k}})$
is still non-degenerate.
To see this, we note that for any object $U$ and $A$-modules $M$ and $N$ we
have $P_{MN}\cir\varphi\eq\varphi$ for all $\varphi\iN\Hom^{(P)}(M\oti N,U)$,
and $\psi\cir P_{MN}\eq\psi$ for all $\psi\iN\Hom^{(P)}(U,M\oti N)$, and that
the maps from $\Hom(M\oti N, U)$ to $\Hom^{(P)}(M\oti N,U)$ and from 
$\Hom(U,M\oti N)$ to $\Hom^{(P)}(U,M\oti N)$ that are obtained by composition 
with $P_{MN}$ from the left and right, respectively, are surjective. As a 
consequence, for any two morphisms $\varphi^-\iN\Hom^{(P)}(U_{\bar{k}},
M_r^{\vee} \Oti\, M_l)$ and $\varphi^+\iN\Hom(M_r^{\vee}\oti M_l,U_{\bar{k}})$, 
we have the equalities $\varphi^+\cir\varphi^- \eq \varphi^+\cir
(P_{M_r^{\vee}M_l}\cir\varphi^-)\eq \tilde\varphi^+\cir\varphi^-$ with
$\tilde\varphi^+\,{:=}\,\varphi^+\cir P_{M_r^{\vee}M_l}$. In pictures,
  \bea \begin{picture}(210,85)(0,35)
  \put(0,0)        {\INcludeourbeautifulpicture 105a }
  \put(85,0)       {\INcludeourbeautifulpicture 105b }
  \put(180,0)      {\INcludeourbeautifulpicture 105a }
  \put(187,79.5)   {\scriptsize $\tilde{\varphi}^+$}
  \put(187,33)     {\scriptsize ${\varphi^-}$}
  \put(100,93)     {\scriptsize ${\varphi^+}$}
  \put(100,20)     {\scriptsize ${\varphi^-}$}
  \put(7,80)       {\scriptsize ${\varphi^+}$}
  \put(7,33)       {\scriptsize ${\varphi^-}$}
  \put(7,119)      {\scriptsize $U_{\bar{k}}$}
  \put(7,-11)      {\scriptsize $U_{\bar{k}}$}
  \put(100,119)    {\scriptsize $U_{\bar{k}}$}
  \put(100,-11)    {\scriptsize $U_{\bar{k}}$}
  \put(187,119)    {\scriptsize $U_{\bar{k}}$}
  \put(187,-11)    {\scriptsize $U_{\bar{k}}$}
  \put(172,52)     {\scriptsize $M_r^\vee$}
  \put(198,52)     {\scriptsize $M_l$}
  \put(68,52)      {\scriptsize $M_r^\vee$}
  \put(125,52)     {\scriptsize $M_l$}
  \put(-9,52)      {\scriptsize $M_r^\vee$}
  \put(17,52)      {\scriptsize $M_l$}
  \put(48,54)      {$ = $}
  \put(150,54)     {$ = $}
  \epicture22 \labl{fig:linform2}
Thus indeed $\Lambda$ is non-degenerate also on these subspaces.
As a consequence, given a basis $\{e_{\alpha}\}$ of $\Hom^{(P)}(M_r^{\vee}\Oti\, 
M_l,U_{\bar{k}})$, we can choose a dual basis $\{\bar e_{\beta}\}$ of
$\Hom^{(P)}(U_{\bar{k}},M_r^{\vee}\Oti\, M_l)$, such
that $\Lambda(e_{\alpha},\bar{e}_{\beta}) \eq \delta_{\alpha,\beta}$.
Define the invertible square matrix $L \,{\equiv}\, L(M_l,M_r,k)$ by
  \be
  f_{\beta} =: \sum_{\gamma}L_{\gamma\beta}\,\bar e_{\gamma} \,.
  \labl{eq:bastr}
Let now $\{\psi^+_\alpha\}$ and $\{\psi^-_\beta\}$ denote bases of 
$\HomA(M_l\oti U_k,M_r)$ and $\HomA(M_r\oti U_{\bar{k}},M_l)$, respectively. 
Using the isomorphisms \erf{fig:modisom1} and \erf{fig:fdef}, we then get
  \bea \begin{picture}(285,69)(0,38)
  \put(145,0)      {\Includeourbeautifulpicture 103 
    \setlength{\unitlength}{.38pt}
    \put(-8,133)       {\scriptsize $M_r^\vee$}
    \put(58,135)       {\scriptsize $M_l$}
    \setlength{\unitlength}{1pt}
    }
  \put(8,54)       {$ \Cbnd_{M_l,M_r,k,\psi^+_{\alpha},\psi^-_{\beta}}
                      \;=~ \dsty \sum_{\gamma}L_{\gamma\beta} $}
  \put(207,54)     {$ =\; \dim(U_k)\, L_{\alpha\beta} \,. $}
  \put(157,77.7)   {\footnotesize $e_{\alpha}$}      
  \put(157,30)     {\footnotesize $\bar e_{\gamma}$}
  \put(190,30)     {\scriptsize $U_{\bar{k}}$}
  \epicture17 \labl{fig:Dstrconst}
In particular, the matrix $\Cbnd_{M_l,M_r,k}$ is invertible.

\subsection{The two-point correlator on the sphere} \label{sec:2ptS2}

$~$

\noindent
Here we present the ribbon graph whose invariant gives
the structure constants for the two-point correlator of bulk fields on the
sphere $S^2$. We start with the world sheet
  \be
  X = S \equiv S(i,j,\phi^+_{\alpha},\phi^-_{\beta})
  \ee
as defined in section \ref{sec:facbulk}. Its double $\hat{S}$ is
the disconnected sum of two spheres with two marked arcs each, differing in
the labels of marked arcs and surface orientation. The identification of
$\hat{S}$ with $S^2_{(2)}\,{\sqcup}\, S^2_{(2)}$ uses the map
$(x,y,z)\,{\mapsto}\, (x,-y,z)$ on one of the components.
The connecting manifold ${\rm M}_{S}$ is embedded in $\mathbb{R}^3$
as a cylinder over $S^2_{(2)}$ centered at the origin. The parameter $t$ in
the construction of the connecting manifold is then identified with the
distance from the origin, such that the boundary component carrying the marked 
points $(U_j,+)$ and $(U_{\jb},+)$ is located at $t\eq 1$, while the component
carrying $(U_i,+)$ and $(U_{\ib},+)$ resides at $t\eq 2$. Straightening a part 
of the spherical shell to the plane, the connecting manifold is illustrated as
  \begin{eqnarray} \begin{picture}(420,460)(20,0)
  \put(0,0)       {\Includeournicelargepicture 110 }
  \put(85,350)    {\scriptsize $(U_{\ib},+)$}
  \put(87,40)     {\scriptsize $(U_{\jb},+)$}
  \put(325,400)   {\scriptsize $(U_i,+)$}
  \put(326,90)    {\scriptsize $(U_j,+)$}
  \put(111,202)   {\scriptsize \begin{turn}{305} $\phi^-_\beta$\end{turn}}
  \put(350,260)   {\scriptsize \begin{turn}{305} $\phi^+_\alpha$\end{turn}}
  \put(32,361)    {\tiny $1$}
  \put(28,346)    {\tiny $2$}
  \put(89,13)     {\tiny $1$}
  \put(67,20)     {\tiny $2$}
  \end{picture} \nonumber\\[-.6em]
  \label{fig:MS2} {} \end{eqnarray}
To insert $A$-ribbons in the embedded world sheet, we have chosen a
triangulation in the following way. One edge runs along the equator of
$\imath(S)$. Starting at the equator, another edge runs along a great
circle through the coupon labelled by $\phi^+_{\alpha}$ and continues until
it hits the equator again. Similarly, for an edge running through the other
coupon, but the start and end points on the equator are translated a small
distance, so as to avoid four-valent vertices. This triangulation is
covered by $A$-ribbons. In the picture, the two loose ends of $A$-ribbons
are identified.
A neater picture is obtained by using the specialness, symmetry and
Frobenius properties of $A$, together with the intertwining properties of
the bimodule morphisms; this allows us to remove a large portion of the
covered triangulation. To simplify the picture further, the
identity~\erf{fig:prm7} is used. These manipulations result in
  \bea \begin{picture}(370,310)(0,20)
  \put(0,0)       {\Includeournicemediumpicture 111 }
  \put(100,37)	{\scriptsize $(U_\jb,+)$}
  \put(107,291)	{\scriptsize $(U_\ib,+)$}
  \put(230,37)	{\scriptsize $(U_j,+)$}
  \put(237,291)	{\scriptsize $(U_i,+)$}
  \put(99,175)	{\scriptsize \begin{turn}{285} $\phi^-_\beta$\end{turn}}
  \put(232,174)	{\scriptsize \begin{turn}{285} $\phi^+_\alpha$\end{turn}}
  \put(15,171)	{\scriptsize \begin{turn}{285} $\eta$\end{turn}}
  \put(299,170)	{\scriptsize \begin{turn}{285} $\eps$\end{turn}}
  \put(23,306)	{\tiny $2$}
  \put(30,320)	{\tiny $1$}
  \put(57,24)	{\tiny $2$}
  \put(84,8)	{\tiny $1$}
  \epicture09 \labl{fig:improved}
To determine the structure constants $(\Cbulk_{i,j})_{\alpha\beta}$
in \erf{eq:S2sc}, we compose the cobordism ${\rm M}_{S}$ with 
$\overline B^+_{i\ib}\,{\sqcup}\,\overline B^+_{j\jb}$.
The resulting cobordism is
  \bea \begin{picture}(90,77)(0,32)
  \put(0,0)       {\Includeourbeautifulpicture 88h }
  \put(6,-2)      {\tiny $\bar{\lambda}^{i\ib}$}
  \put(55,-7)     {\tiny $\bar{\lambda}^{j\jb}$}
  \put(22,73)     {\scriptsize $i$}
  \put(43,73)     {\scriptsize $j$}
  \put(18,10)     {\scriptsize $\ib$}
  \put(66,6)      {\scriptsize $\jb$}
  \put(30,85)     {\scriptsize $\phi_{\alpha}^+$} 
  \put(31,39)     {\scriptsize $\phi_{\beta}^-$} 
  \put(70,100)	  {\Stnbox}
  \put(-36,45)    {$\dsty\frac{1}{S_{0,0}}$}
  \epicture17 \labl{fig:2ptgraph2}
Applying the same procedure also on the right hand side of~\erf{eq:S2sc},
we arrive at the result
  \bea \begin{picture}(185,78)(0,32)
  \put(93,0)      {\Includeourbeautifulpicture 88h }
  \put(99,-2)     {\tiny $\bar{\lambda}^{i\ib}$}
  \put(147,-7)    {\tiny $\bar{\lambda}^{j\jb}$}
  \put(115,73)    {\scriptsize $i$}
  \put(137,73)    {\scriptsize $j$}
  \put(112,12)    {\scriptsize $\ib$}
  \put(161,7)     {\scriptsize $\jb$}
  \put(123,85.5)  {\scriptsize $\phi_{\alpha}^+$}
  \put(123,39)    {\scriptsize $\phi_{\beta}^-$}
  \put(9,50)      {$\dsty \Cbulk_{ij,\alpha\beta} \;=~\frac 1{S_{0,0}} $}
  \put(163,90)    {\Stnbox} 
  \epicture13 \labl{fig:S2strc}
for the structure constants of the bulk two-point correlator.

\dtlLemma{lem:cbulk-nondeg}
The matrix $\Cbulk_{i,j}$ is non-degenerate.
\ENDL

\Proof
The proof is similar to the proof of non-degeneracy for the two-point
correlator on the disk. First, by setting
  \be
  {\rm tr}(\bar{\psi}\circ\varphi) =: \Lambda(\bar{\psi},\varphi)\, \dim(A)
  \labl{eq:nondegpair}
for any $\bar{\psi}\iN \Hom(U\oti A\oti V,A)$ and $\varphi\iN\Hom(A,U\oti A\oti
V)$, we obtain a non-degenerate pairing $\Lambda$ of the vector spaces
$\Hom(A,U\oti A\oti V)$ and $\Hom(U\oti A\oti V,A)$. To see that the
restriction of $\Lambda$ to the subspaces $\HomAA(U\oT+A\ot-V,A)$ and
$\HomAA(A,U\oT+A\ot-V)$ is non-degenerate as well, take $\varphi$ to be a 
bimodule morphism and consider the pairing
with an arbitrary $\bar\psi$. The bimodule property allows for the insertion
of a projector, and taking the trace of the composed morphisms allows to move
the projector to the other morphism. This is shown in the following picture:
  \begin{eqnarray} \begin{picture}(390,198)(0,0)
  \put(0,0)        {\INcludeourbeautifulpicture 114a }
  \put(103,0)      {\INcludeourbeautifulpicture 114b }
  \put(253,0)      {\INcludeourbeautifulpicture 114c }
  \put(70,91)      {$ = $}
  \put(221,91)     {$ = $}
  \end{picture} \nonumber\\
                   \begin{picture}(390,222)(0,0)
  \put(77,0)       {\INcludeourbeautifulpicture 114d }
  \put(277,0)      {\INcludeourbeautifulpicture 114e }
  \put(48,91)      {$ = $}
  \put(241,91)     {$ = $}
  \put(-6,320)     {\scriptsize $U$}
  \put(21,320)     {\scriptsize $V$}
  \put(278,330)    {\scriptsize $U$}
  \put(334,330)    {\scriptsize $V$}
  \put(104,100)    {\scriptsize $U$}
  \put(159,100)    {\scriptsize $V$}
  \put(284,90)     {\scriptsize $V$}
  \put(340,90)     {\scriptsize $U$}
  \put(8,275.5)    {\footnotesize $\phi$}
  \put(8,359)      {\footnotesize $\bar{\psi}$}
  \put(134.5,319)  {\footnotesize $\phi$}
  \put(134.5,381.5){\footnotesize $\bar{\psi}$}
  \put(306.5,296.5){\footnotesize $\phi$}
  \put(306.5,359.5){\footnotesize $\bar{\psi}$}
  \put(131.6,66)   {\footnotesize $\phi$}
  \put(131.5,129)  {\footnotesize $\bar{\psi}$}
  \put(312,59)     {\footnotesize $\phi$}
  \put(312,121.5)  {\footnotesize $\bar{\psi}$}
  \end{picture}  \nonumber\\[-1.5em]
  {} \label{fig:nondegp2} \end{eqnarray}
Thus indeed $\Lambda$ restricts non-degenerately to the bimodule morphisms.
By using the identities \erf{V11} and \erf{fig:prm2}, together with 
associativity, specialness and symmetry, it follows easily that 
${\rm tr}(\bar{\psi}\circ\varphi) \eq \eps\circ\bar\psi\cir\varphi\cir\eta$ 
when $\varphi$ and $\bar{\psi}$ are bimodule morphisms. This relates $\Lambda$ 
to the graph \erf{fig:S2strc}, and hence $\Cbulk_{i,j}$ is non-degenerate.
\qed


\sect{Some properties of (bimodule) morphisms} \label{sec:morphprop}

Here we list a few useful properties of some morphisms -- mainly morphisms of
$A$-bimodules -- that appear in the proof of bulk factorisation and in the
discussion of the two-point correlator on $S^2$.

First, the following results follow immediately by deformation of the graphs:
  \bea \begin{picture}(220,83)(0,22)
  \put(0,0)        {\Includeourbeautifulpicture 79a }
  \put(120,10)     {\Includeourbeautifulpicture 79b }
  \put(84,47)      {$ = $}
  \put(170,47)     {$ =~1 $}
  \put(15,87)	 {\scriptsize $U_i$}
  \put(32,87)	 {\scriptsize $U_{\ib}$}
  \put(111,43)	 {\scriptsize $U_i$}
  \put(139,43)	 {\scriptsize $U_{\ib}$}
  \epicture03 \labl{fig:cid2a}
and
  \bea \begin{picture}(260,88)(0,22)
  \put(0,0)        {\Includeourbeautifulpicture 80a }
  \put(120,10)     {\Includeourbeautifulpicture 80b }
  \put(84,48)      {$ = $}
  \put(170,48)     {$ =~ \theta_i^{-1} \circ \theta_i^{} ~=~1 \,. $}
   \put(111,20)    {\scriptsize $U_i$}
   \put(137,40)    {\scriptsize $U_\ib$}
   \put(11,15)    {\scriptsize $U_\ib$}
   \put(31,15)    {\scriptsize $U_i$}
   \put(11,83)    {\scriptsize $U_\ib$}
   \put(31,83)    {\scriptsize $U_i$}
  \epicture08 \labl{fig:cid2c}

Next we state

\dtlLemma{lem:A3-1}
Every $\phi\iN\HomAA(U\oT+A\ot-V,A)$ satisfies\\[-.4em]~
  \be \hsp{-.2}\begin{array}{rl}
  ({\rm i}) & \phi
  = m\circ (\phi\oti\id_A)\circ (\id_U\oti\eta\oti c_{A,V}^{})
  \,, \\{}\\[-.6em]
  ({\rm ii}) & \phi
  = m\circ (\id_A\oti\phi)\circ (c_{U,A}^{}\oti\eta\oti\id_V)
  \,, \\{}\\[-.6em]
  ({\rm iii}) & \phi
  = ((\eps\cir \phi\cir(\id_U\oti m\oti\id_V))\oti\id_A) \circ
    (\id_{U\oti A}\oti c^{-1}_{A,V}\oti\id_A) \circ
    (\id_{U\oti A\oti V}\oti (\Delta\cir\eta))
  \,, \\{}\\[-.6em]
  ({\rm iv}) & \phi
  = (\id_A\oti ((\eps\cir\phi\cir(\id_U\oti m\oti\id_V)) \circ
    (\id_A\oti c^{-1}_{U,A}\oti \id_{A\oti V}) \circ
    ((\Delta\cir\eta)\oti\id_{U\oti A\oti V})
  \,.  \\[-.7em]~
  \eear \labl{eq:prm1}
Analogous properties hold for $\psi\iN\HomAA(A,U\oT+ A\ot- V)$.
\ENDL

\Proof
({\rm i,\,ii}): Using the defining properties of $\eta$ and the intertwining
properties of $\phi$ we have
  \bea \begin{picture}(420,70)(0,25)
  \put(0,0)          {\Includeourbeautifulpicture 10a }
  \put(90,0)         {\Includeourbeautifulpicture 10b }
  \put(187,0)        {\Includeourbeautifulpicture 10c }
  \put(270,0)        {\Includeourbeautifulpicture 10d }
  \put(365,0)        {\Includeourbeautifulpicture 10e }
  \put(-2,-7.1)      {\scriptsize $U$}
  \put(8,-8.9)       {\scriptsize $A$}
  \put(19,-7.1)      {\scriptsize $V$}
  \put(8.7,94.4)     {\scriptsize $A$}
  \put(8,50.5)         {\footnotesize $\phi$}
  \put(95,-7.1)      {\scriptsize $U$}
  \put(105,-8.9)     {\scriptsize $A$}
  \put(116,-7.1)     {\scriptsize $V$}
  \put(105.7,94.4)   {\scriptsize $A$}
  \put(105.5,61.5)       {\footnotesize $\phi$}
  \put(185,-7.1)     {\scriptsize $U$}
  \put(195,-8.9)     {\scriptsize $A$}
  \put(206,-7.1)     {\scriptsize $V$}
  \put(195.7,94.4)   {\scriptsize $A$}
  \put(195.7,50)       {\footnotesize $\phi$}
  \put(275,-7.1)     {\scriptsize $U$}
  \put(285,-8.9)     {\scriptsize $A$}
  \put(296,-7.1)     {\scriptsize $V$}
  \put(285.7,94.4)   {\scriptsize $A$}
  \put(285.5,61)       {\footnotesize $\phi$}
  \put(374,-7.1)     {\scriptsize $U$}
  \put(383,-8.9)     {\scriptsize $A$}
  \put(395,-7.1)     {\scriptsize $V$}
  \put(383.7,94.4)   {\scriptsize $A$}
  \put(384,50.5)       {\footnotesize $\phi$}
  \put(58,42)        {$ = $}
  \put(150,42)       {$ = $}
  \put(238,42)       {$ = $}
  \put(333,42)       {$ = $}
  \epicture15 \labl{fig:prm1}
({\rm iii,\,iv}): Similarly, with the help of the counit and Frobenius
properties one gets
  \bea \begin{picture}(420,75)(0,26)
  \put(0,0)          {\Includeourbeautifulpicture 11a }
  \put(93,0)         {\Includeourbeautifulpicture 11b }
  \put(187,0)        {\Includeourbeautifulpicture 11c }
  \put(275,0)        {\Includeourbeautifulpicture 11d }
  \put(365,0)        {\Includeourbeautifulpicture 11e }
  \put(63,44)        {$ = $}
  \put(150,44)       {$ = $}
  \put(240,44)       {$ = $}
  \put(332,44)       {$ = $}
  \put(0,-9)         {\scriptsize $U$}
  \put(8,-9)         {\scriptsize $A$}
  \put(18,-9)        {\scriptsize $V$}
  \put(9,100)        {\scriptsize $A$}
  \put(10,51.5)         {\footnotesize $\phi$}
  \put(90,-9)        {\scriptsize $U$}
  \put(99,-9)        {\scriptsize $A$}
  \put(109.9,-9)     {\scriptsize $V$}
  \put(101,100)      {\scriptsize $A$}
  \put(101.5,24)       {\footnotesize $\phi$}
  \put(185,-9)       {\scriptsize $U$}
  \put(193,-9)       {\scriptsize $A$}
  \put(203,-9)       {\scriptsize $V$}
  \put(195,100)      {\scriptsize $A$}
  \put(195.5,49.5)       {\footnotesize $\phi$}
  \put(284.5,-9)     {\scriptsize $U$}
  \put(293,-9)       {\scriptsize $A$}
  \put(304,-9)       {\scriptsize $V$}
  \put(294,100)      {\scriptsize $A$}
  \put(295,24)       {\footnotesize $\phi$}
  \put(375,-9)       {\scriptsize $U$}
  \put(383,-9)       {\scriptsize $A$}
  \put(394,-9)       {\scriptsize $V$}
  \put(385,100)      {\scriptsize $A$}
  \put(385.5,51)       {\footnotesize $\phi$}
  \epicture16 \labp 11
The corresponding relations for $\psi\iN\HomAA(A,U\oti^+ A\oti^- V)$ follow 
analogously:
  \bea \begin{picture}(420,79)(0,26)
  \put(0,0)          {\Includeourbeautifulpicture 12a }
  \put(98,0)         {\Includeourbeautifulpicture 12b }
  \put(187,0)        {\Includeourbeautifulpicture 12c }
  \put(278,0)        {\Includeourbeautifulpicture 12d }
  \put(365,0)        {\Includeourbeautifulpicture 12e }
  \put(2,100)        {\scriptsize $U$}
  \put(10,100)       {\scriptsize $A$}
  \put(19,100)       {\scriptsize $V$}
  \put(8,-9)         {\scriptsize $A$}
  \put(9.5,38)       {\footnotesize $\psi$}
  \put(98,100)       {\scriptsize $U$}
  \put(106,100)      {\scriptsize $A$}
  \put(115,100)      {\scriptsize $V$}
  \put(105,-9)       {\scriptsize $A$}
  \put(106,63)       {\footnotesize $\psi$}
  \put(187,100)      {\scriptsize $U$}
  \put(195,100)      {\scriptsize $A$}
  \put(204,100)      {\scriptsize $V$}
  \put(194,-9)       {\scriptsize $A$}
  \put(195.5,43.5)   {\footnotesize $\psi$}
  \put(278,100)      {\scriptsize $U$}
  \put(286,100)      {\scriptsize $A$}
  \put(295,100)      {\scriptsize $V$}
  \put(285,-9)       {\scriptsize $A$}
  \put(286,63)       {\footnotesize $\psi$}
  \put(375,100)      {\scriptsize $U$}
  \put(383,100)      {\scriptsize $A$}
  \put(392,100)      {\scriptsize $V$}
  \put(382,-9)       {\scriptsize $A$}
  \put(383,38.5)     {\footnotesize $\psi$}
  \put(63,43)        {$ = $}
  \put(150,43)       {$ = $}
  \put(243,43)       {$ = $}
  \put(333,43)       {$ = $}
  \epicture07 \labl{fig:prm2}
and
  \bea \begin{picture}(420,79)(0,26)
  \put(0,0)          {\Includeourbeautifulpicture 13a }
  \put(86,0)         {\Includeourbeautifulpicture 13b }
  \put(187,0)        {\Includeourbeautifulpicture 13c }
  \put(270,0)        {\Includeourbeautifulpicture 13d }
  \put(365,0)        {\Includeourbeautifulpicture 13e }
  \put(58,44)        {$ = $}
  \put(152,44)       {$ = $}
  \put(239,44)       {$ = $}
  \put(338,44)       {$ = $}
  \put(0,100)        {\scriptsize $U$}
  \put(8,100)        {\scriptsize $A$}
  \put(17,100)       {\scriptsize $V$}
  \put(7,-9)         {\scriptsize $A$}
  \put(8,42.5)       {\footnotesize $\psi$}
  \put(88,100)       {\scriptsize $U$}
  \put(96,100)       {\scriptsize $A$}
  \put(105,100)      {\scriptsize $V$}
  \put(96,-9)        {\scriptsize $A$}
  \put(96,16.5)      {\footnotesize $\psi$}
  \put(187,100)      {\scriptsize $U$}
  \put(195,100)      {\scriptsize $A$}
  \put(204,100)      {\scriptsize $V$}
  \put(194,-9)       {\scriptsize $A$}
  \put(195,43.5)     {\footnotesize $\psi$}
  \put(289,100)      {\scriptsize $U$}
  \put(297,100)      {\scriptsize $A$}
  \put(306,100)      {\scriptsize $V$}
  \put(296,-9)       {\scriptsize $A$}
  \put(297,16.5)     {\footnotesize $\psi$}
  \put(376,100)      {\scriptsize $U$}
  \put(384,100)      {\scriptsize $A$}
  \put(394,100)      {\scriptsize $V$}
  \put(384,-9)       {\scriptsize $A$}
  \put(385,42.5)     {\footnotesize $\psi$}
  \epicture02 \labp 13
\mbox{$\quad $}~
\qed

Applying these identities to coupons in the connecting manifold that are
labelled by the respective bimodule morphisms, one can effectively move
$A$-ribbons from one side of the coupon to the other and thereby
substantially reduce the total number of ribbons in the connecting manifold
(not counting the resulting `short' $A$-ribbons that directly connect a
unit or co-unit coupon to coupons for bimodule morphisms).

For convenience, we also list a few special consequences of lemma
\ref{lem:A3-1}. Their proofs are elementary, involving essentially the same 
moves as the proof of the lemma, combined with the symmetry of $A$.
  \bea \begin{picture}(420,70)(3,22)
  \put(0,7)          {\Includeourbeautifulpicture 14a }
  \put(73,7)         {\Includeourbeautifulpicture 14b }
  \put(150,0)        {\Includeourbeautifulpicture 14c }
  \put(267,0)        {\Includeourbeautifulpicture 14d }
  \put(357,0)        {\Includeourbeautifulpicture 14e }
  \put(0,-2)         {\scriptsize $U$}
  \put(8,-2)         {\scriptsize $A$}
  \put(17,-2)        {\scriptsize $V$}
  \put(25,-2)        {\scriptsize $A$}
  \put(10.5,65.5)    {\footnotesize $\phi$}
  \put(71,-2)        {\scriptsize $U$}
  \put(79,-2)        {\scriptsize $A$}
  \put(88,-2)        {\scriptsize $V$}
  \put(96,-2)        {\scriptsize $A$}
  \put(81.5,34.5)    {\footnotesize $\phi$}
  \put(180,-9)       {\scriptsize $U$}
  \put(188,-9)       {\scriptsize $A$}
  \put(197,-9)       {\scriptsize $V$}
  \put(205,-9)       {\scriptsize $A$}
  \put(190.5,22.5)   {\footnotesize $\phi$}
  \put(274,-9)       {\scriptsize $U$}
  \put(282,-9)       {\scriptsize $A$}
  \put(292,-9)       {\scriptsize $V$}
  \put(300,-9)       {\scriptsize $A$}
  \put(284.5,22.5)   {\footnotesize $\phi$}
  \put(376,-9)       {\scriptsize $U$}
  \put(384,-9)       {\scriptsize $A$}
  \put(394,-9)       {\scriptsize $V$}
  \put(401,-9)       {\scriptsize $A$}
  \put(386.5,58)     {\footnotesize $\phi$}
  \put(48,44)        {$ = $}
  \put(125,44)       {$ = $}
  \put(242,44)       {$ = $}
  \put(332,44)       {$ = $}
  \epicture07 \labp 14
%
  \bea \begin{picture}(368,71)(0,26)
  \put(0,7)          {\Includeourbeautifulpicture 15a }
  \put(83,7)         {\Includeourbeautifulpicture 15b }
  \put(168,0)        {\Includeourbeautifulpicture 15c }
  \put(273,0)        {\Includeourbeautifulpicture 15d }
  \put(4,-2)         {\scriptsize $U$}
  \put(11,-2)        {\scriptsize $A$}
  \put(21,-2)        {\scriptsize $V$}
  \put(-6,-2)        {\scriptsize $A$}
  \put(13,65)        {\footnotesize $\phi$}
  \put(91,-2)        {\scriptsize $U$}
  \put(98,-2)        {\scriptsize $A$}
  \put(108,-2)       {\scriptsize $V$}
  \put(81,-2)        {\footnotesize $A$}
  \put(101,34.5)     {\footnotesize $\phi$}
  \put(182,-9)       {\scriptsize $U$}
  \put(189,-9)       {\scriptsize $A$}
  \put(199,-9)       {\scriptsize $V$}
  \put(172,-9)       {\scriptsize $A$}
  \put(192,22.5)     {\footnotesize $\phi$}
  \put(287,-9)       {\scriptsize $U$}
  \put(294,-9)       {\scriptsize $A$}
  \put(304,-9)       {\scriptsize $V$}
  \put(277,-9)       {\scriptsize $A$}
  \put(297,58)       {\footnotesize $\phi$}
  \put(50,44)        {$ = $}
  \put(135,44)       {$ = $}
  \put(240,44)       {$ = $}
  \epicture10 \labp 15
%
  \bea \begin{picture}(355,111)(0,16)
  \put(0,0)          {\Includeourbeautifulpicture 16a }
  \put(80,0)         {\Includeourbeautifulpicture 16b }
  \put(168,0)        {\Includeourbeautifulpicture 16c }
  \put(283,0)        {\Includeourbeautifulpicture 16d }
  \put(49,52)        {$ = $}
  \put(139,52)       {$ = $}
  \put(255,52)       {$ = $}
  \put(1,112)        {\scriptsize $U$}
  \put(22,-11)       {\scriptsize $A$}
  \put(17,112)       {\scriptsize $V$}
  \put(6,-11)        {\scriptsize $A$}
  \put(8.5,57.5)     {\footnotesize $\psi$}
  \put(86,112)       {\scriptsize $U$}
  \put(92,-11)       {\scriptsize $A$}
  \put(102,112)      {\scriptsize $V$}
  \put(108,-11)      {\scriptsize $A$}
  \put(94,31.5)      {\footnotesize $\psi$}
  \put(187,112)      {\scriptsize $U$}
  \put(193,-11)      {\scriptsize $A$}
  \put(203,112)      {\scriptsize $V$}
  \put(209,-11)      {\scriptsize $A$}
  \put(194.7,18.8)   {\footnotesize $\psi$}
  \put(305,112)      {\scriptsize $U$}
  \put(309,-11)      {\scriptsize $A$}
  \put(320,112)      {\scriptsize $V$}
  \put(326,-11)      {\scriptsize $A$}
  \put(312,50)       {\footnotesize $\psi$}
  \epicture10 \labp 16
%
  \bea \begin{picture}(350,105)(0,16)
  \put(0,0)          {\Includeourbeautifulpicture 17a }
  \put(80,0)         {\Includeourbeautifulpicture 17b }
  \put(168,0)        {\Includeourbeautifulpicture 17c }
  \put(283,0)        {\Includeourbeautifulpicture 17d }
  \put(49,52)        {$ = $}
  \put(139,52)       {$ = $}
  \put(255,52)       {$ = $}
  \put(5,112)        {\scriptsize $U$}
  \put(11,-11)       {\scriptsize $A$}
  \put(21,112)       {\scriptsize $V$}
  \put(-6,-11)       {\scriptsize $A$}
  \put(12.5,57.5)    {\footnotesize $\psi$}
  \put(90,112)       {\scriptsize $U$}
  \put(79,-11)       {\scriptsize $A$}
  \put(106,112)      {\scriptsize $V$}
  \put(96,-11)       {\scriptsize $A$}
  \put(97,31.5)      {\footnotesize $\psi$}
  \put(194,112)      {\scriptsize $U$}
  \put(184,-11)      {\scriptsize $A$}
  \put(210,112)      {\scriptsize $V$}
  \put(201,-11)      {\scriptsize $A$}
  \put(202.5,19)     {\footnotesize $\psi$}
  \put(298,112)      {\scriptsize $U$}
  \put(288,-11)      {\scriptsize $A$}
  \put(314,112)      {\scriptsize $V$}
  \put(305,-11)      {\scriptsize $A$}
  \put(306.8,50)       {\footnotesize $\psi$}
  \epicture10 \labp 17
%
  \bea \begin{picture}(355,75)(0,21)
  \put(0,0)          {\Includeourbeautifulpicture 18a }
  \put(84,0)         {\Includeourbeautifulpicture 18b }
  \put(172,0)        {\Includeourbeautifulpicture 18c }
  \put(283,0)        {\Includeourbeautifulpicture 18d }
  \put(56,41)        {$ = $}
  \put(143,41)       {$ = $}
  \put(244,41)       {$ = $}
  \put(0,-11)        {\scriptsize $U$}
  \put(8,90)         {\scriptsize $A$}
  \put(14,-11)       {\scriptsize $V$}
  \put(22,-11)       {\scriptsize $A$}
  \put(8.5,34)         {\footnotesize $\phi$}
  \put(90,-11)       {\scriptsize $U$}
  \put(98,90)        {\scriptsize $A$}
  \put(103,-11)      {\scriptsize $V$}
  \put(113,-11)      {\scriptsize $A$}
  \put(98.2,60)        {\footnotesize $\phi$}
  \put(177,-11)      {\scriptsize $U$}
  \put(186,90)       {\scriptsize $A$}
  \put(191,-11)      {\scriptsize $V$}
  \put(201,-11)      {\scriptsize $A$}
  \put(186.5,62.5)   {\footnotesize $\phi$}
  \put(292,-11)	     {\scriptsize $U$}
  \put(300,90)	     {\scriptsize $A$}
  \put(306,-11)	     {\scriptsize $V$}
  \put(316,-11)	     {\scriptsize $A$}
  \put(301,41.5)     {\footnotesize $\phi$}
  \epicture13 \labp 18
%
  \bea \begin{picture}(350,88)(0,11)
  \put(0,0)          {\Includeourbeautifulpicture 19a }
  \put(88,0)         {\Includeourbeautifulpicture 19b }
  \put(180,0)        {\Includeourbeautifulpicture 19c }
  \put(275,0)        {\Includeourbeautifulpicture 19d }
  \put(58,39)        {$ = $}
  \put(149,39)       {$ = $}
  \put(241,39)       {$ = $}
  \put(20.5,34)    {\footnotesize $\phi$}
  \put(106,60.5)   {\footnotesize $\phi$}
  \put(194.5,63)   {\footnotesize $\phi$}
  \put(289,41.5)   {\footnotesize $\phi$}
  \put(20,90)  	   {\scriptsize $A$}
  \put(2.5,-9)     {\scriptsize $A$}
  \put(13,-9)	   {\scriptsize $U$}
  \put(27,-9)	   {\scriptsize $V$}
  \put(106,90)	   {\scriptsize $A$}
  \put(88.5,-9)	   {\scriptsize $A$}
  \put(98.5,-9)	   {\scriptsize $U$}
  \put(113,-9)	   {\scriptsize $V$}
  \put(195,90)	   {\scriptsize $A$}
  \put(177,-9)	   {\scriptsize $A$}
  \put(187.5,-9)   {\scriptsize $U$}
  \put(202,-9)	   {\scriptsize $V$}
  \put(289,90)	   {\scriptsize $A$}
  \put(271,-9)	   {\scriptsize $A$}
  \put(281.5,-9)   {\scriptsize $U$}
  \put(296,-9)	   {\scriptsize $V$}
  \epicture07 \labp 19
%
  \bea \begin{picture}(350,89)(0,11)
  \put(0,0)          {\Includeourbeautifulpicture 21a }
  \put(94,0)         {\Includeourbeautifulpicture 21b }
  \put(180,0)        {\Includeourbeautifulpicture 21c }
  \put(267,0)        {\Includeourbeautifulpicture 21d }
  \put(61,39)        {$ = $}
  \put(148,39)       {$ = $}
  \put(233,39)       {$ = $}
  \put(15,23.5)	   {\footnotesize $\phi$}
  \put(103,52.5)   {\footnotesize $\phi$}
  \put(189.5,57)   {\footnotesize $\phi$}
  \put(285.5,37)   {\footnotesize $\phi$}
  \put(23,90)	   {\scriptsize $V$}
  \put(7,90)	   {\scriptsize $U$}
  \put(15,90)	   {\scriptsize $A$}
  \put(29.5,-9)	   {\scriptsize $A$}
  \put(112,90)	   {\scriptsize $V$}
  \put(95,90)	   {\scriptsize $U$}
  \put(103,90)	   {\scriptsize $A$}
  \put(118,-9)	   {\scriptsize $A$}
  \put(198,90)	   {\scriptsize $V$}
  \put(182,90)	   {\scriptsize $U$}
  \put(189,90)	   {\scriptsize $A$}
  \put(203.5,-9)   {\scriptsize $A$}
  \put(294,90)	   {\scriptsize $V$}
  \put(278,90)	   {\scriptsize $U$}
  \put(285,90)	   {\scriptsize $A$}
  \put(299.5,-9)   {\scriptsize $A$}
  \epicture07 \labp 21
%
  \bea \begin{picture}(350,85)(0,12)
  \put(0,0)          {\Includeourbeautifulpicture 22a }
  \put(94,0)         {\Includeourbeautifulpicture 22b }
  \put(180,0)        {\Includeourbeautifulpicture 22c }
  \put(267,0)        {\Includeourbeautifulpicture 22d }
  \put(61,39)        {$ = $}
  \put(148,39)       {$ = $}
  \put(233,39)       {$ = $}
  \put(18,24)        {\footnotesize $\phi$}
  \put(107.5,52.5)   {\footnotesize $\phi$}
  \put(193,57)       {\footnotesize $\phi$}
  \put(280.5,37.3)   {\footnotesize $\phi$}
  \put(26.5,90)	     {\scriptsize $V$}
  \put(11,90)	     {\scriptsize $U$}
  \put(17.5,90)	     {\scriptsize $A$}
  \put(1,-9)         {\scriptsize $A$}
  \put(116,90)       {\scriptsize $V$}
  \put(100,90)	     {\scriptsize $U$}
  \put(107,90)	     {\scriptsize $A$}
  \put(91,-9)	     {\scriptsize $A$}
  \put(202,90)	     {\scriptsize $V$}
  \put(186,90)	     {\scriptsize $U$}
  \put(193,90)	     {\scriptsize $A$}
  \put(176.5,-9)     {\scriptsize $A$}
  \put(289,90)	     {\scriptsize $V$}
  \put(273,90)	     {\scriptsize $U$}
  \put(280,90)	     {\scriptsize $A$}
  \put(263.5,-9)     {\scriptsize $A$}
  \epicture13 \labp 22

Recall from definition I:5.5 that a morphism $\phi\iN\Hom(A\oti U,V)$ is
called a {\em local\/} morphism iff $\phi\cir P_U\eq \phi$, with $P_U$ the 
idempotent defined in (I:5.34). Similarly, a morphism $\phi\iN\Hom(V,A\oti U)$ 
is called local iff $P_U \cir \phi\eq \phi$. Further, we can use the dualities 
of \C\ to obtain isomorphisms
  \bea
  \Hom(A\oti X,Y) \cong \Hom(Y^\vee\oti A\oti X,\one) \cong \Hom(A,Y\oti X^\vee)
  \quad{\rm and}\\[.3em]
  \Hom(Y,A\oti X) \cong \Hom(\one,Y^\vee\oti A\oti X) \cong \Hom(Y\oti X^\vee,A)
  \,. \eear
  \labl{eq:some-isos}
We refer to a morphism that via these isomorphisms is mapped to a local morphism
in $\Hom(A\oti X,Y)$ and $\Hom(Y,A\oti X)$ again as being local. Local morphisms 
satisfy properties analogous to those of bimodule morphisms given above (compare 
proposition I:5.7). Here we are interested in the following particular case.

\dtlLemma{lem:A3-2}
For $\phi_1,\,\phi_2,\,\phi_3,\,\phi_4$ arbitrary morphisms in
the vector spaces $\,\Hom(U\oti A\oti V,
    \linebreak[0]
\one)$, $\Hom(U\oti V,A)$,
$\Hom(A,U\oti V)$ and $\,\Hom(\one,U\oti A\oti V)$, respectively, the
following morphisms are local morphisms in the respective spaces:
  \bea \begin{picture}(420,87)(0,28)
  \put(0,0)          {\Includeourbeautifulpicture 23a }
  \put(115,0)        {\Includeourbeautifulpicture 23b }
  \put(230,0)        {\Includeourbeautifulpicture 23c }
  \put(335,0)        {\Includeourbeautifulpicture 23d }
  \put(33,82)        {\footnotesize $\phi_1$}
  \put(143.5,42)     {\footnotesize $\phi_2$}
  \put(261,68)       {\footnotesize $\phi_3$}
  \put(364.5,30)     {\footnotesize $\phi_4$}
  \put(40.5,-9)	     {\scriptsize $V$}
  \put(25,-9)	     {\scriptsize $U$}
  \put(32,-9)	     {\scriptsize $A$}
  \put(151,-9)	     {\scriptsize $V$}
  \put(136,-9)	     {\scriptsize $U$}
  \put(144,117)	     {\scriptsize $A$}
  \put(269,117)	     {\scriptsize $V$}
  \put(255,117)	     {\scriptsize $U$}
  \put(260,-9)	     {\scriptsize $A$}
  \put(372.5,117)    {\scriptsize $V$}
  \put(357,117)	     {\scriptsize $U$}
  \put(364,117)	     {\scriptsize $A$}
  \epicture18 \labl{fig:prm5}
\ENDL

\Proof
The statement is a straightforward consequence of the fact that the
morphisms corresponding to the closed $A$-loops in these pictures
are idempotents, just as $P_U$ in (I:5.34). The required manipulations
are very similar to those in the proof of proposition I:5.7, and we
refrain from spelling the out explicitly.
\qed

\dtlLemma{lem:A3-3}
For any pair of morphisms $\phi\iN\HomAA(U\oT+A\ot-V,A)$ and
$\psi\iN\HomAA(A,
     $\linebreak[0]$
U\oT+A\ot-V)$ we have
  \be
  \Tr \big( \phi\cir(\id_U\oti (\eta\cir\eps)\oti\id_V)\cir\psi \big)
  = \eps\circ\phi\circ\psi\circ\eta \,.
  \labl{eq:prm6}
Graphically,
  \bea \begin{picture}(155,90)(0,19)
  \put(0,0)          {\Includeourbeautifulpicture 24a }
  \put(97,0)         {\Includeourbeautifulpicture 24b }
  \put(65,53)        {$ = $}
  \put(7.5,77.5)     {\footnotesize $\phi$}
  \put(7.5,29.5)     {\footnotesize $\psi$}
  \put(-4,52)        {\scriptsize $U$}
  \put(19,52)        {\scriptsize $V$}
  \put(105.5,76.5)   {\footnotesize $\phi$}
  \put(105.5,26)     {\footnotesize $\psi$}
  \put(94,51)	     {\scriptsize $U$}
  \put(117,51)	     {\scriptsize $V$}
  \epicture-1 \labl{fig:prm7}
\ENDL

\Proof
The following equalities hold:
  \bea \begin{picture}(365,93)(0,42)
  \put(0,0)          {\Includeourbeautifulpicture 25a }
  \put(93,0)         {\Includeourbeautifulpicture 25b }
  \put(217,0)        {\Includeourbeautifulpicture 25c }
  \put(330,0)        {\Includeourbeautifulpicture 25d }
  \put(63,64)        {$ = $}
  \put(185,64)       {$ = $}
  \put(299,64)       {$ = $}
  \put(7.5,91)	     {\footnotesize $\phi$}
  \put(7.5,35.5)     {\footnotesize $\psi$}
  \put(-4,62)	     {\scriptsize $U$}
  \put(19,62)	     {\scriptsize $V$}
  \put(109.5,101)    {\footnotesize $\phi$}
  \put(109.5,23.5)   {\footnotesize $\psi$}
  \put(86,61)	     {\scriptsize $U$}
  \put(112,61)	     {\scriptsize $V$}
  \put(225.5,99)     {\footnotesize $\phi$}
  \put(225.5,25.5)   {\footnotesize $\psi$}
  \put(214,63)	     {\scriptsize $U$}
  \put(228,63)	     {\scriptsize $V$}
  \put(339,91)	     {\footnotesize $\phi$}
  \put(339,34.5)     {\footnotesize $\psi$}
  \put(327,62)	     {\scriptsize $U$}
  \put(350,62)	     {\scriptsize $V$}
  \epicture22 \labl{fig:prm8}
The first step follows from lemma \ref{lem:A3-1}(iii) and the first identity
in~\erf{fig:prm2}, the second step uses the unit property and symmetry,
and the final equality follows by the unit, counit and Frobenius properties.
\qed

\smallskip

Next we define two linear maps
  \be
  f{:}\quad \HomAA(U_i\oT+ A\ot- U_j,A)\To \HomAA(A,U_\ib\oT+ A\ot- U_\jb)
  \ee 
and 
  \be
  g{:}\quad \HomAA(A,U_\ib\oT+ A\ot- U_\jb)\To \HomAA(U_i\oT+ A\ot- U_j,A) 
  \ee
by 
  \bea \begin{picture}(225,105)(70,66)
  \put(0,0)		{\INcludeourbeautifulpicture 115a }
  \put(90,0)		{\INcludeourbeautifulpicture 115b }
  \put(210,0)		{\INcludeourbeautifulpicture 115c }
  \put(300,0)		{\INcludeourbeautifulpicture 115d }
  \put(55,100)          {$ := $}
  \put(265,100)	        {$ := $}
  \put(4,102)           {\scriptsize $f(\phi)$}
  \put(107,102)         {\scriptsize $\phi$}
  \put(214,101.5)       {\scriptsize $g(\bar \phi)$}
  \put(316,101.5)       {\scriptsize $\bar \phi$}
  \put(-3,160)          {\scriptsize $U_\ib$}
  \put(8,160)           {\scriptsize $A$}
  \put(7,45)            {\scriptsize $A$}
  \put(16,160)          {\scriptsize $U_\jb$}
  \put(106,160)         {\scriptsize $A$}
  \put(87,160)          {\scriptsize $U_\ib$}
  \put(120,160)         {\scriptsize $U_\jb$}
  \put(105,45)          {\scriptsize $A$}
  \put(218,160)         {\scriptsize $A$}
  \put(206,45)          {\scriptsize $U_i$}
  \put(227,45)          {\scriptsize $U_j$}
  \put(218,45)          {\scriptsize $A$}
  \put(314.8,160)       {\scriptsize $A$}
  \put(297,45)          {\scriptsize $U_i$}
  \put(331,45)          {\scriptsize $U_j$}
  \put(314,45)          {\scriptsize $A$}
  \epicture-3 \labl{pic:bimtr}

\dtlLemma{lem:A3-4}
The functions $f$ and $g$ satisfy
  \bea
  g\circ f = \Frac{1}{\dim(U_i)\dim(U_j)}\, \id_{\HomAA(U_i\oT+A\ot-U_j,A)}
  \qquad{\rm and} \\{}\\[-.4em]
  f\circ g = \Frac{1}{\dim(U_i)\dim(U_j)}\, \id_{\HomAA(A,U_\ib\oT+A\ot-U_\jb)}
  \,, \eear \ee
\ENDL

\Proof
The composition $g\circ f$ applied to a bimodule morphism $\phi$ gives

  \bea \begin{picture}(155,94)(0,66)
  \put(0,0)			{\INcludeourbeautifulpicture 115c }
  \put(110,0)		{\Includeourbeautifulpicture 116 }
  \put(65,100)		{$ = $}
  \put(2,102.5)		{\tiny $g\! \cir\!\! f(\phi)$}
  \put(133,102.5)         {\scriptsize $\phi$}
  \put(8,159)		{\scriptsize $A$}
  \put(-3,45)		{\scriptsize $U_i$}
  \put(17,45)		{\scriptsize $U_j$}
  \put(7.6,45)	        {\scriptsize $A$}
  \put(131,159)		{\scriptsize $A$}
  \put(107,45)		{\scriptsize $U_i$}
  \put(161,45)		{\scriptsize $U_j$}
  \put(131,45)		{\scriptsize $A$}
  \epicture08 \labl{pic:btpr1}
where the small coupons stand for the morphisms $\lambda_{i\ib}$
and $\bar\lambda^{i\ib}_{}$ (respectively $\lambda_{j\jb}$
and $\bar\lambda^{j\jb}_{}$) which were defined in \erf{fig:llbar}.
Since $U_i$ and $U_j$ are simple objects, we have 
  \bea \begin{picture}(185,86)(50,76)
  \put(0,0)		{\INcludeourbeautifulpicture 117a }
  \put(91,0)		{\INcludeourbeautifulpicture 117c }
  \put(200,0)		{\INcludeourbeautifulpicture 117b }
  \put(291,0)		{\INcludeourbeautifulpicture 117c }
  \put(50,115)		{$ =~~ \mu_i$}
  \put(139,115)		{and}
  \put(250,115)		{$ =~~ \nu_j$}
  \put(-3,72)		{\scriptsize $U_i$}
  \put(17,156)          {\scriptsize $U_i$}
  \put(88,72)		{\scriptsize $U_i$}
  \put(88,156)          {\scriptsize $U_i$}
  \put(219,72)          {\scriptsize $U_j$}
  \put(198,156)         {\scriptsize $U_j$}
  \put(287,72)          {\scriptsize $U_j$}
  \put(288,156)         {\scriptsize $U_j$}
  \epicture-3 \labl{pic:btpr2}
for some scalars $\mu_i$ and $\nu_j$. By taking the trace on both sides of 
the equalities, simple manipulations show $\mu_i\eq \dim(U_i)^{-1} \eq \nu_i$.
Similar manipulations give the result for $f\cir g$.
\qed

It is useful to express this lemma in terms of specific bases.
Let us select bases $\{\phi_\alpha^{ij}\}$ of $\HomAA(U_i\oT+ A\ot- U_j,A)$ and
$\{\bar\phi_{\alpha}^{ij}\}$ of $\HomAA(A,U_i\oT+ A\ot- U_j)$ that are dual in 
the sense of \erf{fig:dualbimorph}. Define matrices $\Delta$ and $\Omega$ by
  \be
  f(\phi_\alpha^{ij}) = \sum_{\gamma}\Delta_{\alpha\gamma}\,
  \bar\phi_{\gamma}^{\ib\jb},
  \qquad g(\bar\phi_\alpha^{\,\ib\jb}) = \sum_{\gamma}\Omega_{\alpha\gamma}\,
  \phi_\gamma^{ij}.
  \labl{eq:bimmatr}
In terms of these matrices, lemma \ref{lem:A3-4} takes the form
  \be
  \sum_{\gamma} \Delta_{\alpha\gamma}\,\Omega_{\gamma\beta}
  = \sum_{\gamma} \Omega_{\alpha\gamma}\,\Delta_{\gamma\beta}
  = \Frac{1}{\dim(U_i)\dim(U_j)}\,\delta_{\alpha,\beta},
  \labl{eq:bimtrmat}
thus showing that $\dim(U_i)\dim(U_j)\,\Omega \eq \Delta^{-1}$.

\vfill

\newcommand\wb{\,\linebreak[0]} \def\wB {$\,$\wb}
 \newcommand\Bi[1]    {\bibitem{#1}}
 \renewcommand\J[6]   {{\sl #6\/}, {#1} {#2} ({#3}) {#4--#5} }
 \newcommand\K[7]     {{\sl #7\/}, {#1} {#2} ({#3}) {#4--#5} {\tt[#6]} }
 \newcommand\BOOK[4]  {{\sl #1\/} ({#2}, {#3} {#4})}
 \def\jf    {J.\ Fuchs}
 \def\dim   {dimension}
 \def\adma  {Adv.\wb Math.}
 \def\comp  {Com\-mun.\wb Math.\wb Phys.}
 \def\cpma  {Com\-pos.\wb Math.}
 \def\gatm  {Geom.\wB and\wB Topol.\wb Monogr.}
 \def\jomp  {J.\wb Math.\wb Phys.}
 \def\nupb  {Nucl.\wb Phys.\ B}
 \def\phlb  {Phys.\wb Lett.\ B}
 \def\phrl  {Phys.\wb Rev.\wb Lett.}
 \def\topo  {Topology}
 \def\AMS    {{American Mathematical Society}}
 \def\AP     {{Academic Press}}
 \def\CUP    {{Cambridge University Press}}
 \def\IPC    {{International Press Company}}
 \def\KLU    {{Kluwer Academic Publishers}}
 \def\PL     {{Plenum Press}}
 \def\SV     {{Sprin\-ger Ver\-lag}}
 \def\WI     {{Wiley Interscience}}
 \def\WS     {{World Scientific}}
 \def\Be     {{Berlin}}
 \def\Ca     {{Cambridge}}
 \def\Do     {{Dordrecht}}
 \def\PR     {{Providence}}
 \def\Si     {{Singapore}}
 \def\NY     {{New York}}

\end{document}